\documentclass[12pt,preprint]{aastex}	

\usepackage{color}					
\usepackage{amsmath}
\definecolor{black}{rgb}{0,0,0}			
\definecolor{red}{rgb}{1,0,0}			%
\definecolor{blue}{rgb}{0,0,1}			%

\renewcommand{\aap}{Astron. Astrophys. }

\slugcomment{Last updated \today}

\shorttitle{Structure of Chariklo' rings}
\shortauthors{B\'erard et al.}

\begin{document}


\title{The structure of Chariklo's rings from stellar occultations}


\author{%
D. B\'erard\altaffilmark{1},
B. Sicardy\altaffilmark{1},
}%

\author{%
J. I. B. Camargo\altaffilmark{24,25},
J. Desmars\altaffilmark{1},
F. Braga-Ribas\altaffilmark{27,24,25},
}

\author{%
J.-L. Ortiz\altaffilmark{3},
R. Duffard\altaffilmark{3},
N. Morales\altaffilmark{3},
}

\author{%
E. Meza\altaffilmark{1},
R. Leiva\altaffilmark{1,6},
}%

\author{%
G. Benedetti-Rossi\altaffilmark{24,25},
R. Vieira-Martins\altaffilmark{4,23,24,25},
A.-R. Gomes J\'unior\altaffilmark{23},
M. Assafin\altaffilmark{23},  
}

\author{%
F. Colas\altaffilmark{4},
J.-L. Dauvergne\altaffilmark{41},
P. Kervella\altaffilmark{1,26},
J. Lecacheux\altaffilmark{1}, 
L. Maquet\altaffilmark{4},
F. Vachier\altaffilmark{4},
S. Renner\altaffilmark{51},
B. Monard\altaffilmark{52},
}

\author{
A. A. Sickafoose\altaffilmark{35,36},
H. Breytenbach\altaffilmark{35,49},
A. Genade\altaffilmark{35,49}
}

\author{
W. Beisker\altaffilmark{10,44},
K.-L. Bath\altaffilmark{10,44},
H.-J. Bode\altaffilmark{10,44,\footnote{Deceased on 16 July 2017}},
M. Backes\altaffilmark{50}
}

\author{
V. D. Ivanov\altaffilmark{14,15},
E. Jehin\altaffilmark{5},
M. Gillon\altaffilmark{5}
J. Manfroid\altaffilmark{5}
J. Pollock\altaffilmark{7},
G. Tancredi\altaffilmark{20},
S. Roland\altaffilmark{19},
R. Salvo\altaffilmark{19},
L. Vanzi\altaffilmark{2},
}

\author{
D. Herald\altaffilmark{11,12,18},
D. Gault\altaffilmark{11,17},
S. Kerr\altaffilmark{11,28},
H. Pavlov\altaffilmark{11,12},
K. M. Hill\altaffilmark{29},
J. Bradshaw\altaffilmark{12,13},
M. A. Barry\altaffilmark{11,30},
A. Cool\altaffilmark{33,34},
B. Lade\altaffilmark{32,33,34},
A. Cole\altaffilmark{29},
J. Broughton\altaffilmark{11},
J. Newman\altaffilmark{18},
R. Horvat\altaffilmark{17},
D. Maybour\altaffilmark{31},
D. Giles\altaffilmark{17,31},
L. Davis\altaffilmark{17},
R.A. Paton\altaffilmark{17},
}

\author{
B. Loader\altaffilmark{11,12},
A. Pennell\altaffilmark{11,48},
P.-D. Jaquiery\altaffilmark{47,48},
}

\author{
S. Brillant\altaffilmark{15},
F. Selman\altaffilmark{15},
C. Dumas\altaffilmark{53},
C. Herrera\altaffilmark{15},
G. Carraro\altaffilmark{43},
L. Monaco\altaffilmark{40},
A. Maury\altaffilmark{21},
}

\author{
A. Peyrot\altaffilmark{42},
J.-P. Teng-Chuen-Yu\altaffilmark{42},
}

\author{
A. Richichi\altaffilmark{46},
P. Irawati\altaffilmark{37},
}

\author{%
C. De Witt\altaffilmark{10},
P. Schoenau\altaffilmark{10},
R. Prager\altaffilmark{44},
}%

\author{
C. Colazo\altaffilmark{8,9},
R. Melia\altaffilmark{9},
J. Spagnotto\altaffilmark{22},
A. Blain\altaffilmark{39},
}

\author{
S. Alonso\altaffilmark{16},
A. Rom\'an\altaffilmark{38},
P. Santos-Sanz\altaffilmark{3},
J.-L. Rizos\altaffilmark{3},
J.-L. Maestre\altaffilmark{45},
}

\author{
D. Dunham\altaffilmark{12}
}

\affil{\newpage}
\affil{$^{1}$ LESIA, Observatoire de Paris, PSL Research University, CNRS, Sorbonne 
Universit\'es, UPMC Univ. Paris 06, Univ. Paris Diderot, Sorbonne Paris 
Cit\'e}
\affil{$^{2}$  Department of Electrical Engineering and Center of Astro-Engineering, Pontificia Universidad Cat\'olica de Chile, Av. Vicu\~na Mackenna 4860, Santiago, Chile}
\affil{$^{3}$Instituto de Astrof\'isica de Andaluc\'ia, CSIC, Apt. 3004,18080 Granada, Spain}
\affil{$^{4}$IMCCE, Observatoire de Paris, PSL Research University, CNRS, Sorbonne Universités, UPMC Univ. Paris 06, 77 av. Denfert-Rochereau, 75014, Paris, France}
\affil{$^{5}$Institut d'Astrophysique de l'Universit\'e de Li\`ege, All\'ee du 6 Aoû\^ut 17, B-4000 Li\`ege, Belgique}
\affil{$^{6}$Instituto de Astrof\'isica, Facultad de F\'isica, Pontificia Universidad Cat\'olica de Chile, Av. Vicu\~na Mackenna 4860, Santiago, Chile}
\affil{$^{7}$Physics and Astronomy Department, Appalachian State Univ., Boone, NC 28608, USA}
\affil{$^{8}$Ministerio de Educaci\'on de la Provincia de C\'ordoba, C\'ordoba, Argentina}
\affil{$^{9}$Observatorio Astron\'omico, Universidad Nacional de C\'ordoba, C\'ordoba, Argentina}
\affil{$^{10}$IOTA/ES, Barthold-Knaust-Strasse 8, D-30459 Hannover, Germany} 
\affil{$^{11}$Occultation Section of the Royal Astronomical Society of New Zealand (RASNZ), Wellington, New Zealand}
\affil{$^{12}$International Occultation Timing Association (IOTA), PO Box 7152, Kent, WA 98042, USA}
\affil{$^{13}$Samford Valley Observatory, QLD, Australia}
\affil{$^{14}$ ESO, Karl-Schwarzschild-Str. 2, 85748 Garching bei, M\"unchen, Germany}
\affil{$^{15}$ESO, Alonso de Cordova 3107, Casilla 19001, Santiago 19, Chile}
\affil{$^{16}$Software Engineering Department, University of Granada, Spain}
\affil{$^{17}$Western Sydney Amateur Astronomy Group (WSAAG), Sydney, NSW, Australia}
\affil{$^{18}$Canberra Astronomical Society, Canberra, ACT, Australia}
\affil{$^{19}$Observatorio Astron\'omico Los Molinos, DICYT, MEC, Montevideo, Uruguay}
\affil{$^{20}$Dpto. Astronomia, Facultat de Ciencias, Uruguay}
\affil{$^{21}$San Pedro de Atacama Celestial Explorations, Casilla 21, San Pedro de Atacama, Chile}
\affil{$^{22}$Observatorio El Catalejo, Santa Rosa, La Pampa, Argentina}
\affil{$^{23}$Observat\'orio do Valongo/UFRJ, Ladeira Pedro Antonio 43, RJ 20.080-090 Rio de Janeiro, Brazil}
\affil{$^{24}$Observat\'orio Nacional/MCTIC, R. General Jos\'e Cristino 77, RJ 20921-400 Rio de Janeiro, Brazil}
\affil{$^{25}$Laborat\'orio Interinstitucional de e-Astronomia - LIneA, Rua Gal. Jos\'e Cristino 77, Rio de Janeiro- RJ 20921-400, Brazil}
\affil{$^{26}$Unidad Mixta Internacional Franco-Chilena de Astronom\'ia (CNRS UMI 3386), Departamento de Astronom\'ia, Universidad de Chile, Camino El Observatorio 1515, Las Condes, Santiago, Chile}
\affil{$^{27}$Federal University of Technology- Paran\'a (UTFPR/DAFIS), Rua Sete de Setembro, 3165, CEP 80230-901, Curitiba, PR, Brazil}
\affil{$^{28}$ Astronomical Association of Queensland, QLD, Australia}
\affil{$^{29}$School of Physical Sciences, University of Tasmania, Private Bag 37, Hobart, TAS 7001, Australia}
\affil{$^{30}$Electrical and Information Engineering Department, University of Sydney, Camperdown, NSW 2006, Australia}
\affil{$^{31}$Penrith Observatory, Western Sydney University, Sydney, NSW, Australia}
\affil{$^{32}$Stockport Observatory, Astronomical Society of South Australia, Stockport, SA, Australia}
\affil{$^{33}$Defence Science \& Technology Group, Edinburgh, South Australia }
\affil{$^{34}$The Heights Observatory, Modbury Heights, South Australia}
\affil{$^{35}$South African Astronomical Observatory, PO Box 9, Observatory, 7935, South Africa}
\affil{$^{36}$Department of Earth, Atmospheric, and Planetary Sciences, Massachusetts Institute of Technology Cambridge, MA 02139-4307, United States}
\affil{$^{37}$National Astronomical Research Institute of Thailand, Siriphanich Building, Chiang Mai 50200 - Thailand}
\affil{$^{38}$Sociedad Astron\'omica Granadina, Granada, Spain}
\affil{$^{39}$Asociaci\'on Argentina Amigos de la Astronom\'ia, Av. Patricias Argentinas 550, Buenos Aires, Argentina}
\affil{$^{40}$Departamento de Ciencias Fisicas, Universidad Andres Bello, Fernandez Concha 700, Santiago, Chile }
\affil{$^{41}$ Ciel \& Espace, Paris, France}
 \affil{$^{42}$ Makes Observatory,La R\'eunion, France}
 \affil{$^{43}$ Dipartimento di Fisica e Astronomia, Universita di Padova, Italy} 
  \affil{$^{44}$ Internationale Amateursternwarte e. V., IAS, Hakos/Namibia and Bichlerstr. 46, D-81479 M\"unchen (Munich), Germany.}
  \affil{$^{45}$  Observatorio Astron\'omico de Albox, Apt. 63, 04800 Albox (Almeria), Spain}
  \affil{$^{46}$  INAF - Osservatorio Astrofisico di Arcetri, Largo E. Fermi 5, 50125 Firenze, Italy}
    \affil{$^{47}$ Royal Astronomical Society of New Zealand (RASNZ), Wellington, New Zealand}
\affil{$^{48}$    Dunedin Astronomical Society, Dunedin, New Zealand}
\affil{$^{49}$ University of Cape Town, Department of Astronomy, Rondebosch , Cape Town, Western Cape, South Africa, 7700}
\affil{$^{50}$ Department of Physics, University of Namibia, 340 Mandume Ndemufayo Ave, P/Bag 13301, Windhoek, Namibia}
\affil{$^{51}$IMCCE, Observatoire de Paris, CNRS UMR 8028, Universit\'e Lille 1, Observatoire de Lille 1 impasse de l'Observatoire, 59000 Lille, France }
\affil{$^{52}$Kleinkaroo Observatory, Calitzdorp, St. Helena 1B, P.O. Box 281, 6660 Calitzdorp, Western Cape, South Africa}
\affil{$^{53}$TMT International Observatory, 100 West Walnut Street, Suite 300, Pasadena, CA 91124, USA}

\affil{\newpage}







\begin{abstract}
%
Two narrow and dense rings (called C1R and C2R) were discovered around the Centaur object (10199) Chariklo 
during a stellar occultation observed on June 3, 2013 \citep{bra14}.
%
Following this discovery, we have planned observations of several occultations by Chariklo's system 
in order to better characterize the ring and main body physical properties.
Here, we use 12 successful Chariklo's occultations observed between 2014 and 2016.
They provide ring profiles (physical width, opacity, edge structure)
and constraints on their radii and pole position.
Our new observations are currently consistent with 
the circular ring solution and pole position,
to within the $\pm 3.3$~km formal uncertainty for the ring radii, derived by \cite{bra14}.
%
The six resolved C1R profiles reveal significant width variations from $\sim$~5.5 to 7~km.
The width of the fainter ring C2R is less constrained, and may vary between
 0.1 and 1 km.
The inner and outer edges of C1R are consistent with infinitely sharp boundaries,
with typical upper limits of one kilometer for the transition zone between the ring and empty space.
No constraint on the sharpness of C2R's edges is available.
A $1\sigma$ upper limit of $\sim20$~m is derived for the equivalent width of narrow 
(physical width $<4$~km) rings up to distances of 12,000~km, counted in the ring plane.
\end{abstract}





\section{Introduction}

The asteroid-like body (10199) Chariklo is a Centaur object orbiting between Saturn and Uranus. 
It probably moved recently ($\sim$~10~Myr ago) from the Trans-Neptunian region
to its present location, and will leave it on a similar short time scale, due to perturbations by Uranus \citep{hor04}.
With 	a radius of $119 \pm 5$~km, estimated from thermal measurements \citep{for14},  
it is the largest Centaur known to date, but still remains very modest in size compared to the telluric or giant planets.
On June 3, 2013, a ring system was discovered around this small object during a stellar occultation. 
Two dense and narrow rings, 2013C1R and 2013C2R (C1R and C2R for short) were detected.
They are separated by about 15~km and orbit close to 400~km from Chariklo's center
(see \cite{bra14} for details).

Until 2013, rings were only known around the giant planets.
This discovery was thus surprising, and is a key to better understanding of the planetary rings, 
since they now appear to be more common than previously thought.
In particular, the two rings being dense, narrow and (at least for C1R) sharp-edged, 
they look like several of the dense ringlets seen around Saturn and Uranus \citep{ellnic84,fre91,fre16}. 
In that context, there was a strong incentive for planning more occultation campaigns, 
first to unambiguously confirm the existence of Chariklo's rings and 
second, to obtain more information on their physical properties.
%

While the discovery occultation of June 3, 2013 provided the general ring physical parameters 
(width, orientation, orbital radius, optical depth,...), 
several questions are still pending, some of them being addressed in this work:
do the rings have inner structures that give clues about collisional processes?
How sharp are their edges?
What are the general shapes of C1R and C2R? 
Do they consist of solidly precessing ellipses like some Saturn's or Uranus' ringlets?
Do they have more complex proper modes with higher azimuthal wave numbers?
Are there other fainter rings around Chariklo? 
What is the shape of the object itself and its role for the ring dynamics?
Based on new results, what can we learn about their origin and evolution, 
which remains elusive \citep{sic16}?

This study is made in a context where material has also been detected around 
the second largest Centaur, Chiron (again using stellar occultations).
The nature of this material is still debated and it could be interpreted
as a ring system \citep{ort15} or a dust shell associated with Chiron's cometary activity \citep{rup15}.
%
%
%
Since Chariklo is presently moving close to the galactic plane, 
stellar occultations by this body are much more frequent than for Chiron,
hence a more abundant amount of information concerning its rings.
The spatial resolution achieved during occultations reaches the sub-km level,
impossible to attain with any of the current classical imaging instruments.
This said, the very small angular size subtended by the rings 
(0.08~arcsec tip to tip, as seen from Earth)
have made occultation predictions difficult in the pre-Gaia era.

In spite of those difficulties, 
we could observe 13 positive stellar occultations (including the discovery one) between 2013 and 2016,
from a total of 42 stations distributed worldwide 
(in Brazil, Argentina, Australia, Chile, La R\'eunion Island, Namibia, New Zealand, South Africa, Spain, 
Thailand and Uruguay).
%
Here, we focus on the ring detections (a total of 11 chords recorded after the discovery).
We also obtained a total of 12 occultation chords by the main body from 2014 to 2016. 
Their timings are derived here, but their implications concerning Chariklo's size and shape
will be presented elsewhere (Leiva et al., 2017, in preparation).
%
%
In Section~2, we present our observations and data analysis.
In Section~3 we concentrate on the ring structures (width, inner structures, edge sharpness) and 
geometry (radius and orbital pole).
The ring integral properties (equivalent width and depth) are derived in Section~4, before
concluding remarks in Section~5. 
 
\section{Observations and Data Analysis}

Following the ring discovery of June 3, 2013, 
we predicted and observed 12 positive stellar occultations by Chariklo and/or its rings 
between 2014 and 2016.
In the following list, we mark in italic the events that led to multi-chord ring detections
(thus providing constraints on the ring orientation, as discussed latter).
Four occultations were observed in 2014, on  
February 16 (rings), 
March 16 (rings), 
\textit{April 29 (rings and body)} and 
June 28 (rings and body). 
In 2015, only two positive detections were recorded on April 26 (rings) and May 12 (body), 
while six occultations were recorded in 2016: 
July 25 (body), 
August 8 (rings and body), 
August 10 near 14h UT (body), 
August 10  near 16h UT (body), 
August 15 (body) and 
\textit{October 1 (rings and body)}.

\subsection{Predictions}

Predicting stellar occultations by Chariklo and its rings is a difficult task, as the
main body subtends about 25~milliarcsec (mas) as seen from Earth, 
while the rings have a span of  about 80~mas.
Thus, to be effective, predictions require accuracies of a few tens of mas
on both Chariklo's ephemeris and the star position.
To meet this requirement, 
we used a bootstrapping approach, in which each new detection of occultation is used to improve
Chariklo's ephemeris, thus providing a better prediction for the next occultation.
This continuous update results in the so-called NIMA ephemeris
(Numerical Integration of the Motion of an Asteroid, \citealt{des15})
accessible online\footnote{see http://lesia.obspm.fr/lucky-star/nima/Chariklo/}.

The candidate stars for events in 2014 and in 2016 
were identified during a systematic search for occultations by TNOs
using the \textit{Wide Field Imager} (WFI) at the ESO/MPG 2.2m telescope \citep{cam14}, 
with typical accuracies of $\sim30$ mas. 
However, for the 2015 season, 
the candidate stars were observed using only the IAG 0.6m telescope at OPD/LNA in Brazil,
with lower accuracy than WFI, resulting in a larger number of missed events (two successes out of six attempts).

In the majority of the cases, the occulted star
was imaged a few days or weeks prior the event in order
to improve the astrometry.
If possible the observations were made when Chariklo and the star 
were in the same Field Of View in order to cancel systematic errors.
In those cases, accuracy of the predictions was estimated down to 
$\sim 20$ mas.

The last occultation in our list (October 1$^{\rm st}$, 2016) is special as its prediction was based
on the new GAIA DR1 catalog released on September 15, 2016 \citep{gaia16}.
However, the J2000 DR1 star position $\alpha= 18\rm h 16\rm m 20.0796\rm s$, $\delta=-33\degr 01\arcmin 10.756\arcsec$ (at epoch 2015.0) does not account for proper motion.
We estimated the latter by using the UCAC4 star position (under the name UCAC4 285-174081)
 at epoch 2000 and obtained proper motions in right ascension (not weighted by $\cos(\delta)$) and declination of
$$
\begin{tabular}{ll}
$\mu_\alpha$ =      & $-0.43 \pm 0.008$~ms/yr \\
$\mu_\delta$ =    & $-2.02  \pm 1.05$~mas/yr.\\
\end{tabular}
$$
This provides a star position of  
$\alpha= 18\rm h 16\rm m 20.0789\rm s$, $\delta=-33\degr 01\arcmin 10.760\arcsec$
at the epoch of occultation.
Combining this result with the NIMA ephemeris 	(version 9)
finally provided a prediction that agreed
to within 5 mas perpendicular to the shadow track and 20 seconds
in terms of timing, and lead to a multi-chord ring and body detection. 

\subsection{Observations}
\label{sec_observations}

The circumstances of the observations (telescope, camera, set up, observers, site coordinates, star information)
that lead to ring or main body detections are listed in Table~\ref{tab_circumstances_positives}.
Conversely,  the circumstances of negative observations (no event observed) are provided in Table~\ref{tab_circumstances_negatives}.
Note that observations were made with both small portable telescopes and larger, fixed instruments. 
Each detection will be designated herein by the name of the station or by the name of the telescope, if well known. 

From the timings of the star disappearance (or ``ingress") and re-appearance (``egress") behind Chariklo
 and/or the rings, the geometry of each occultation 
was reconstructed, as illustrated in Fig.~\ref{fig_geometry}. 
Currently, Chariklo's size and shape are not known well enough to 
reconstruct the occultation geometries from the events involving the main body.
So, we used instead the ring events (even single-chord) to retrieve those geometries. 
As a starting point, 
we assume that the rings are circular with fixed orientation in space,
and with the orbital parameters derived by \cite{bra14}, namely a J2000 pole position of
$\alpha_p=10\rm h 05\rm m 11.0016\rm s$, $\delta_p=+41\degr 28\arcmin 32.4891\arcsec$ 
and respective radii $a_{C1R} =390.6$ ~km and $a_{C2R}=404.8$~km for the two rings.
The reconstructed geometry allows us to derive the observed position
of Chariklo center (reported in Table~\ref{tab_circumstances_positives}).
If the star position was perfect, this derived position must coincide with the
occulted star position. The difference between the two positions
is the offset between the predicted and the observed Chariklo's position.
This offset is implemented in NIMA after each occultation, in order
to improve Chariklo's ephemeris.

If the rings are \textit{not} circular, this will impact their pole position and will eventually
be visible as discrepancies between observations and predictions. 
The pole position problem is discussed further in Section~\ref{section_pole}.

Note that some stations did not detect any ring occultations, 
whereas they should have considering the occultation geometry, 
see Reedy Creek on May 12, 2015 and Sydney on August 10, 2016. 
Data analysis shows that those non-detections are actually consistent with the low 
signal-to-noise-ratio (SNR) obtained at those stations.
Thus, secondary events have always been detected if SNR was high enough.
This leads us to conclude that C1R (which always dominates the profile)
is continuous.
Same conclusion on C2R is more ambiguous as C2R
was usually blended together with C1R.
Nevertheless, we will assume that C2R is continuous in this paper.

\subsection{Data Reduction}

After a classical data processing that included dark subtraction and flat fielding, 
aperture photometry provided the stellar flux as a function of time 
(the date of each data point corresponding to mid-exposure time),
the aperture being chosen to maximize SNR. 
The background flux was estimated near the target and nearby reference stars, 
and then subtracted, so that the zero flux corresponds to the sky level. 
The total flux from the unocculted star and Chariklo was normalized to unity after fitting the light curve
by a third or forth-degree polynomial before and after the event.
In all cases, a reference star (brighter than the target) was used to correct for low frequency 
variations of the sky transparency. 

The light curves are displayed in Fig.~\ref{fig_fits_rings_1} and \ref{fig_fits_rings_2}, each of them 
providing a one-dimensional scan across Chariklo's system, as projected in the sky plane.
In some cases, the readout time between two frames caused a net loss of information
as photon acquisition was interrupted during those ``dead time" intervals.
The flux statistics provides the standard deviation of the signal, which defines the $1\sigma$ error bar 
on each data point which was used latter for fitting diffraction models to ingress and egress events. 
Note that during an occultation by the main body, the stellar flux drops to zero, 
but the flux in the light curve is not zero, as it contains Chariklo's contribution, 
and in one case, the flux from a nearby companion star, see below.

\subsection{The case of the double star of April 29, 2014}
\label{section_etoile}

This event, observed from South Africa (see Table~\ref{tab_circumstances_positives}), revealed that the occulted star was a binary.
As seen from Springbok, the primary star (``A") was occulted by C1R and C2R (but missed the main body), 
while the fainter companion star (``B") disappeared behind Chariklo along an essentially diametric chord 
at Springbok (Fig.~\ref{fig_geometry}). 
Because the component B was about 9 times fainter than A (see below), 
and considering the drop of A caused by C1R at Springbok, 
we expect a short drop in the light curve of only 8\%
due to the disappearance of component B behind C1R.
This is too small to be detected, in view of the SNR of about 7 per data point obtained at that station
(Fig.~\ref{fig_fits_rings_2}).

Meanwhile in Gifberg, we obtained only a grazing occultation of the primary star by C2R (Fig.~\ref{fig_geometry}).
This provides the best profile of that ring ever recorded (see Section~\ref{section_profile}). 
Finally, at the South African Astronomical Observatory (SAAO), 
only the component B was occulted by the rings, while the main star missed both
the rings and the main body (Fig.~\ref{fig_geometry}).
However, due to the high SNR obtained at that station, the partial drop caused
by the rings on component B has about the same useful SNR as
the drop of component A as seen from the smaller telescope at Springbok.

For the Springbok light curve, 
we can estimate the flux ratio $\Phi_A/\Phi_B$ between the two stars
by considering the drop of component B caused by Chariklo. 
In doing so, we can neglect Chariklo's contribution to the total flux. 
From Chariklo's absolute magnitude, $H_V=7.0$ in 2014 \citep{duf14},
and heliocentric and geocentric distances of 14.8 au and 14.1 au during the event, 
respectively, we obtain a Chariklo apparent magnitude $\sim 18.6$.
This is 5.6 magnitudes fainter than the star, which has V=13.0
(NOMAD catalog\footnote{See http://vizier.u-strasbg.fr/viz-bin/VizieR}),
meaning that Chariklo contributed to the total flux of less than 0.6\%, 
a negligible value at our level of accuracy.

The fractional drop observed during the occultation of B by Chariklo provides 
its partial contribution to the total stellar flux, $\Phi_B= 0.1036 \pm 0.0075$
(Fig.~\ref{fig_fits_body}).
This implies a flux ratio $(\Phi_A/\Phi_B)_{\rm TC247}= 8.65 \pm 0.65$,  
as measured by the Texas Instruments TC247 array used at Springbok (in broad band mode, no filter). 
This directly provides the baseline level for the occultations of A by the rings
(Figs.~\ref{fig_fits_body} and \ref{fig_profile_C2R_centre}), 
i.e. the level that corresponds to a total disappearance of component A.

A similar calibration is not possible for the SAAO ring events, 
as that station did not record an occultation by the main body.
Moreover, the ratio $(\Phi_A/\Phi_B)_{\rm TC247}$ cannot be used, 
as the SHOC instrument (see \citealt{cop13}) used at SAAO (also in broad band mode) 
has a different spectral response, 
so that the ratio depends on the color of the two stars.

To proceed forward, 
we have used the B, V, K magnitudes of the star 
(taken from the Vizier page, in NOMAD catalog). 
We have generated combined synthetic spectra energy distribution of the two components,
 and using various (and separate) effective temperature $T_{\rm eff}$ for A and B.
The effect of interstellar reddening has been parametrized using the color excess $E(B-V)$. 
We adopted the classical total to selective extinction parameter $R_V = 3.1$ for Milky Way dust from \cite{fit99}.
The relative contributions of each component were adjusted 
in order to fit both the observed magnitude of the star and the flux ratio 
as observed with the TC247 array.
Finally, accounting for the spectral response of the Andor array, 
we can then estimate the ratio $(\Phi_A/\Phi_B)_{\rm Andor}$ for that detector.

A difficulty stems from the fact that there is a degeneracy between the effective 
temperatures assumed for the two components,  $T_{\rm eff}(A)$ and $T_{\rm eff}(B)$.
The star B cannot be much cooler than A, otherwise its diameter would be larger and 
strong signatures in the near IR would appear in the composite spectrum. 
We have opted for a difference $T_{\rm eff}(A) - T_{\rm eff}(B) \sim 1000$~K, and
assume that the two stars are on the main sequence.
We find a good fit to the observed magnitudes with  
$T_{\rm eff}(A)= 5000$~K and $T_{\rm eff}(B) = 4000$~K,
and then a ratio $(\Phi_A/\Phi_B)_{\rm Andor}= 7.66$, 
corresponding to a contribution to the total flux of $\Phi_A = 0.885 \pm 0.025$ for component A,
where the error bar is estimated from the typical possible ranges for
$T_{\rm eff}(A)$ and $T_{\rm eff}$. 

Finally, we can estimate the apparent diameter 
of each component projected at Chariklo's distance:
$\theta_A=0.199\pm0.015$ km and $\theta_B=0.092\pm0.015$ km.
Those values will be used latter when fitting the ring profiles with models of 
diffracting, semi-transparent bands.

Assuming the ring radii and pole orientation of \cite{bra14},
see also Section~\ref{sec_observations}, 
and using the ring detections in Springbok, Gifberg and SAAO, 
we deduce that star B was at angular distance 20.6 mas from star A as
projected in the sky plane, with position angle $P=209.8\degr$ relative to the latter 
(where $P$ is counted positively from celestial North towards celestial East).
 
\section{Ring events analysis}

\subsection{Profiles fitting}
\label{section_diffraction}

In order to determine accurate and consistent timings of the ring occultations,
we use a ``square-well model'' in which each ring is modeled 
as a sharp-edged, 
semi-transparent band of 
apparent opacity $p'$ (along the line of sight) and 
apparent width (in the sky plane) $W_{\perp}$.
We use the numerical schemes described in \cite{roq87} to account for 
Fresnel diffraction, 
stellar diameter projected at Chariklo's distance, 
finite bandwidth of the CCD, and
finite integration time of the instrument. 
Finally, considering projection effects, we can derive the ring physical parameters
(radial width, normal opacity, etc...) and orbital elements, see Appendix for details. 

For sake of illustration, we give various parameters of interest in the case of the April 29, 2014 occultation.
The Fresnel scale $F=\sqrt{\lambda D / 2}$ for Chariklo's geocentric distance at epoch, 
$D = 2.11 \times 10^{9}$ km is 0.83 km, for a typical wavelength of $\lambda=0.65$~$\mu$m. 
The projected stellar diameters have been estimated above to 
$0.199 \pm 0.015$~km and $0.092 \pm 0.015$~km for the primary star and secondary star, 
respectively (see Section~\ref{section_etoile}). 
The smallest cycle time used during that campaign was 0.04~s (at SAAO),
corresponding to 0.5~km traveled by the star relative to Chariklo in the celestial plane. 
Consequently, the light curves are dominated by Fresnel diffraction, 
but the effects of stellar diameters and finite integration time remain comparable. 
Similar calculations for the other twelve occultations show that the effect of 
finite integration time dominated in all those cases.

%
The synthetic ring profiles are then fitted to observations so as to minimize the classical $\chi^{2}$ function:
\begin{equation}
\label{eq_chi2}
\chi^{2}=\sum_i \frac{(\Phi_{i,\rm obs}-\Phi_{i,\rm calc})^{2}}{\sigma^{2}_{i}}
\end{equation}
where $\Phi$ is the flux, 
$i$ refers to the $i^{\rm th}$ data point, 
``obs" refers to observed, 
``calc" refers to calculated, and 
$\sigma$ the $1\sigma$-level error of the $i^{\rm th}$ data point.
The free parameters of the model are described in the next subsection. 
The $1\sigma$ error bar on each parameter is estimated by varying this particular parameter 
to increase $\chi^{2}$ from the best value $\chi^{2}_{min}$ to $\chi^{2}_{min}+1$, 
the others parameters are set free during this exploration. 

\subsection{Mid-times and widths of the rings}
\label{section_width_rings}

The best fitting square-well model described above provides relevant parameters 
that depend on the occulting object. 
Three cases are possible: occultations by 
(1) main body;
(2) resolved rings; 
(3) unresolved rings. 
The relevant parameters in each case are respectively 
(1) the times of ingress and the egress of the star behind the body; 
(2) the mid-time of the occultation $t_0$, 
the radial width \emph{reprojected in the plane of the rings}, $W_{r}$, 
and the local normal opacity $p_N$ for each ring (see Appendix for details); 
(3) the mid-time of the occultation.
Those parameters are listed in 
Table~\ref{tab_param_rings} (resolved ring events), 
Table~\ref{tab_param_rings_unresolved} (unresolved ring events), and in 
Table~\ref{tab_param_body} (main body events). 
The best fits for each occultation are plotted in 
Fig.~\ref{fig_fits_rings_1} and \ref{fig_fits_rings_2} (ring occultations) and in 
Fig.~\ref{fig_fits_body} (main body occultations).

The grazing occultation by C2R recorded in Gifberg (Fig.~\ref{fig_geometry}) requires a special analysis. 
In this geometry, the radial velocity of the star relative to the ring changes significantly during the event
(while it is assumed to be constant for all the other events). 
To account for this peculiarity,
we first converted the light curve (time, flux) into a profile ($\Delta r$, flux), 
where $\Delta r$ is the radial distance to the point of closest approach 
to Chariklo's center (in the sky plane).
Then we can apply the square-well model as explained in section~\ref{section_diffraction}, 
except that the flux is now given in terms of $\Delta r$, instead of time. 
Best fits for ingress and egress are plotted in Fig.~\ref{fig_fits_gifberg}.

%
Table~\ref{tab_param_rings} summarizes the values of $W_r$ for each resolved profile.
Fig.~\ref{fig_W_vs_L} shows $W_{r}$ vs. the true longitude $L$ counting from the ascending node.
Accounting for the most constraining events, 
$W_{r}$ varies between 5~km and 7.5~km in C1R
 and between 0.05~km and 1~km in C2R (at $1\sigma$-level). 
Fig~\ref{fig_W_vs_L} could constrain the rings proper mode. 
Unfortunately, the true longitude $L$ plotted in Fig.~\ref{fig_W_vs_L} (and latter in Fig.~\ref{fig_E_A_C1R})
is not the correct quantity to use in order to detect $m=1$ proper modes (the true anomaly 
$L-\varpi$ should be used instead of $L$, where $\varpi$ is the longitude of periapse).
As the precession rates of the rings are unknown, no conclusion can be made. 
Nevertheless, those width variations are observed both 
for a given occultation at different longitudes and 
for different occultations at different dates, see Fig.~\ref{fig_W_vs_L}.
Implications are discussed in Section~\ref{section_discussion}.

\subsection{Ring inner structures} 
\label{section_profile}

Fig.~\ref{fig_profile_C2R_centre} shows the best radial profiles of the rings that we have obtained so far,
taken from the discovery observation of June 3, 2013 and the April 29, 2014 event.
They are currently the only profiles that clearly resolve C1R from C2R, 
and in the case of the April 29, 2014 event, the only profiles that resolve C1R.
A W-shape structure inside C1R is clearly seen at egress in the Springbok and SAAO profiles, 
and marginally detected in the Springbok ingress profile, while being absent (to within the noise)
in the SAAO  ingress profile.
Note that small (2-4~km) variations of radial distances between the two rings are 
visible in Fig.~\ref{fig_profile_C2R_centre}.
The average gap distance between the two rings on the six profiles is thus 14.8~km.

Since the origin of radial distance has been fixed \textit{arbitrarily} on the center of C2R,
it is not possible to attribute those variations to an eccentricity of C1R, C2R or both.
Note also that the April 29 profiles are montages obtained by juxtaposing 
the profiles of C1R recorded at Springbok and SAAO and the profile of C2R recorded in Gifberg.
So, they scan different rings longitudes, and conclusions based on this plot can only be qualitative. 

%
%

\subsection{Ring pole}
\label{section_pole}


By analogy with their Uranian counterparts, 
we expect that Chariklo's ring orbits have essentially elliptical shapes,
corresponding to a normal mode with a $m=1$ azimuthal harmonic number.
Moreover, other modes with higher values of $m$ are possible and the two rings may not be coplanar.
However, data on Chariklo's rings are currently too scarce to reach those levels of details.
Instead, we have to simplify our approach, considering the observational constraints at hand.

The simplest hypothesis is to assume that the two are circular, concentric and coplanar.
Then, their projections in the sky plane are ellipses characterized by $M=5$ adjustable parameters: 
the apparent semi-major axis $a'$, 
the coordinates of the ellipse's center $(f_c, g_c)$, 
the apparent oblateness $\epsilon'= (a'-b')/a'$ (where $b'$ is the apparent semi-minor axis), 
and the position angle $P$ of the semi-minor axis $b'$. 
For circular rings, $\epsilon'=1-\sin(B)$, where $B$ is the ring opening angle 
($B=0$ and $B=90\degr$ corresponding to edge-on and pole-on geometries, respectively).
%

Note that $(f_c, g_c)$ is related to
the offsets in right ascension and declination between the predicted and observed 
positions of the object, relative to the occulted star. 
The positions of Chariklo deduced from $(f_c, g_c)$ 
-- at prescribed times and for given star positions --
are listed in Table~\ref{tab_circumstances_positives}.
They can be used to improve Chariklo's ephemeris, once the star positions are improved,
using the DR1 Gaia catalog and its future updates.

This circular ring model requires at least  $N \ge M=5$ data points in order to provide a unique solution for 
the ring radius $a$ (coincident with $a'$) and its J2000 pole position $(\alpha_p,\delta_p)$. 
Only the June 3, 2013 discovery observation with 7 chords (and thus $N=14$ data points
corresponding to the chord extremities) has sufficient constraints to provide unambiguous ring orbits.
More precisely, as only one instrument (Danish telescope) could resolve the rings C1R and C2R in 2013, 
this multi-chord event mainly determines the orbit of C1R, 
which largely dominates the usually blended ring profiles.
Then we assumed that C2R is coplanar with C1R and separated radially from it 
by a constant distance $\Delta a = 14.2 \pm 0.2$~km \citep{bra14}.
 
The April 29, 2014 event provides two chords ($N=4$ data points) on C2R.
This allows us to definitely eliminate one of the pole positions derived from the 2013 event.
Actually, determining the angles $B$ and $P$ at a given date provides
\textit{two} possible pole positions, 1 and 2, depending on which part of the rings, as seen in the sky plane,
is the ``near arm" or the ``far arm", see \cite{bra14} for details.
The C2R chord observed at Springbok turned out to be longer than the longest possible length 
allowed by solution 2, thus confirming that the preferred solution 1 of \cite{bra14},
based on the long-term photometric behavior of Chariklo (see also below), 
was actually the correct one. 

In order to constrain the pole position, even with $N< M$, we vary the couple ($P,B$) in a predetermined grid, 
while the other three parameters are adjusted in order to 
minimize the radial residuals in the sky plane relative to the ring center.
Since the pole position is given by two parameters $(\alpha_P,\delta_P)$, the 68.7\% confidence domain 
(called $1\sigma$-level here) is obtained by allowing variations of the $\chi^2$ function 
from $\chi^2_{\rm min}$ to $\chi^2_{\rm min} +2.3$ \citep{pre92}, 
and by selecting values of $a$ to $\pm 3.3$~km, the nominal error on the C1R and C2R radii:
$a_{C1R} \sim 391$~km and  $a_{C2R} \sim 405$~km \citep{bra14} .
The pole position derived from the April 29, 2014 occultation is displayed in Fig.~\ref{fig_pole_circular}.
Note that it is consistent with but less accurate than the pole determined in 2013. 
 
Finally, the October 1$^{\rm st}$, 2016 event also provided two chords ($N=4$ data points)
across the rings, but without resolving C1R from C2R (Fig.~\ref{fig_fits_rings_2}).
Thus, we assumed that the profiles are dominated by C1R, and derived 
the pole position displayed in Fig.~\ref{fig_pole_circular}. 
It is again consistent with the poles of 2013 and 2014, but with larger error bars 
due to the ill-configured chord geometry (nearly diametric) that permits
more freedom on the pole position (Fig.~\ref{fig_geometry}).


 
Further constraints are in principle provided by the long-term photometric behavior 
of Chariklo's system between 1997 and 2014, as compiled by \cite{duf14}, see their Fig.~1.
The observed photometric variations can be explained by the changing viewing geometry of the rings, 
linked itself to the pole orientation.
%
Contrary to the occultation data, the photometric variations do not depend
on the particular shape of the rings (e.g. circular vs. elliptic).
Fitting for the pole position and accounting for the error bars taken from \cite{duf14},
we obtain the possible domain shown in Fig.~\ref{fig_pole_circular}.
Note that it is consistent with but less accurate than all our occultation results.

From Fig.~\ref{fig_pole_circular} we can conclude that our current data set 
(spanning the 3-year interval 2013-2016) 
is consistent with circular rings that maintain a fixed pole in space, 
and to within the current formal error bar on the semi-major axis $a$ ($\pm 3.3$~km). 
Note that the extensions of the error domains for the pole position
(colored regions in Fig.~\ref{fig_pole_circular})
are dominated by the errors in the data (i.e. the timings of the ring occultations),
not by the formal error for $a$ quoted above.
In other words, even if the ring shape were known perfectly, 
the pole position would not be significantly improved compared to
the results shown in  Fig.~\ref{fig_pole_circular}.
A Bayesian approach could be used to estimate the probability that the rings are
elliptic, considering the data at hand and assuming a random orientation for the
ring apsidal lines. Considering the paucity of data and the large number of degrees
of freedom, this task remains out of the scope of the present paper. 
In any case, new observations will greatly help into this approach by adding more
constraints on the ring shapes and orientations.

For all the other ring single-chord detections ($N=2$ data points), 
neither the rings radii nor their pole position can be constrained.
Instead, assuming the pole orientation of \cite{bra14}, we determined the ring center, 
also assumed to coincide with Chariklo's center of mass.
Having only one ring chord introduces an ambiguity as two solutions (North or East of
the body center) are possible. However, in all cases but one (August 8, 2016) it was possible to resolve this
ambiguity as the absence of detections made by other stations eliminated 
one of the two solutions.
For the August 8, 2016 event, the ambiguity remains, and we give the two possible Chariklo's positions,
see Table~\ref{tab_circumstances_positives}.

None of the single chords are longer than the longest chord expected from the \citealt{bra14}'s solution,
and thus remain fully consistent with that solution.

\subsection{Sharpness of C1R edges}
\label{section_sharpness}

A striking feature of the resolved C1R profiles from the April 29, 2014 event
is the sharpness of both its inner and outer edges. 
This is reminiscent of the Uranian rings \citep{ell84,fre91}, 
and might stem from confining mechanisms caused by 
nearby, km-sized shepherding moonlets \citep{bra14}.
In order to assess the sharpness of C1R's edges, we use a simple model,
where each edge has a stepwise profile, as illustrated in Fig.~\ref{fig_sharpness}. 
Instead of having an abrupt profile that goes from apparent opacity 0 to $p'$, 
we add an intermediate step of radial width in the ring plane $ \Delta w_r$
 and opacity $p'/2$ around the nominal  ingress or egress times, 
as deduced from the square-well model described before, see also Table~\ref{tab_param_rings}.
With that definition, $ \Delta w_r$ is a measure of the typical edge width,
i.e. the radial distance it takes to go from no ring material to a significant optical depth.

We explored values of $ \Delta w_r$ by varying the $\chi^2$ function (Eq.~\ref{eq_chi2}) from
its minimum value $\chi_{\rm min}^2$ to $\chi_{\rm min}^2 + 1$. 
The results are listed in Table~\ref{tab_sharpness} and illustrated in Fig.~\ref{fig_sharpness}. 
Note that all edges are consistent with infinitely sharp edges ($ \Delta w_r=0$) to within the $1\sigma$ level
and that upper limits for $ \Delta w_r$ are typically 1~km.
No significant differences are noticeable between the inner and the outer edges, 
contrary to, e.g., some Uranian rings \citep{fre91}.

Note finally that the width of C2R, as derived from the grazing event in Gifberg (Fig.~\ref{fig_fits_gifberg})
is slightly smaller ($\sim 0.7$~km) than the Fresnel scale ($\sim 0.8$~km).
As such, it is not possible to assess the sharpness of its edges.


\section{Integral properties of rings: equivalent width and depth}

We now turn to the measure of the ring's equivalent width $E_p$ and equivalent depth $A_\tau$, 
two quantities defined  and discussed by \cite{ell84} and \cite{fre91}, as detailed in the Appendix. 
Those quantities are physically relevant, as they are related to the amount of material 
present in a radial cut of the ring, in the extreme cases of monolayer and polylayer rings, respectively.

The values of $E_p$ are given 
in Table~\ref{tab_param_rings} (resolved events) and in
Table~\ref{tab_param_rings_unresolved} (unresolved events). 
For the resolved profiles, we have plotted $E_p$ against the radial width $W_r$ in Fig.~\ref{fig_E_A_C1R}. 
Implications in terms of mono- versus polylayer models will be discussed in Section~\ref{section_discussion}.
For the profiles that resolve C1R from C2R (and where both rings were detected), and those
where the two profiles are blended (the majority of our observations), 
we have plotted the \textit{integrated} $E_p(1+2)$ against the true longitude $L$ 
(counted from the J2000 ring plane ascending node) in  Fig.~\ref{fig_E_A_C1R}. 
%
From that figure, 
we see that the values of $E_p(1+2)$ lie in the interval 1-3~km, 
with no significant differences between the various measurements.
In other words, no significant variations of $E_p(1+2)$ with time and/or longitude 
are detected in our data set.

%

%

In this \textit{preliminary} study, the rings are considered as one entity C1R + C2R 
but further studies should treat them independently to derive conclusions on the structure of each of them.



\section{Search for faint ring material}

The best light curve available in terms of photometric quality is from the Danish Telescope.
It was acquired at a rate of 10 frames per second during the 30 minutes bracketing the occultation 
of June 3, 2013 \citep{bra14}.
It can be used to search for additional material orbiting Chariklo, 
assuming semi-transparent, uninterrupted, and permanent rings
coplanar to C1R and C2R.


For this purpose, we consider the equivalent width $E_p(i)$
of the putative ring material intercepted during the acquisition interval $\Delta t(i)$ 
corresponding to the $i^{\rm th}$ data point, and counted radially in the ring plane.
Using the results of the Appendix (see also \citealt{boi14} for details), we obtain
\begin{equation}
\label{eq_Er}
E_p(i)=\frac{|\sin(B)|}{2}[1-\phi(i)]\Delta r(i)
\end{equation}
where $\Delta r(i)$ is the radial interval travelled by the star during $\Delta t(i)$ (projected in the ring plane), 
and where $\phi(i)$ is the normalized stellar flux.
Due to projection effects, the value of $\Delta r(i)$ varied between the extreme values of 3 to 4~km
during the acquisition interval, which sets the radial resolution of this particular data set.

The values of $E_p(i)$ vs. radial distance $r$ is displayed in Fig.~\ref{fig_r_Er}.
Note that the light curve probes radial distances of up to $\sim 12,000$~km, about 30 times the ring radii.
Using bins of width 60 km, we evaluate the variance 
of the \textit{difference} between two consecutive points in each box,
thus eliminating low frequency variations of $E_p(i)$.
Dividing this variance by two (to account for the fact that the data points are uncorrelated) 
and taking the square root, we obtain the
$1\sigma$ level, standard deviation of $E_p(i)$, denoted $E_p(1\sigma)$, 
see the red line in Fig.~\ref{fig_r_Er}. 
The value of $E_p(1\sigma)$ remains stable in the entire range considered here,
with typical values of $20$~m.
%
%
Thus, at the $1\sigma$ level, 
we do not detect narrow ($W_r <$~3-4~km) rings coplanar with C1R and C2R
with equivalent width larger than about 20 meters. This is about ten times fainter than
the equivalent width of C2R (Fig.~\ref{fig_E_A_C1R}).
Note that this limit corresponds to extreme cases of 
either opaque rings with width  $\sim$~20~m, 
or semi-transparent rings of width $\sim$~3-4~km and normal opacity 0.007-0.005,
and all the intermediate solutions that keep $E_p(i)$ at 20~m.

\section{Concluding remarks}
\label{section_discussion}

We detected Chariklo and/or its rings during a total of thirteen 
stellar occultations between 2013 and 2016.
They demonstrate beyond any doubt that this Centaur is surrounded by a system of
two flat rings, C1R and C2R.
All the observations at hand are consistent with the circular ring solution of \cite{bra14},
with C1R orbiting at $391 \pm 3$~km from Chariklo center and 
with C2R orbiting outside C1R at an average distance of 14.8~km (Fig.~\ref{fig_profile_C2R_centre}).
This definitely rules out interpretations of the initial observation of June 3, 2013 by
a 3D dust shell, or a set of cometary-type jets being ejected from
the surface of the body.
In fact, 
the changing aspect of the rings seen during  the occultations is entirely attributable
the changing position of Chariklo relative to Earth, 
with a ring pole position that remains fixed in space (Fig.~\ref{fig_pole_circular}).

Our best resolved observation (April 29, 2014) reveals 
a W-shaped structure inside the main ring C1R (Fig.~\ref{fig_profile_C2R_centre}).
Moreover, the radial width $W_r$ of C1R measured on the best profiles exhibits 
significant variations with longitude, with a peak to peak variation 
of $\delta W_r \sim 2.5$~km between 5 and 7.5~km,
see Table~\ref{tab_param_rings} and Fig.~\ref{fig_W_vs_L}.
All the resolved profiles of C1R exhibit edges that are consistent 
with infinitely sharp boundaries, once diffraction and star diameter effects are accounted for.
The typical 1$\sigma$ upper limits for the edge transition zones is about one kilometer
(Table~\ref{tab_sharpness} and Fig.~\ref{fig_sharpness}).
Note finally that none of our observations permits to resolve the profile of ring C2R, 
whose width is constrained between 100~m to 1~km (Fig.~\ref{fig_E_A_C1R}).

Remarkably, C1R properties (W-shaped profile, variation of width with longitude and sharp edges) 
are reminiscent of the narrow eccentric ringlets found around Saturn \citep{fre16} or Uranus \citep{ellnic84,fre91}.
The maintenance of apse alignment could be due to self-gravity \citep{gol79},
viscous effects at the edges \citep{cha00}, or a
combination of self-gravity and viscous effects \citep{mos02}.
If validated, those models may provide insights into the ring physical parameters.
For instance,
the overdensities of material at some 100's~m from the edges (as seen in Fig.~\ref{fig_profile_C2R_centre})
is predicted by viscous models and deserve more detailed observational support in the case of Chariklo.
Also, the measure of the eccentricity gradient across the rings, $q_e$, 
could be related to the surface density of the ring material,
once Chariklo's dynamical oblateness $J_2$ is known \citep{pan16}.
However, our current data set is too fragmentary for drawing any reliable conclusions in that respect,
since both a comprehensive ring orbit model and the knowledge of Chariklo's $J_2$ are missing. 
 
In their simplest forms,
the Saturn or Uranus ringlets are described as sets of nested elliptical streamlines,
with a width that varies as $W_r = [1-q_e \cos(f)] \delta a$, where 
$f$ is  the true anomaly, 
$q_e= a \delta e/\delta a$ measures the eccentricity gradient across the ring, 
$\delta a$ and $\delta e$ being the changes of the semi-major axis $a$ and eccentricity $e$ across that ring. 
Consequently, 
the interpretation of Fig.~\ref{fig_W_vs_L} remains ambiguous, since 
only the true longitude corresponding to the events is currently known, 
while the true anomaly $f$ is unknown. 
In fact, any (expected) apse precession between observations impairs a correct
interpretation of that figure.
At this point, only
the total eccentricity variation across the ring can be estimated, i.e.
$\delta e =~\delta W_r/2a \sim~0.003$ from the estimations of $W_r$ and $a$ given above. 
This sets a lower limit of the same order for $e$, close to the eccentricity of Uranus' 
$\epsilon$ ring, 0.008 \citep{fre91}.

A much better case for modeling the rings would be to derive
$W_r$ vs. the ring radial excursion $r-a$ relative to the mean radius $r$.
The formula above predicts a linear behavior, u.
Unfortunately, the ring center is currently undetermined: 
we assume on the contrary a circular ring to derive it, and determine its pole.
The fact that the circular hypothesis provides satisfactory fits to our data, 
to within the accuracy of C1R's radius determination (some $\pm 3$~km), 
suggests that $r-a$ should also vary by a few kilometers at most.
In any case, the degeneracy between the ring eccentricity and its pole position can be lifted by
obtaining several multi-chord occultations and more accurate pole positions 
than shown in Fig.~\ref{fig_pole_circular} (and thus distinguish between projection
and eccentricity effects). 
Also, as expected apsidal precession rates are of the order of a couple of months \citep{sic16},
observations closer than that in time should be done to derive Chariklo's $J_2$. 
 

Turning now to the integral properties of the rings, we have determined the equivalent widths $E_p$
of C1R and C2R, when resolved, and the sum of the two when unresolved (Fig.~\ref{fig_E_A_C1R}).
We see that C1R, with $E_p(C1R) \sim$~2~km, contains about ten times more material than 
C2R, $E_p(C2R) \sim$~0.2~km. 
On one hand, if the equivalent width is constant within radial width, 
the ring can be considered as monolayer \citep{fre86},
as no shadowing by neighboring particules occurs (except in nearly edge-on view).
On the other hand, if the ring is polylayer, the equivalent depth is independent from $W_r$.
In that latter case, the equivalent width can be expressed 
as a function of ring width $W_r$ and the constant value of equivalent depth $\bar A_\tau$:
\begin{equation}
\label{eq:Ep_Wr}
E_p(W_r)=W_r (1- {\rm e}^{-2\bar{A_\tau}/W_r})
\end{equation}
(this equation, based on the work of \citealt{fre86} has been corrected by the factor 2 
in optical depth due to the diffraction by ring particules - see Appendix).
Fig.~\ref{fig_E_A_C1R} shows $E_p$ vs. $W_r$ 
assuming several values of $\bar A_\tau$ between 1.15 and 2~km for C1R
and between 0.15 and 0.4 km for C2R
(no real measurement of this parameter has been made in this work,
the lines show the expected trends - see Appendix).
Contrary to \cite{fre86}, the data do not allow any discrimination  
between $E_p$ or $A_\tau$ constant within the radial width.
Thus, no choice between the mono- or polylayer models can be made.
%

Finally, we have searched for faint material ring around the already discovered rings.
The best data set at hand provides 1$\sigma$ upper limits of $\sim20$~m for the equivalent width 
of narrow ($<$~3-4~km physical width) rings coplanar with C1R and C2R, up to distances of $12,000$~km
(counted in the ring plane). 
Note that in 2015, direct images of Chariklo have been recorded
using HST and SPHERE (\citealt{sic15b,sic15}).
The goal was to image the rings and/or look for possible shepherd 
satellite(s) and jets.
Considering material of same albedo as the rings (p=0.1), following limits have been inferred:
(1) no satellite bigger than $\sim 2$ km (being brighter than $V \sim 26.1$) up to 6400 km ($\sim 8$ times the ring size) 
from Chariklo center.
(2) no satellite bigger than $\sim1$ km  ($V\sim 27.5$) up to 8 arcsec.
For comparison the Hill radius is 7.5 arcsec.
(3) no jet, coma or material brighter than $V\sim28$ corresponding to jets of width $\sim10$~km
or material of optical depth of around $2 \times 10^{-5}$ per pixel.
Note that HST resolution did not allow to look closer than 1000 km from Chariklo's center,
so the rings were not detected.

Future observations will benefit greatly from the Gaia catalog.
A flavor of it has been provided by the Gaia-based prediction of the October 1$^{\rm st}$, 2016 occultation, 
which turned out to be correct to within 5 mas in declination 
(respectively 9 mas in right ascension), corresponding to about $50$ km (respectively 90 km).
The improvement of Chariklo's orbit stemming from successful occultation observations and
the sub-mas accuracy of forthcoming Gaia catalogs will provide predictions accurate to the 
few-kilometer level.
This will allow a much better distribution of stations (using portable instruments), 
with an optimal ring longitude coverage aimed at improving the ring orbital models.
It will also be possible to plan multi-wavelength observations to constrain the ring particle sizes. 
Multi-wavelength instruments are rare, and difficult to obtain unless a strong case is made, 
based on reliable predictions.
Higher SNR light curves will also be obtained in order to calculate the equivalent depths of both rings 
and definitely answer if the rings are monolayer or polylayer. 
Finally, the Gaia catalog will allow a much better coverage of Chariklo's limb,
which is currently poorly mapped. 
The general shape and local irregularities of the body will in turn have important 
consequences for a better understanding of the ring dynamics.

\appendix

\section{Appendix: Equivalent width and equivalent depth definitions}
\label{appendix}

We define $p'$ as the \textit{apparent} opacity of the ring. 
It measures the fractional drop of stellar flux $1-I/I_{0}$ as observed from Earth
(where $I_0$ and $I$ are the incident and transmitted fluxes, respectively).
Thus, 
$p'=0$ means a transparent ring and 
$p'=1$ means an opaque ring.
By ``apparent", we mean here as observed from Earth in the plane of the sky.
The apparent quantities will be primed hereafter to distinguish them for
the actual  quantities at the level of the ring, see below.
The apparent ring optical depth is defined as $\tau'= -\ln(1-p')$. 

Appropriate transformations, accounting for the ring opening angle $B$ and distance $D$ to the ring,
must be applied to derive the opacity $p_N$ and optical depth $\tau_N$ at the ring level,
where ``N" means normal to the ring plane.
Once this is done, one may define the 
equivalent width $E_p$ and 
equivalent depth $A_\tau$ 
of the ring as the integrals of $p_N$ and $\tau_N$, respectively, 
over the ring radial profile of width $W_{r}$ (measured radially in the plane of the ring):
\begin{equation}
\displaystyle
E_p= \int_{W_{r}} (v_r p_N) dt
\label{eq_E}
\end{equation}
\begin{equation}
A_\tau= \int_{W_{r}} (v_r \tau_N) dt,
\label{eq_A}
\end{equation}
where $v_r$ is the radial velocity of the star relative to Chariklo in the ring plane.

The quantities $E_p$ and $A_\tau$ are relevant for two extreme cases of ring structures.
One is a monolayer ring, in which case $p_N= |\sin(B)| \cdot p$ (for $|\sin(B)| \leq 1/p$),
where $p$ is the ring opacity as seen under an opening angle $B$.
The other model is a polylayer ring (where the ring thickness is much larger than the particle sizes),
in which case $\tau_N= |\sin(B)| \cdot \tau$, 
where $\tau$ is the ring optical depth, seen again under an angle $B$, see details in \cite{ell84}.

In principle, $E_p$ and and $A_\tau$ can be determined by numerically performing the integrations 
$|\sin(B)| \cdot \int (v_r p) dt$ and
$|\sin(B)| \cdot \int (v_r \tau) dt$ over the observed profiles.
Since the convolutions of the profiles by both Fresnel diffraction and stellar diameter 
conserve energy, those integrations provide the correct values of 
$E_p$ and and $A_\tau$.
Those two quantities are eventually measures of the amount of material (per unit length) 
contained along a radial cut of the ring, in their respective domains of validity (monolayer vs. polylayer),
see \cite{fre91}.

However, complications arise because of two effects: 
(1) the ring is not an uniform screen of opacity $p$, 
but rather a set of many particles that cover a fractional surface area $p$ of the ring,
while individually diffracting the incoming wavefront, and
(2) in several cases, the ring profiles are not resolved, 
i.e. the entire stellar drop occurs inside an individual acquisition interval, thus ``diluting"
the opacity $p$ over that interval.
We now comment these points in turn.

First, individual ring particles of radius $r$ 
diffract the incoming wave (with wavelength $\lambda$) over an Airy scale $F_A \sim (\lambda/2r)D$,
as seen by the observer at distance $D$ from the rings.
With $r \sim$ of a few meters and 
$D \sim 2 \times 10^{9}$~km, and using wavelengths in the visible range, 
we obtain $F_A >\sim 500$~km, which is significantly larger than typical values of a few kilometers for 
$W_{\perp}$, the width of ring as seen in the sky plane.
%
This results in a loss of light in the occultation profiles,
making the rings appear more opaque than they actually are.
It can be shown that the ring apparent optical depth $\tau'$ (in the sky plane) 
is actually twice as large as its actual value $\tau$,
i.e. the one  would have for an observer close to the ring: $\tau'= 2\tau$, see \cite{cuz85}. 
An equivalent way to describe that effect is to note that
the actual ring opacity $p$ is related to $p'$ by $(1-p)^2= 1-p'$.
Thus, the ring acts as a screen of amplitude for the incoming wave, 
instead of screen of intensity, see details in \cite{roq87}.

If the ring profile is resolved, it is enough to estimate numerically the integrals:
\begin{equation}
\displaystyle
E_p= |\sin(B)| \cdot v_r \int_{\rm profile} (1-\sqrt{1-p'}) dt 
\label{eq_E_square}
\end{equation}
\begin{equation}
A_\tau= -\frac{|\sin(B)|}{2} \cdot v_r \int_{\rm profile} \ln(1-p') dt 
\label{eq_A_square}
\end{equation}

The second point to examine is the fact that the ring profile may not be resolved
during the integration time $\Delta t$.
In this case $p'$ is not known, and the integrals above cannot be evaluated 
without an independent piece of information.
Let us consider the simple case of a uniform opacity $p$ across the ring profile (square-well model).
Then, the apparent equivalent width $E'= p' W_{\perp}$ 
(where $W_{\perp}$ is the width of the ring as observed in the sky plane)
can be evaluated from energy conservation by $E'= f' v_{\perp} \Delta t$, 
where $v_{\perp}$ is the velocity of the star normal to the ring in the sky plane,
and $f'$ is the fractional stellar drop during $\Delta t$.
From the definition of $E_p$ above (Eq.~\ref{eq_E_square}) and from $(1-p)^2= 1-p'$,
one obtains:
\begin{equation}
E_p= |\sin(B)| \frac{v_r}{v_{\perp}} \cdot \frac{E'}{2-p}  
\label{eq_EN_p}
\end{equation}
Since $0 \leq p \leq 1$, we have:
\begin{equation}
 |\sin(B)|\cdot \frac{v_r}{v_{\perp}} \cdot \frac{E'}{2} \leq E_p \leq |\sin(B)|\cdot \frac{v_r}{v_{\perp}} \cdot E',
 \label{eq_EN_limites_Eprime}
\end{equation}
i.e. a uncertainty factor of two, depending on the assumption on $p$.

For unresolved events, the fit of the best square-well model to the data allows measurements of $E_p$.
 The problem is that $p$ is badly constrained ($0 \leq p \leq 1$) by the fits.
  Eq.~\ref{eq_EN_limites_Eprime} shows that error bars will be much larger than for resolved events.
It could be possible to solve that problem by noting that $p' = E'/W_{\perp} = (E'/W_r) (v_r/v_\perp)$. 
As we know $W_r$ we can constrain $p'$, and thus $E_p$.
Assuming  that $W_{r,C1R+C2R}$ lies between 3 and 14 km (see Table~\ref{tab_param_rings}), 
the error bars values of $E_p$ remain similar to those without the width constraint. 
As we are not certain that 3 and 14 km are the width minimum and maximum, we choose not to use this constraint.

Note that the case of $A_\tau$ is in general harder to solve.
Even when the profile is resolved, the densest parts of the ring have 
high opacities $p' \sim 1$, and thus large uncertainties on $\tau'= -\ln(1-p')$
stemming from the data noise and uncertainties on the baseline levels (Fig.~\ref{fig_profile_C2R_centre}).
Consequently, we have not attempted to derive $A_\tau$ for our current data set.

\acknowledgments
 
\textit{Acknowledgments:}

The authors acknowledge support from the French grants 
``Beyond Neptune"  ANR-08-BLAN-0177 and
``Beyond Neptune II"  ANR-11-IS56-0002.
Part of the research leading to these results has received funding from
the European Research Council under the European Community's H2020
(2014-2020/ERC Grant Agreement no. 669416 ``LUCKY STAR").
This work is partly based on observations performed at the European Southern Observatory (ESO), 
proposals 092.C-0186(B) and 092.C-0186(C), and 
on observations made at the South African Astronomical Observatory (SAAO).
This work has made use of data obtained at the Thai National Observatory on Doi Inthanon, operated by NARIT.
Technical support was provided by 
G. Hau and P. Kabath for the February 16, 2014 observation at ESO/VLT, and
by G. Rom\'an for the July 25, 2016 observation with the Dobson 60-cm telescope
at Granada.
A. Maury acknowledges the use of Caisey Harlingten's 50cm telescope for the February 16, 2014 occultation.
The 50 cm telescopes used for the Hakos observations belong to the IAS observatory at Hakos/Namibia. 
E. Jehin is a FNRS Research Associate.
TRAPPIST is a project funded by the Belgian Fund
for Scientific Research (Fonds National de la Recherche Scientifique, F.R.S.-FNRS) under grant FRFC 2.5.594.09.F
A. Pennell and P.-D. Jaquiery thank the Dunedin Astronomical Society.
E. Meza acknowledges support from the Contrato de subvenci\'on 205-2014 Fondecyt - Concytec, Per\'u.
M. Assafin thanks the CNPq (Grants 473002/2013-2 and
308721/2011-0) and FAPERJ (Grant E-26/111.488/2013).
G.Benedetti-Rossi acknowledges for the support of the CAPES (203.173/2016) and FAPERJ/PAPDRJ (E26/200.464/2015 - 227833) grants.
R.Vieira-Martins thanks grants: CNPq-306885/2013, Capes/Cofecub-2506/2015, Faperj:  
PAPDRJ-45/2013 and E-26/203.026/2015.
The research leading to these results has received funding from the European
Union's Horizon 2020 Research and Innovation Programme, under Grant
Agreement No. 687378, project SBNAF.
The authors acknowledges the use of Sonja Itting-Enke's C14 telescope and the facilities at the 
Cuno Hoffmeister Memorial Observatory (CHMO),
the use of the Skywatcher 16" telescope of the Deutsche H\"ohere Privatschule (DHPS)
in Windhoek
and the use of the Meade 14 telescope of Space Observation Learning (Rob Johnstone).
Funding from Spanish grant
AYA-2014-56637-C2-1-P is acknowledged, as is the Proyecto de
Excelencia de la Junta de Andaluc\'ia, J. A. 2012-FQM1776.
J.I.B. Camargo acknowledges CNPq grants 308489/2013-6 and 308150/2016-3.
The research leading to these results has received funding from the European Union?s Horizon 2020
 Research and Innovation Programme, under Grant Agreement N. 687378, project SBNAF.

\clearpage

\bibliographystyle{Icarus}
\bibliography{biblio}

\clearpage

\begin{deluxetable}{llllll}
\tablecolumns{6}
\tablewidth{0pc}
\tabletypesize{\scriptsize}
\tablecaption{Circumstances of positive detections (main body and/or rings)
\label{tab_circumstances_positives}}
\tablehead{
\multicolumn{6}{c}{Date}\\
\multicolumn{6}{c}{Rmag (NOMAD catalog), $(\alpha,\delta)$ star coordinates, $\theta_\star$ stellar diameter$^{(a)}$}\\
\multicolumn{6}{c}{$(\alpha_{Ck},\delta_{Ck})$ derived Chariklo's geocentric coordinates at specified date}\\
\hline
\hline
\colhead{Site} & \colhead{Longitude}   & \colhead{Telescope} & \colhead{Instrument}           & \colhead{Observers} & \colhead{Results}\\
        &  \colhead{Latitude}&                  &   \colhead{Exposure Time (s)}  & & \\
        &   \colhead{ Altitude (m)} &                  &  & &         
}
\startdata
 \multicolumn{6}{c}{\textbf{June 3, 2013}}\\
\multicolumn{6}{c}{$R=12.070$, $\alpha=16\rm h 56\rm m 06.4876\rm s$, $\delta=-40\degr 31\arcmin 30.205\arcsec$, $\theta_\star=2.18$~km  }\\
\multicolumn{6}{c}{at 06:25:30 UT: $\alpha_{Ck}=16\rm h  56\rm m 06.3202\rm s$, $\delta_{Ck}=-40\degr  31\arcmin 30.2803\arcsec$}\\
\hline
\hline 
\\
\multicolumn{6}{c}{ See details in \cite{bra14}}\\
\\
\hline
\hline  
 
\multicolumn{6}{c}{\textbf{February 16, 2014}}\\
\multicolumn{6}{c}{$R=16.$, $\alpha=17\rm h 35\rm m 55.3333\rm s$, $\delta=-38\degr 05\arcmin 17.184\arcsec$, $\theta_\star=0.265$~km  }\\
\multicolumn{6}{c}{at 07:45:35 UT: $\alpha_{Ck}=17\rm h  35\rm m 54.980\rm s$, $\delta_{Ck}=-38\degr  05\arcmin 17.449\arcsec$}\\
\hline
\hline 

Paranal           & 24 37 31.      S       & UT4  8.2 m           & HAWK-I           & F. Selman, C. Herrera      &C1R and C2R \\
Chile                        & 70  24 07.95  W    &       H-filter                           &  0.25        & G. Carraro, S. Brillant         &  partially  \\
                        & 2635.43                   &                            &               & C. Dumas, V. D. Ivanov            &  resolved \\

San Pedro Atacama   & 22 57 12.3      S    & 50 cm                &      APOGEE U42 & A. Maury                        &   Main body     \\
Chile                                       & 68 10 47.6  W       &               &         10              & N. Morales   &     \\
                                       & 2397                     &                              &               &                                        &         \\

\hline
\hline 
\multicolumn{6}{c}{\textbf{March 16, 2014}}\\
\multicolumn{6}{c}{$R=15.45$, $\alpha=17\rm h 40\rm m 39.8690\rm s$, $\delta=-38\degr 25\arcmin 46.887\arcsec$, $\theta_\star=0.121$~km  }\\
\multicolumn{6}{c}{at 20:31:45 UT: $\alpha_{Ck}=17\rm h  40\rm m 39.7743\rm s$, $\delta_{Ck}=-38\degr  25\arcmin 46.4198\arcsec$}\\
\hline
\hline 
Doi Inthanon &  18  34  25.41 N& TNT 2.4 m &ULTRASPEC & P. Irawati & C1R and C2R \\
Thailand &98 28 56.06 E  &R'-filter &3.3 & A. Richichi& unresolved\\
& 2450 & & & & \\
\hline
\hline 
\multicolumn{6}{c}{\textbf{April 29, 2014}}\\
\multicolumn{6}{c}{$R_A^{(b)}=12.72$, $\alpha_A^{(b)}=17\rm h 39\rm m 02.1336\rm s$, $\delta_A^{(b)}=-38\degr 52\arcmin 48.801\arcsec$, $\theta_{\star A}^{(b)}=0.199$~km  }\\
\multicolumn{6}{c}{at 23:14:12 UT: $\alpha_{Ck}=17\rm h  39\rm m 01.7943\rm s$, $\delta_{Ck}=-38\degr  52\arcmin 48.858\arcsec$}\\
\hline
\hline

SAAO        & 32 22 46.0 S    & 1.9 m             &  SHOC          &   H. Breytenbach        &    C1R and C2R   \\
Sutherland                   & 20 48  38.5 E    &                      &   0.0334                &   A. A. Sickafoose       &  resolved  \\
South Africa                            &         1760                    &                      &              &     &    Main body     \\

Gifberg            & 31  48  34.6   S        & 30 cm            & Raptor Merlin 127   &  J.-L. Dauvergne  &  Grazing C2R \\
South Africa                         & 18  47  0.978 E       &                        & 0.047         &   P. Schoenau    &                  \\
                         & 338                           &                         &            &                &          \\

Springbok      & 29 39 40.2 S       & 30 cm                           & Raptor Merlin 127                   & F. Colas    & C1R and C2R     \\
South Africa                        & 17 52 58.8  W       &                                     & 0.06                          &  C. de Witt   &sharp and resolved  \\
                        & 900           &                                     &                                   &   &          \\

\hline
\hline 
\multicolumn{6}{c}{\textbf{June 28, 2014}}\\
\multicolumn{6}{c}{$R=13.65$, $\alpha=17\rm h 24\rm m 50.3821\rm s$, $\delta=-38\degr 41\arcmin 05.609\arcsec$, $\theta_\star=0.167$~km  }\\
\multicolumn{6}{c}{at 22:24:35 UT: $\alpha_{Ck}=17\rm h  24\rm m 50.2954\rm s$, $\delta_{Ck}=-38\degr  41\arcmin 05.7445\arcsec$}\\
\hline
\hline 
Hakos & 23 14 11  S& 50 cm AK3 & Raptor Merlin 127 &K.-L. Bath & C1R and C2R\\
Namibia & 16  21 41.5  E& &0.2 & & unresolved\\
                     & 1825                 &                               &                        &          &                       \\

Kalahari       &26 46 26.91 S& 30 cm  & Raptor Merlin 127 & L. Maquet&Main body\\
South Africa       & 20 37 54.258 E& &0.4 & &\\
 & 861 & & & &\\
Twee Rivieren& 26 28 14.106 S & 30 cm   &  Raptor Merlin 127 &J.-L. Dauvergne &Main body\\
South Africa       & 20 36 41.694 E& &0.4 & &\\
 &883 & & & & \\*    
\hline
\hline 
\multicolumn{6}{c}{\textbf{April 26, 2015}}\\
\multicolumn{6}{c}{$R=12.04$, $\alpha=18\rm h 10\rm m 46.1450\rm s$, $\delta=-36\degr 38\arcmin 56.368\arcsec$, $\theta_\star=0.361$~km  }\\
\multicolumn{6}{c}{at 02:11:58 UT: $\alpha_{Ck}=18\rm h  10\rm m 45.9676\rm s$, $\delta_{Ck}=-36\degr  38\arcmin 56.608\arcsec$}\\
\hline
\hline 
Los Molinos        & 34 45 19.3 S    &      OALM            &   FLI CCD   & S. Roland    & CR and C2R \\
 Uruguay    & 56 11 24.6 W    &      46 cm     &      0.8          &  R. Salvo               & unresolved   \\
     & 130     &        &             &  G. Tancredi & \\
     \hline
\hline 
\multicolumn{6}{c}{\textbf{May 12, 2015}}\\
\multicolumn{6}{c}{$R=15.93$, $\alpha=18\rm h 08\rm m 29.2962\rm s$, $\delta=-36\degr 44\arcmin 56.814\arcsec$, $\theta_\star=0.219$~km  }\\
\multicolumn{6}{c}{at 17:55:40 UT: $\alpha_{Ck}=18\rm h  08\rm m 29.2447\rm s$, $\delta_{Ck}=-36\degr  44\arcmin 56.7965\arcsec$}\\
\hline
\hline
Samford Valley        &  27 22 07.00 S                   &       35 cm    &   G-star  & J. Bradshaw   & Main Body \\
Australia                          & 152 50 53.00 E                  &                              &      0.32      &      & Emersion of unresolved  \\
                            &			80	&				&			& &rings only \\ 
\hline
\hline 
\multicolumn{6}{c}{\textbf{July 25, 2016}}\\
\multicolumn{6}{c}{$R=14.02$, $\alpha=18\rm h 20\rm m 35.3645\rm s$, $\delta=-34\degr 02\arcmin 29.590\arcsec$, $\theta_\star=0.234$~km  }\\
\multicolumn{6}{c}{at 23:59:00 UT: $\alpha_{Ck}=18\rm h  20\rm m 35.3640\rm s$, $\delta_{Ck}=-34\degr  02\arcmin 29.0378\arcsec$}\\
\hline
\hline
Liverpool Telescope            &    28 45 44.8 N     &           2 m  &   RISE   & J.-L. Ortiz   & Main Body  \\
 Canary Islands                      &  17 52 45.2 W     &                        &    0.6      &   N. Morales  &            \\
                         &               2363             &                         &            &            &           \\
\hline
\hline 
\multicolumn{6}{c}{\textbf{August 08, 2016}}\\
\multicolumn{6}{c}{$R=13.67$, $\alpha=18\rm h 18\rm m 03.6927\rm s$, $\delta=-33\degr 52\arcmin 28.392\arcsec$, $\theta_\star=0.204$~km  }\\
\multicolumn{6}{c}{at 19:57:00 UT: $\alpha_{Ck}=18\rm h  18\rm m 03.8297\rm s$, $\delta_{Ck}=-33\degr  52\arcmin 28.181\arcsec$}\\
\multicolumn{6}{c}{or $\alpha_{Ck}=18\rm h  18\rm m 03.8449\rm s$, $\delta_{Ck}=-33\degr  52\arcmin 28.196\arcsec$}\\
\hline
\hline
Windhoek (CHMO)       & 22  41 54.5  S       & 35 cm           &       ZWO / ASI120MM       & H.-J. Bode   &  Main Body\\
Namibia                        & 17  06  32.0  E        &                        & 1                &     &    C1R and C2R   \\
                        & 1920           &                        &                          &  &  unresolved \\

\hline
\hline 
\multicolumn{6}{c}{\textbf{August 10, 2016}}\\
\multicolumn{6}{c}{$R=16.53$, $\alpha=18\rm h 17\rm m 47.3492\rm s$, $\delta=-33\degr 51\arcmin 02.516\arcsec$, $\theta_\star=0.053$~km  }\\
\multicolumn{6}{c}{at 14:23:00 UT: $\alpha_{Ck}=18\rm h  17\rm m 47.3089\rm s$, $\delta_{Ck}=-33\degr  51\arcmin 02.478\arcsec$}\\
\hline
\hline
Murrumbateran   &  34 57 31.50 S    & 40 cm                 & WATEC 910BD         &   D. Herald & Main Body\\
Australia                      &   148 59 54.80 E    &                            & 0.64                        &    &  \\
                       &         594     &                            &                                              &                     &             \\
\hline
\hline 
\multicolumn{6}{c}{\textbf{August 10, 2016}}\\
\multicolumn{6}{c}{$R=16.22$, $\alpha=18\rm h 17\rm m 46.4827\rm s$, $\delta=-33\degr 50\arcmin 57.826\arcsec$, $\theta_\star=0.083$~km  }\\
\multicolumn{6}{c}{at 16:43:00 UT: $\alpha_{Ck}=18\rm h  17\rm m 46.4457\rm s$, $\delta_{Ck}=-33\degr  50\arcmin 57.523\arcsec$}\\
\hline
\hline
Les Makes   & 21  11  57.4     S    & 60 cm                 & Raptor Merlin 127         &   F. Vachier & Main Body\\
 La R\'eunion                      & 55  24  34.5  E      &                            & 2                          &    &  \\
                       & 972             &                            &                                              &                     &             \\
\hline
\hline 
\multicolumn{6}{c}{\textbf{August 15, 2016}}\\*
\multicolumn{6}{c}{$R=14.64$, $\alpha=18\rm h 17\rm m 06.2228\rm s$, $\delta=-33\degr 46\arcmin 56.315\arcsec$, $\theta_\star=0.103$~km  }\\*
\multicolumn{6}{c}{at 11:38:00 UT: $\alpha_{Ck}=18\rm h  17\rm m 06.1638\rm s$, $\delta_{Ck}=-33\degr  46\arcmin 56.513\arcsec$}\\*
\hline
\hline
 Darfield       & 43 28 52.90 S                   &  25 cm                & WATEC 910 BD     & B. Loader  & Main Body \\*
New Zealand         &172 06 24.40 E                  &                             &      2.56          &       &             \\*
                            & 210                                  &                             &             &  &\\
\hline
\hline 
\multicolumn{6}{c}{\textbf{October 1, 2016}}\\
\multicolumn{6}{c}{$R=15.36$, $\alpha=18\rm h 16\rm m 20.0796\rm s$, $\delta=-33\degr 01\arcmin 10.756\arcsec$, $\theta_\star=0.119$~km  }\\
\multicolumn{6}{c}{at 10:10:00 UT: $\alpha_{Ck}=18\rm h  16\rm m 20.0324\rm s$, $\delta_{Ck}=-33\degr  01\arcmin 10.841\arcsec$}\\
\hline
\hline
Rockhampton         & 23 16 09.00 S                    &           30 cm        & WATEC 910BD     & S. Kerr   & Main Body  \\
 Australia                          & 150 30 00  E                  &                             &       0.320         &          & C1R and C2R         \\
                            & 50                                  &                             &             & &unresolved\\

Adelaide         & 34 48 44.701 S    & 30 cm             &     QHY 5L11      &   A. Cool            & Main body     \\
Heights School                  & 138 40 56.899 E    &                      &   1                &   B. Lade    & C1R and C2R        \\
 Australia                           &    167                          &                      &             &       &  unresolved  \\
\hline
\multicolumn{6}{l}{$^{(a)}$ projected at Chariklo's distance (using \citealt{van99}, except for the April 29, 2014, see text for details).}\\
\multicolumn{6}{l}{$^{(b)}$ The index $A$ refers to the primary of the binary star.}\\
\enddata \\
\end{deluxetable}

\clearpage


\begin{deluxetable}{lllll}
\tablecolumns{5}
\tablewidth{0pc}
\tabletypesize{\scriptsize}
\tablecaption{Circumstances of observations that detected no event or during which no data was acquired
\label{tab_circumstances_negatives}}
\tablehead{
\colhead{Site} & \colhead{Longitude}   & \colhead{Telescope} & \colhead{Instrument}           & \colhead{Observers} \\
        &  \colhead{Latitude}&                  &   \colhead{Exposure Time (s)}  &  \\
        &   \colhead{ Altitude (m)} &                  &  &         
}
\startdata
\multicolumn{5}{c}{\textbf{February 16, 2014}}\\
\hline
\hline 
Cerro Tololo   & 30 10 03.36 S      &   0.4 m  & PROMPT     &   J. Pollock     \\
Chile                         & 70 48 19.01  W    &  4 telescopes       &      6.0/2.0             &                         \\
                         & 2207                     &                               &                   &                          \\

La Silla           & 29 15 16.59 S    &   TRAPPIST  &      FLI PL3041-BB                   & E. Jehin     \\
Chile                        & 70  44 21.82  W  &       60 cm                          &     4.5             &                      \\
                        & 2315                    &                                 &                              &                      \\

La Silla           & 29 15 32.1 S    & NTT 3.55 m           & SOFI             & L. Monaco    \\
Chile                        & 70  44 01.5  W  &       H-filter                         &  0.05     & + visitor team           \\
                        & 2375                    &                             &                    &                                  \\
\hline
\hline 
\multicolumn{5}{c}{\textbf{April 29, 2014}}\\
\hline
\hline 
Hakos           & 23 14 50.4  S          & 50 cm AK3            & Raptor Merlin 127            & K.-L. Bath      \\
Namibia                      & 16  21 41.5  E          &                               &  0.075              &          \\
                      & 1825                 &                               &                        &                                  \\

Hakos           & 23 14 50.4 S          & 50 cm RC50            & i-Nova             & R. Prager      \\
 Namibia                     & 16  21 41.5  E          &                                &  1.                    &          \\
                      & 1825                &                               &                        &                                 \\

Windhoek (CHMO)       & 22  41 54.5  S       & 35 cm           & Raptor Merlin 127              & W. Beisker     \\
Namibia                        & 17  06  32.0  E        &                        & 0.1                &           \\
                        & 1920           &                        &                          &              \\

\hline
\hline 
\multicolumn{5}{c}{\textbf{June 28, 2014}}\\
\hline
\hline 

Les Makes   & 21  11  57.4     S    & 60 cm                 & WATEC 910HX         &  A. Peyrot \\
 La R\'eunion                      & 55  24  34.5  E      &                            & 0.4                          &  J-P. Teng    \\
                       & 972             &                            &                                              &                                         \\
        
\hline
\hline 
\multicolumn{5}{c}{\textbf{April 26, 2015}}\\
\hline
\hline
Bigand        & 33 26 11 S   &     15 cm       &  Canon Ti         &   S. Bilios              \\
Provincia Santa F\'e                  & 61 08 24 W  &                        &   5               &           \\
Argentina                            &  90           &                      &            &            \\

Bigand        & 33 26 11 S   &     15 cm       &  Canon EOS        &   J. Nardon             \\
Provincia Santa F\'e & 61 08 24 W  &      &   3.2    &            \\
Argentina       &  90           &                &            &            \\

La Silla        & 29 15 16.6 S    &    TRAPPIST           &    FLI PL3041-BB         &   E. Jehin            \\
Chile             & 70 44 21.8 W    &         60 cm             &     4.5       &            \\
                             & 2315        &                      &           &                  \\

Bosque Alegre            & 31 35 54.0 S    & 76 cm            & QHY6     &  R. Melia    \\
Argentina                         & 64 32 58.7 W     &                        & 1.2        &   C. Colazo                      \\
                         & 1250                          &                         &           &                         \\

Santa Rosa      & 36 38 16 S       & 20 cm                        &      Meade DSI-I                & J. Spagnotto      \\
 Argentina                       & 64 19 28  W       &                                     & 3                          &      \\
                        & 182           &                                     &                                   &    \\

Santa Martina           & 33 16 09.0 S          & 40 cm           & Raptor Merlin 127            & R. Leiva Espinoza    \\
 Chile                     & 70 32 04.0 W          &                                &          0.5            &          \\
                      & 1450                 &                               &                        &                                \\

Buenos Aires (AAAA)       & 34 36 16.94 S       & 25 cm          & ST9e              & A. Blain    \\*
 Argentina                & 58 26 04.37 W     &                        & 4                &           \\*
                        & 0           &                        &                          &             \\
\hline
\hline 
\multicolumn{5}{c}{\textbf{May 12, 2015}}\\
\hline
\hline
Reedy Creek        &  28 06 30.4 S    &  25 cm &    WATEC 120N+       &   J. Broughton                \\
Australia                  & 153 23 52.90 E   &                       &     0.64       &            \\
                            &      66      &                      &          &             \\

\hline
\hline 
\multicolumn{5}{c}{\textbf{July 25, 2016}}\\
\hline
\hline

Granada        &     37 00 38.49 N               &  60 cm                 & Raptor Merlin 127     & S. Alonso     \\
 Spain             &      03 42 51.39 W             &                             &         0.4       &  A. Rom\'an                  \\
                            &        1043                           &                             &             &  \\
                            
Albox        &   37 24 20.0 N &       40 cm       &     Atik 314L+       &       J.-L. Maestre            \\
Spain                  &   02 09 6.5 E  &                      &     3             &          \\
                            &        493                           &                             &             &    \\
\hline
\hline 
\multicolumn{5}{c}{\textbf{August 10, 2016 - 14h UT}}\\
\hline
\hline

Blue Mountains            & 33 39 51.9 S       &    30 cm         &   WATEC 910BD   &  D. Gault    \\
 Australia                        &150 38 27.9 E       &                        &    5.12     &                      \\
                         & 286                          &                         &            &                       \\
                         
Samford Valley           & 27 22 07.00 S       &    35 cm         &   WATEC 910BD   &  J. Bradshaw    \\
 Australia                        &152 50 53.00 E       &                        &    0.64     &                      \\
                         & 80                          &                         &            &                       \\

Rockhampton            & 23 16 09.00 S        &     30 cm        &  WATEC 910BD    &  S. Kerr     \\
 Australia                        & 150 30 00.00 E&                        &   1.28    &                      \\
                         & 50                           &                         &            &                   \\

Dunedin      &   45 52 20.83   S   &            36 cm                &        Raptor Merlin 127    & F. Colas        \\
New Zealand                        & 170 29 29.90 E      &                                     &            2.               & A.  Pennell   \\
                        &    154       &                                     &                                   &   P.-D. Jaquiery  \\

   Sydney        &  33 48 35.04 S         & 36 cm          &     Raptor Merlin 127         &  H. Pavlov   \\
 Australia                     & 150 46 36.90 E&                               &         2.2       &          \\
                      & 37                 &                               &                        &                                \\

\hline
\hline 
\multicolumn{5}{c}{\textbf{August 15, 2016}}\\
\hline
\hline
Canberra         & 35 11 55.30 S    &    40 cm         &    WATEC 910BD        &   J. Newman                 \\
Australia                  &149 02 57.50 E    &                      &      2.56       &              \\
                            &        610                     &                      &             &            \\

Murrumbateran         &   34 57 31.50 S  &    40 cm         &    WATEC 920BD         &   D. Herald                 \\
Australia                  &  148 59 54.80 E  &                      &      0.32         &              \\
                            &          594                    &                      &             &            \\
                            
Greenhill Observatory        & 42 25 51.8 S    &       1.3 m       &      Raptor Merlin 127      &   K. Hill                \\
Tasmania               & 147 17 15.8 E    &                      &     0.5       &      A. Cole       \\
                             &641      &                      &           &                                                 \\

Rockhampton            & 23 16 09.00 S        &     30 cm        &  WATEC 910BD    &  S. Kerr     \\
 Australia                        & 150 30 00.00 E&                        &   1.28    &                      \\
                         & 50                          &                         &            &                   \\

Linden Observatory      &    33 42 27.3 S   &            76 cm                &         Grasshopper     & D. Gault        \\
Australia                        &150 29 43.5 E      &                                     &            Express with ADVS               & R. Horvat    \\
                        & 574         &                                     &	0.533	&    R.A. Paton   \\
                        & & & &L. Davis \\

WSU Penrith Observatory           &33 45 43.31 S         & 62 cm            &     Raptor Merlin 127         &  H. Pavlov   \\
 Sydney                     & 150 44 30.30 E&                               &         2       &   D. Giles       \\
Australia                      & 60                &                               &                        &         D. Maybour                       \\
                      &                  &                               &                        &        M. Barry                       \\
\hline
\hline 
\multicolumn{5}{c}{\textbf{October 1, 2016}}\\
\hline
\hline
Blue Mountains            & 33 39 51.9 S       &    30 cm         &   WATEC 910BD   &  D. Gault    \\
 Australia                        &150 38 27.9 E       &                        &    0.64     &                      \\
                         & 286                          &                         &            &                       \\

Linden Observatory      &33 42 27.3 S      &        76 cm                    &         Grasshopper                & M. Barry        \\
Australia                        & 150 29 43.5 E      &                                     &           Express with ADVS                &     \\
                        & 574           &                                     &                    0.27               &          \\

Miles           & 26 39 20.52 S         &     25 cm       &      WATEC 120N+        & D. Dunham     \\
Australia                      & 150 10 19.44 E         &                               &        0.64     &   J. Dunham       \\
                      & 277                 &                               &                        &                                  \\
            
Reedy Creek        &  28 06 30.4 S    &  25 cm &    WATEC 120N+       &   J. Broughton                \\
Australia                  & 153 23 52.90 E   &                       &    1.28       &            \\
                            &      66      &                      &          &             \\
                            
Samford Valley           & 27 22 07.00 S       &    35 cm         &   WATEC 910BD   &  J. Bradshaw    \\
 Australia                        &152 50 53.00 E       &                        &    0.16     &                      \\
                         & 80                          &                         &            &       \\
\hline
\multicolumn{5}{l}{The following stations were cloudy or had technical failure, no data were acquired:}\\
\multicolumn{5}{l}{February 16, 2014: Santa Martina (Chile), Bosque Alegre (Argentina)}\\
\multicolumn{5}{l}{April 29, 2014 : Rodrigues, Sainte Marie, Les Makes (La R\'eunion) ; Calitzdorp and LCOGT (South Africa)}\\
\multicolumn{5}{l}{April 26, 2015 : Cerro Tololo (Chile)}\\
\multicolumn{5}{l}{July 25, 2016: TRAPPIST Nord (Marocco), TAD (Canary Islands), Teide Observatory (Canary Islands)}\\
\multicolumn{5}{l}{August 8, 2016 : Les Makes (La R\'eunion)}\\
\multicolumn{5}{l}{August 15, 2016: Mount John Observatory, Dunedin, Bootes-3, Wellington (New Zealand)}\\
\multicolumn{5}{l}{October 1, 2016: Murrumbateran, Canberra (Australia)}\\
\enddata \\
\end{deluxetable}

\normalsize
\clearpage

\begin{table}
\caption{Sharpness of C1R edges, $\Delta w_r$, from April 29, 2014 events
\label{tab_sharpness}}
\begin{center}
\begin{tabular}{ccc}
\\
\hline
\hline
Event & Inner edge (km) & Outer edge (km) \\
&\multicolumn{2}{c}{ ($1\sigma$ level)}\\
\hline
\hline
Springbok Ingress &
1.1 & 1.1 \\
Springbok Egress &
1.2 & 1.5 \\
\hline
SAAO Ingress &
0.6 & 0.9 \\
SAAO Egress &
0.8 & 0.4 \\
\hline
\end{tabular}
\end{center}
\end{table}

\newpage


\begin{deluxetable}{llccccccc}
\tablecolumns{9}
\tablewidth{0pc}
\tabletypesize{\scriptsize}
\tablecaption{Ring occultation timings and derived physical parameters (resolved events)
\label{tab_param_rings}}
\tablehead{
\colhead{Date} &\colhead{Event} & \colhead{ $t_{0}$ UT$^{(a)}$} & \colhead{$v_\perp^{(b)}$}&\colhead{$v_r^{(b)}$}& \colhead{$L^{(c)}$} & \colhead{$W_r^{(d)}$}& \colhead{ $E_p^{(e)}$}& \colhead{$p_N^{(e)}$ }\\
& &  & \colhead{(km/s)}&\colhead{(km/s)} &\colhead{(deg.)}& \colhead{(km)} &  \colhead{(km)} &
}
\scriptsize
\startdata
\hline
\multicolumn{9}{c}{C1R} \\
\hline
\hline
Jun 3, 2013 &Danish ingress$^{(f)}$ & 06:25:21.166$\pm$0.0007    & 20.345 & 36.113 &  341.76   &  $6.16\pm0.11$           
& $1.90 \pm 0.022$ & $0.308 \pm 0.003$  \\ 
&Danish egress$^{(f)}$ & 06:25:40.462$\pm$0.0012      & 22.031 & 36.504  &  124.38     &  $7.14 \pm 0.04$  
& $1.73 \pm 0.023$ & $0.24 \pm 0.004$ \\
\hline
Feb 16, 2014 &VLT ingress &07:45:25.541$_{-0.004}^{+0.010}$&19.532&28.794&183.37&$5.316_{-1.916}^{+0.868}$& $1.996^{+0.092}_{-0.031}$ & $0.375^{+0.125}_{-0.025}$\\
&VLT egress &07:45:45.133$_{-0.332}^{+0.313}$&21.293&29.602&300.99&$4.833_{-0.476}^{+1.667}$&$2.04^{+0.36}_{-0.14}$ & $0.443^{+0.078}_{-0.103}$\\
\hline
Apr 29, 2014 &Springbok ingress  & 23:14:25.884$\pm$0.007  & 13.432 & 16.493 & 287.42   &  $5.575\pm0.398$           
& $1.80_{-0.143}^{+0.122} $ & $0.3125^{+0.024}_{-0.027}$  \\ 
&Springbok egress  & 23:15:04.362$\pm$0.006    & 10.720 & 16.655  &  157.83     &  $6.75^{+0.48}_{-0.21}$  
& $2.595_{-0.166}^{+0.148}$ & $0.33^{+0.017}_{-0.033}$ \\ 
\hline
&SAAO ingress  & 23:13:56.191$\pm$0.007    & 12.756 & 13.895 &   266.656   &  $5.68\pm0.2$           
& $1.88^{+0.22}_{-0.12}$ & $0.32^{+0.037}_{-0.021}$  \\ 
&SAAO egress& 23:14:28.964$\pm$0.008      & 9.260 & 14.249  &   198.899     &  $6.625\pm0.2$  
& $1.695^{+0.175}_{-0.115}$ & $0.241^{+0.024}_{-0.022}$ \\ 
\hline
\hline
\multicolumn{9}{c}{C2R} \\
\hline
\hline
Jun 3, 2013 &Danish ingress$^{(f)}$  & 06:25:20.765$\pm$0.011     & 20.412 & 36.283 & 341.76   &  $3.380^{+1.424}_{-1.797}$           
& $0.168 \pm 0.02$ & $0.05^{+0.05}_{-0.01}$  \\ 
&Danish egress$^{(f)}$ & 06:25:40.847$\pm$0.006   & 22.029 & 36.632  & 124.38     &  $3.231^{+0.899}_{-1.124}$  
& $0.228 \pm 0.02$ & $0.07^{+0.03}_{-0.01}$ \\ 
\hline
Feb 16, 2014 &VLT ingress &07:45:25.285$_{-0.033}^{+0.057}$&19.532&28.794&183.37&$5.053_{-2.385}^{+1.000}$& $0.491^{+0.445}_{-0.227}$ & $0.091^{+0.495}_{-0.00}$ \\
&VLT egress &07:45:45.473$_{-0.053}^{+0.037}$&21.293&29.602&300.99&$3.333_{-1.333}^{+1.667}$&$0.522^{+0.078}_{-0.050}$ & $0.119^{+0.609}_{-0.118}$\\
\hline
Apr 29, 2014 &Springbok ingress    & 23:14:24.990$\pm$0.020                &      13.430                   &  16.460 &   287.42 & $0.34^{+1.37}_{-0.24}$ & $0.125_{-0.064}^{+0.076} $  & $0.368^{+0.632}_{-0.288}$ \\ 
&Springbok egress   & 23:15:5.324$\pm$0.019               &      10.722                   &  16.620 &   157.832 & $0.6^{+1.7}_{-0.1}$ & $0.253_{-0.069}^{+0.079}$  & $0.582^{+0.32}_{-0.45}$ \\ 
\hline
&Gifberg ingress$^{(g)}$ & 23:14:30.109$^{+0.015}_{-0.008}$  & $(g)$& $(g)$& 227.190 &  $0.522^{+0.227}_{-0.399}$&   $0.090^{+0.039}_{-0.000}$ &  $0.186^{+0.814}_{-0.043}$\\
&Gifberg egress$^{(g)}$ & 23:14:33.750$\pm$0.008 &$(g)$ &$(g)$& 217.761 &  $0.181^{+0.008}_{-0.091}$&  $0.129^{+0.000}_{-0.039}$ &  $0.814^{+0.186}_{-0.671}$\\
\hline
\enddata
\tablenotetext{(a)}{ $t_{0}$ is the mid-time of the event in hours:min:sec. The error bars quoted are given at $1\sigma$ level. }
\tablenotetext{(b)}{$v_\perp$ and $v_r$ are respectively the perpendicular velocity in the sky plane and the radial velocity in the ring plane.}
\tablenotetext{(c)}{ $L$ is the true longitude counted from the J2000 ring plane ascending node.}
\tablenotetext{(d)}{ $W_{r}$ is the radial width, measured in the plane of the rings.}
\tablenotetext{(e)}{ $E_{p}$ is the equivalent width: $E_{p}=W_{r} \cdot p_{N}$ where $p_{N}$ is the normal opacity in the plane of the rings.}
\tablenotetext{(f)}{Timings given by \cite{bra14}.}
\tablenotetext{(g)}{As the occultation was grazing, the velocity changes consequently between ingress and egress and to give fixed values is not relevant in this case (see text section~\ref{section_width_rings})}

\end{deluxetable}

\clearpage

\begin{deluxetable}{llcccccc}
\tablecolumns{7}
\tablewidth{0pc}
\tabletypesize{\scriptsize}
\tablecaption{Ring occultation timings and derived physical parameters (unresolved events)
 \label{tab_param_rings_unresolved}}
\tablehead{
\colhead{Date} &\colhead{Event} & \colhead{ $t_{0}$ UT$^{\dagger}$} & \colhead{$v_\perp$(km/s)$^{\dagger}$}&\colhead{$v_r$ (km/s)$^{\dagger}$} & \colhead{$L$(deg)$^{\dagger}$}& \colhead{ $E_p$ (km)$^{\dagger}$} \\
}
\scriptsize
\startdata
\hline
\hline
June 3, 2013 & Iguacu ingress &06:24:17.5$\pm$1.7$^{(a)}$&18.059&28.899&2.44&$7.602_{-5.195}^{+2.198}$ \\
 &Iguacu egress &06:24:34.1$\pm$2.0$^{(a)}$&21.246&30.446&104.20&$2.580_{-1.713}^{+3.712}$\\
\hline
 &Bosque Alegre 154 egress &06:25:11.44$\pm$0.14$^{(a)}$&18.889&32.663&176.11&$3.806_{-2.198}^{+1.199}$\\
\hline
 &Ponta Grossa ingress &06:23:58.6$\pm$2.5$^{(a)}$&19.781&34.398&348.04&$9.600^{+0.400}_{-4.795} $\\
 &Ponta Grossa egress &06:24:18.0$\pm$2.5$^{(a)}$&21.965&35.520&120.30&$4.605^{+3.596}_{-3.340}$\\
\hline
 &PROMPT ingress &06:25:20.0.46$\pm$0.011$^{(a)}$&21.373&38.269&326.56&$2.208^{+3.196}_{-0.200} $\\
\hline
 &Santa Martina ingress &06:25:21.03$\pm$0.29$^{(a)}$&17.556&18.537&264.53&$2.208^{+2.997}_{-0.200} $\\
&Santa Martina egress &06:25:31.811$\pm$0.025$^{(a)}$&14.605&22.124&200.07&$2.408^{+5.394}_{-0.200} $\\
\hline
&SOAR ingress &06:25:18.8$\pm$1.3$^{(a)}$&21.444&38.320&325.16&$2.208^{+4.196}_{-0.400} $\\
 &SOAR egress &06:25:38.4$\pm$1.4$^{(a)}$&21.660&38.310&140.37&  $5.205^{+0.799}_{-3.596} $\\
\hline
&Bosque Alegre C11 ingress &06:24.55.45$\pm$1.85$^{(a)}$&21.522&30.340&287.37&$4.206_{-2.597}^{+3.297}$\\
 &Bosque Alegre C11 egress &06:25:09.45$\pm$1.75$^{(a)}$&17.882&30.062&183.50&$4.206_{-2.597}^{+3.396}$\\
\hline
 &TRAPPIST ingress &06:25:20.9$\pm$1.9$^{(a)}$&20.293&36.229&341.31&  $4.605_{-2.198}^{+3.796}$\\
\hline
March 16, 2014 &Thailand ingress &20:31:37.640$\pm$1.33&3.656&3.821&95.06&$1.856_{-1.197}^{+0.948}$\\
 & Thailand egress &20:31:53.885$\pm$0.175&3.990&4.290&60.85&$1.856_{-0.801}^{+0.150}$\\
\hline
June 28, 2014 & Hakos ingress &22:24:25.796$\pm$0.041&$19.127$&$28.619$&5.064&$1.472_{-0.517}^{+0.455}$\\
 &  Hakos egress &22:24:44.218$\pm$0.035&20.971&29.744&117.061&$1.983_{-0.508}^{+0.598}$\\
\hline
April 26, 2015 & Los Molinos ingress&02:11:45.707$\pm$0.058 & 3.503 & 3.513 & 238.857&$2.914_{-0.149}^{+0.151}$ \\
&Los Molinos egress&02:12:09.195$\pm$0.070&2.957&3.989&199.749&$2.400_{-0.320}^{+0.28}$ \\
\hline
May 12, 2015 &Brisbane egress & 17:55:56.823$\pm$0.012& 11.823 & 16.567 & 357.23&$2.707^{+2.398}_{-1.198}$ \\
\hline
Aug 8, 2016 & Windhoek (CHMO) ingress & 19:57:18.209 $\pm$ 0.249 & 15.920 & 21.878 & 332.963$^{(b)}$& $2.043^{+2.762}_{-0.734}$\\
 & Windhoek (CHMO) egress & 19:57:51.870 $\pm$ 0.382 & 15.216 & 21.950 & 180.196$^{(b)}$ & $3.806^{+2.598}_{-2.297}$\\
\hline
Oct 1, 2016 & Rockhampton ingress & 10:12:26.284 $\pm$ 0.072 & 10.795 & 13.121 & 123.960& $2.523^{+3.481}_{-0.615}$\\
& Rockhampton egress & 10:13:22.928 $\pm$ 0.049 & 12.573 & 13.146 & 278.911 & $2.586^{+3.818}_{-0.778}$\\
\hline
& Adelaide ingress & 10:10:19.826$\pm$ 0.186 & 12.421 & 12.597 &91.347 & $1.867^{+4.073}_{-0.539}$\\
 & Adelaide egress & 10:11:14.558 $\pm$ 0.218 & 9.942 & 12.651 &311.914 & $2.047^{+3.534}_{-0.719}$\\
\hline
\hline
\enddata
\tablenotetext{\dagger}{ Same parameters as Table~\ref{tab_param_rings}}
\tablenotetext{a}{Timings given by \cite{bra14}.}
\tablenotetext{b}{Two geometries are possible for this occultation. We choose arbitrary the closest to the prediction. The other geometry provides different longitudes: $L_{ingress}=0.154$ and $L_{egress}=152.948$}
\end{deluxetable}

\clearpage

\begin{deluxetable}{llcc}
\tablecolumns{8}
\tablewidth{0pc}
\tabletypesize{\scriptsize}
\tablecaption{Occultation timings for the main body.
\label{tab_param_body}}
\tablehead{
\colhead{Date} &\colhead{Event} & \colhead{ $t_{ingress}$ UT} & \colhead{ $t_{egress}$ UT} \\
}
\scriptsize
\startdata
\hline
June 3, 2013 & Danish &06:25:27.861$\pm$0.014& 06:25:33.188$\pm$0.014\\
&PROMPT  &06:25:24.835$\pm$0.009&06:25:35.402$\pm$0.015\\
&TRAPPIST &06:25:27.893$\pm$0.019&06:25:33.155$\pm$0.007\\
&SOAR  &06:25:24.34$\pm$0.59&06:25:34.597$\pm$0.009\\
February 16, 2014 & San Pedro de Atacama& 07:45:27.450$\pm$0.6&07:45:31.125$\pm$0.57 \\
June 28, 2014 &Kalahari  &22:24:07.383$\pm$0.126&22:24:14.854$\pm$0.096\\
&Twee Rivieren &22:24:06.689$\pm$0.093&22:24:16.481$\pm$0.105\\
April 29, 2014 &Springbok  &23:14:30.02$\pm$0.075&23:14:48.03$\pm$0.075\\
May 12, 2015 &Brisbane & 17:55:35.530$\pm$0.010&17:55:44.135$\pm$0.075 \\
July 25, 2016 &Liverpool Telescope & 23:59:05.494  $\pm$0.054 & 23:59:12.310  $\pm$ 0.054 \\
August 8, 2016 &Windhoek (CHMO)  &  19:57:28.469 $\pm$ 0.042 &19:57:41.886 $\pm$ 0.045 \\
August 10, 2016 - 14h UT&Murrumbateran &14:18:35.030 $\pm$ 0.3 & 14:18:45.145 $\pm$ 0.125 \\
August 10, 2016 - 16h UT& Les Makes & 16:42:51.305  $\pm$ 0.530 &16:43:07.917 $\pm$ 0.848 \\
August 15, 2016 &Darfield & 11:38:27.465  $\pm$ 0.385& 11:38:38.019  $\pm$ 0.873\\
October 1, 2016 &Rockhampton & 10:12:44.664 $\pm$ 0.041 & 10:13:03.199 $\pm$ 0.051\\
&Adelaide & 10:10:41.818 $\pm$ 0.118 & 10:10:54.102 $\pm$ 0.064\\
\hline
\multicolumn{4}{c}{The error bars quoted are given at $1\sigma$ level.}\\
\enddata

\end{deluxetable}

\begin{figure}[!htb]
\centering
\includegraphics[angle=0,scale=0.3,trim=10cm 0 1.8cm 0]{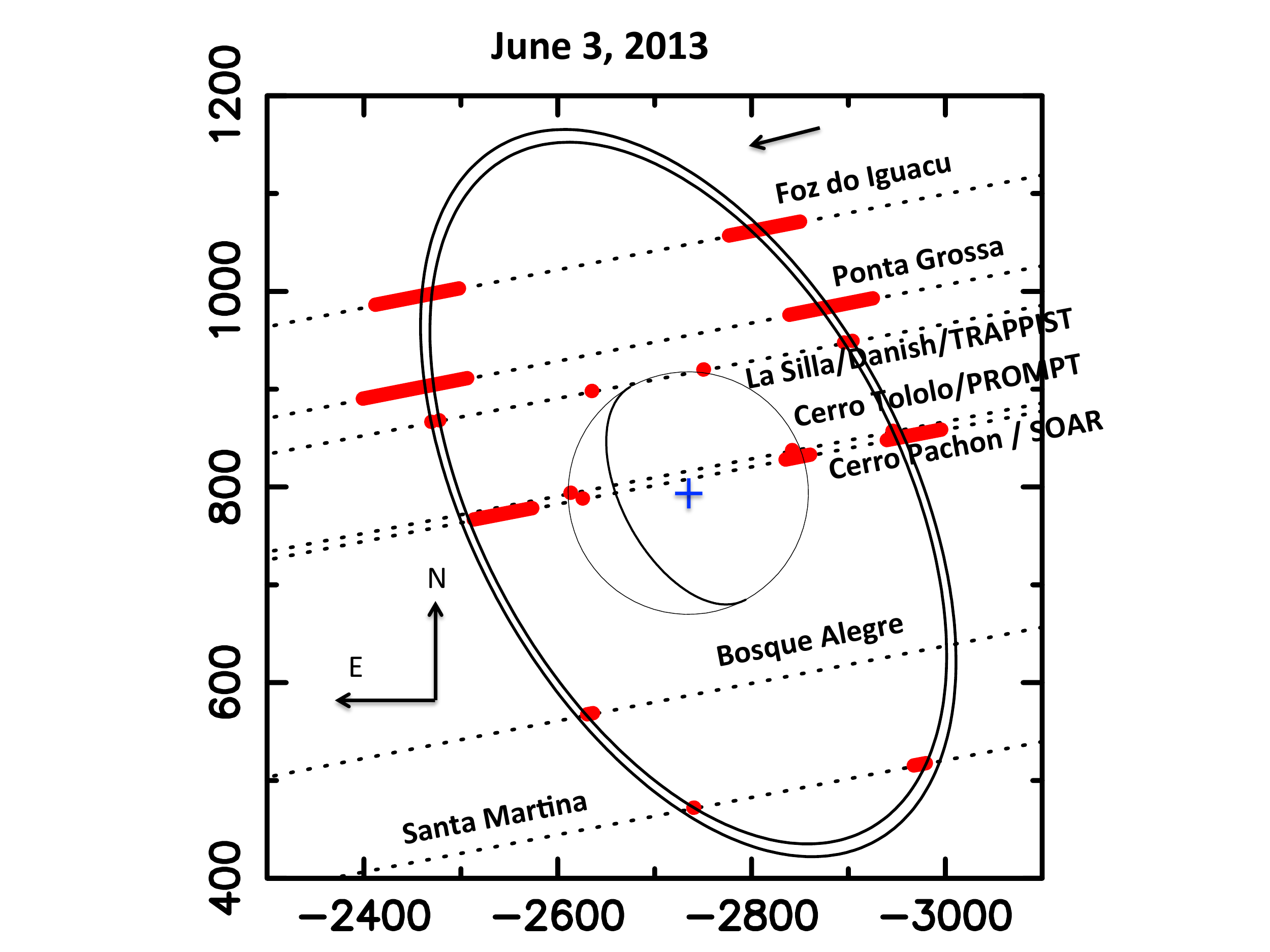} 
\includegraphics[angle=0,scale=0.3,trim=2cm 0 2cm 0]{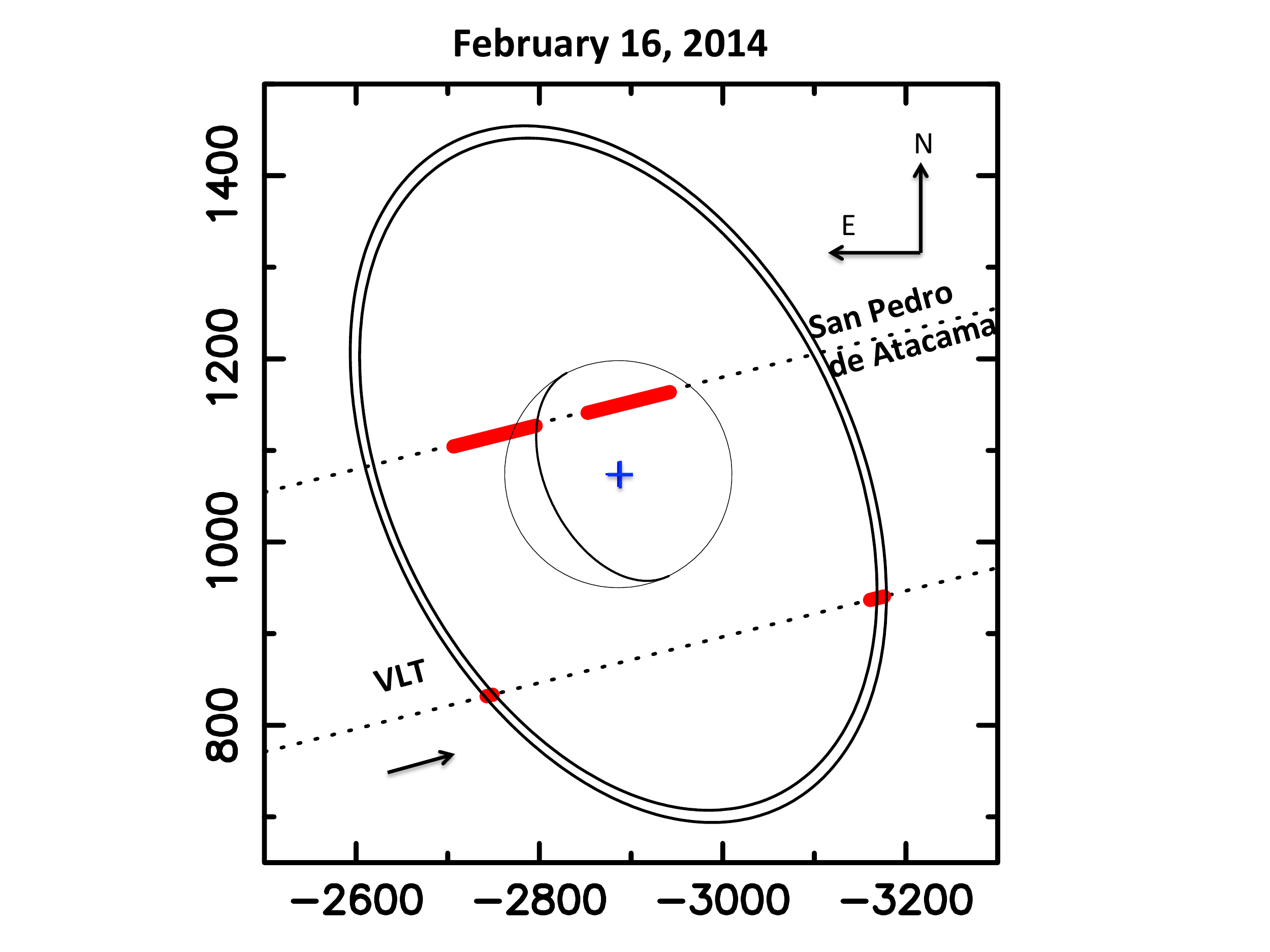}
\includegraphics[angle=0,scale=0.3,trim=2cm 0 10cm 0]{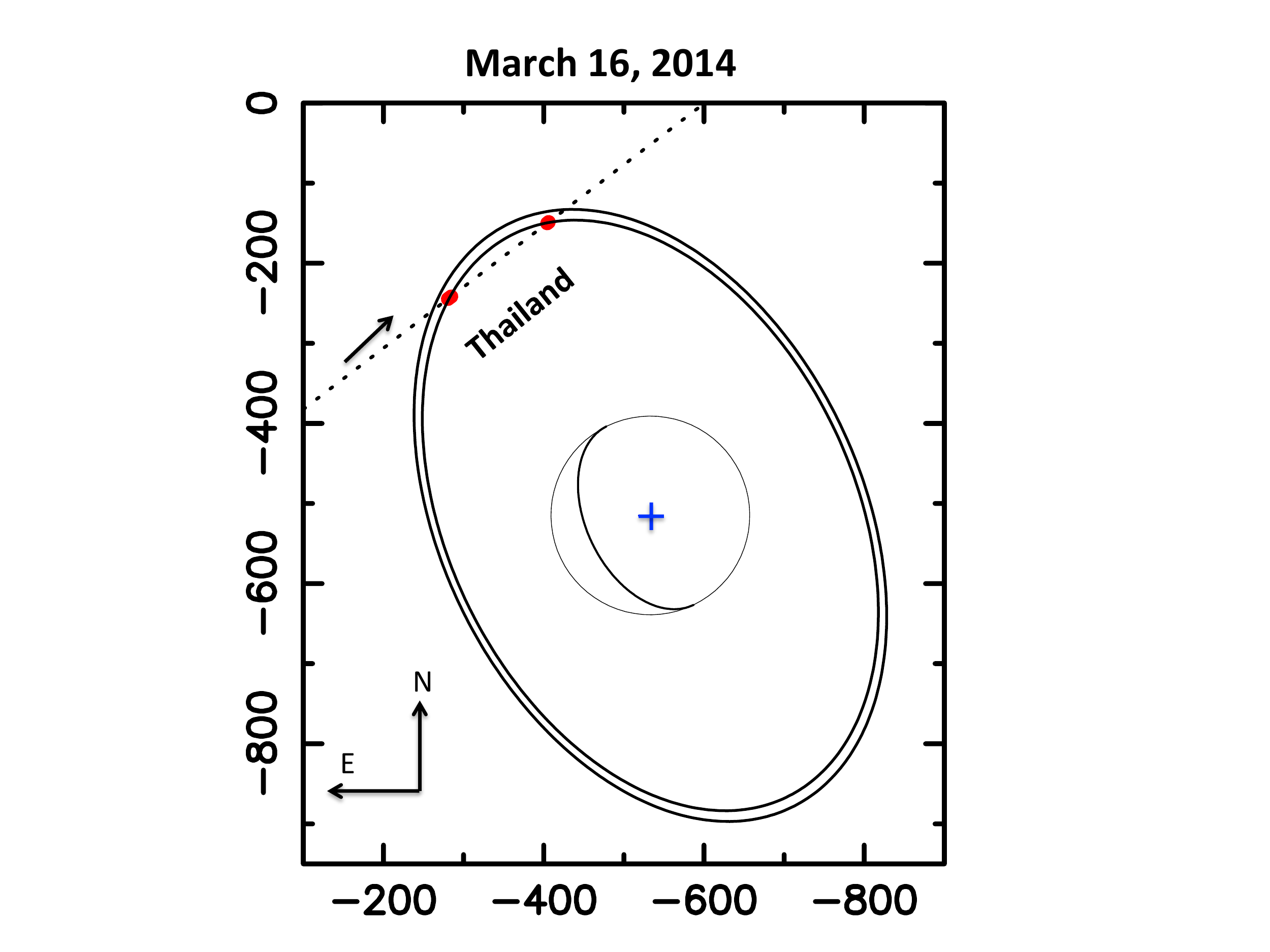}\\
\includegraphics[angle=0,scale=0.3,trim=10cm 0 2cm 0]{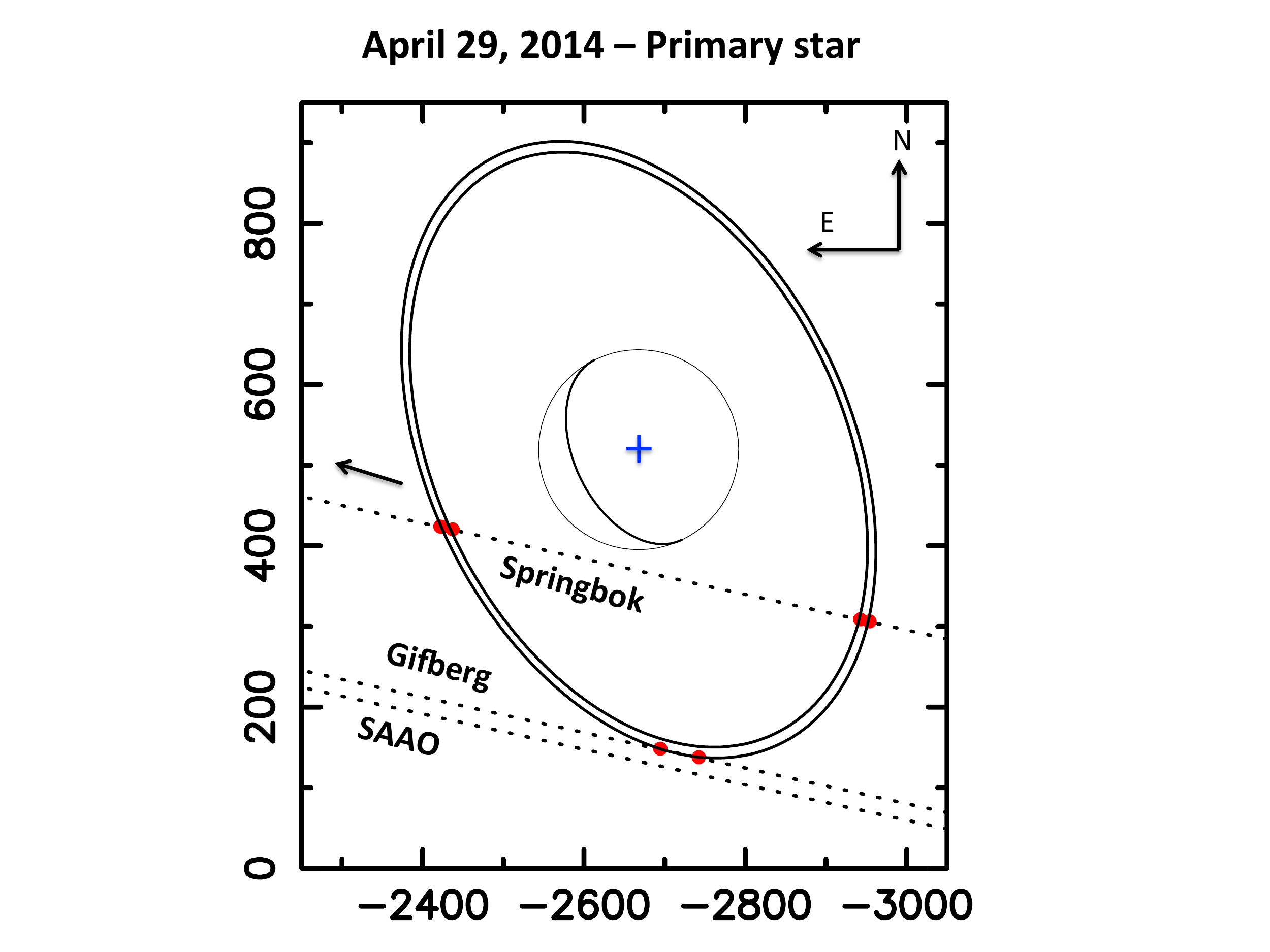} 
\includegraphics[angle=0,scale=0.3,trim=3cm 0 2cm 0]{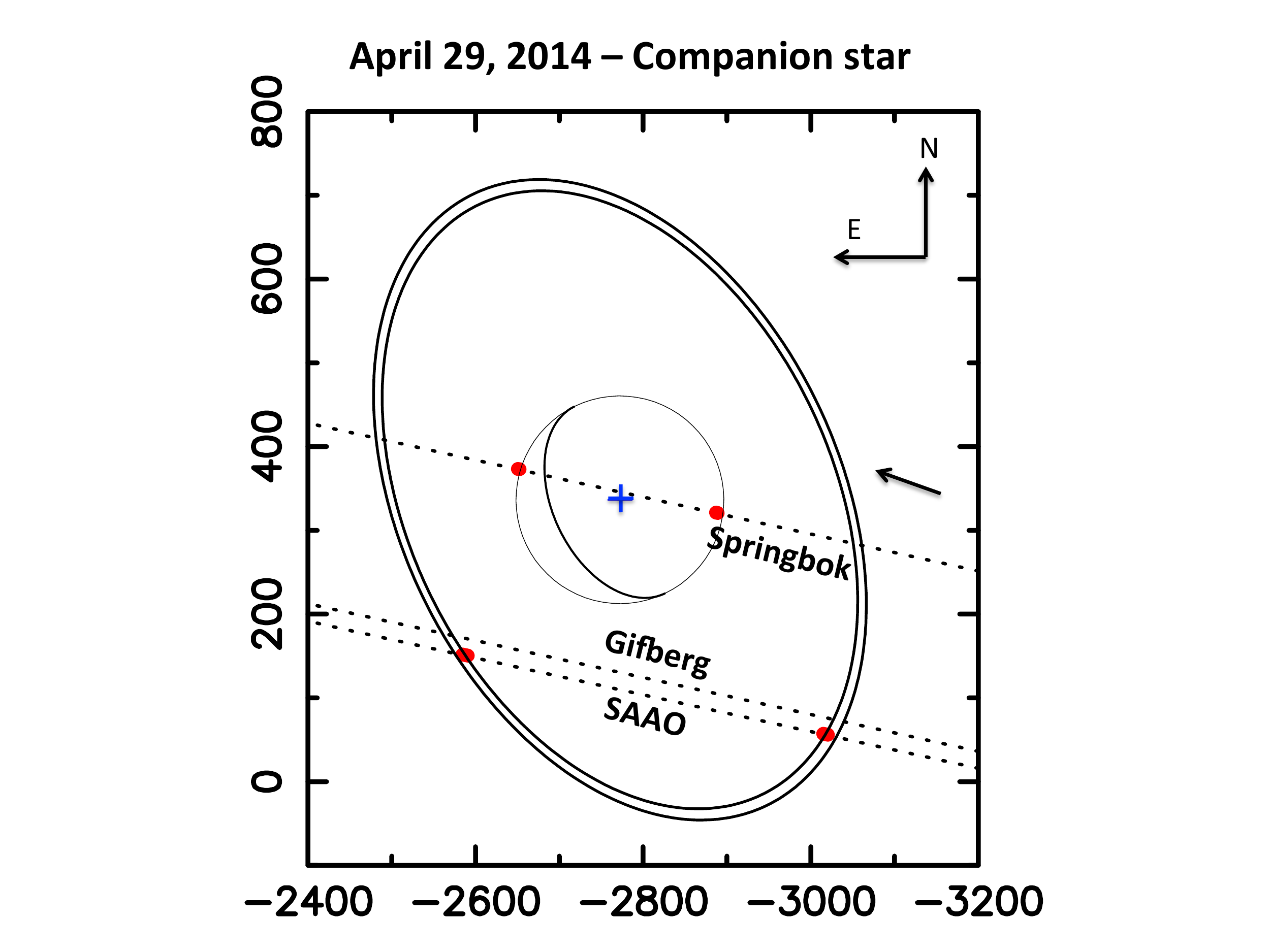}
\includegraphics[angle=0,scale=0.3,trim=3cm 0 10cm 0]{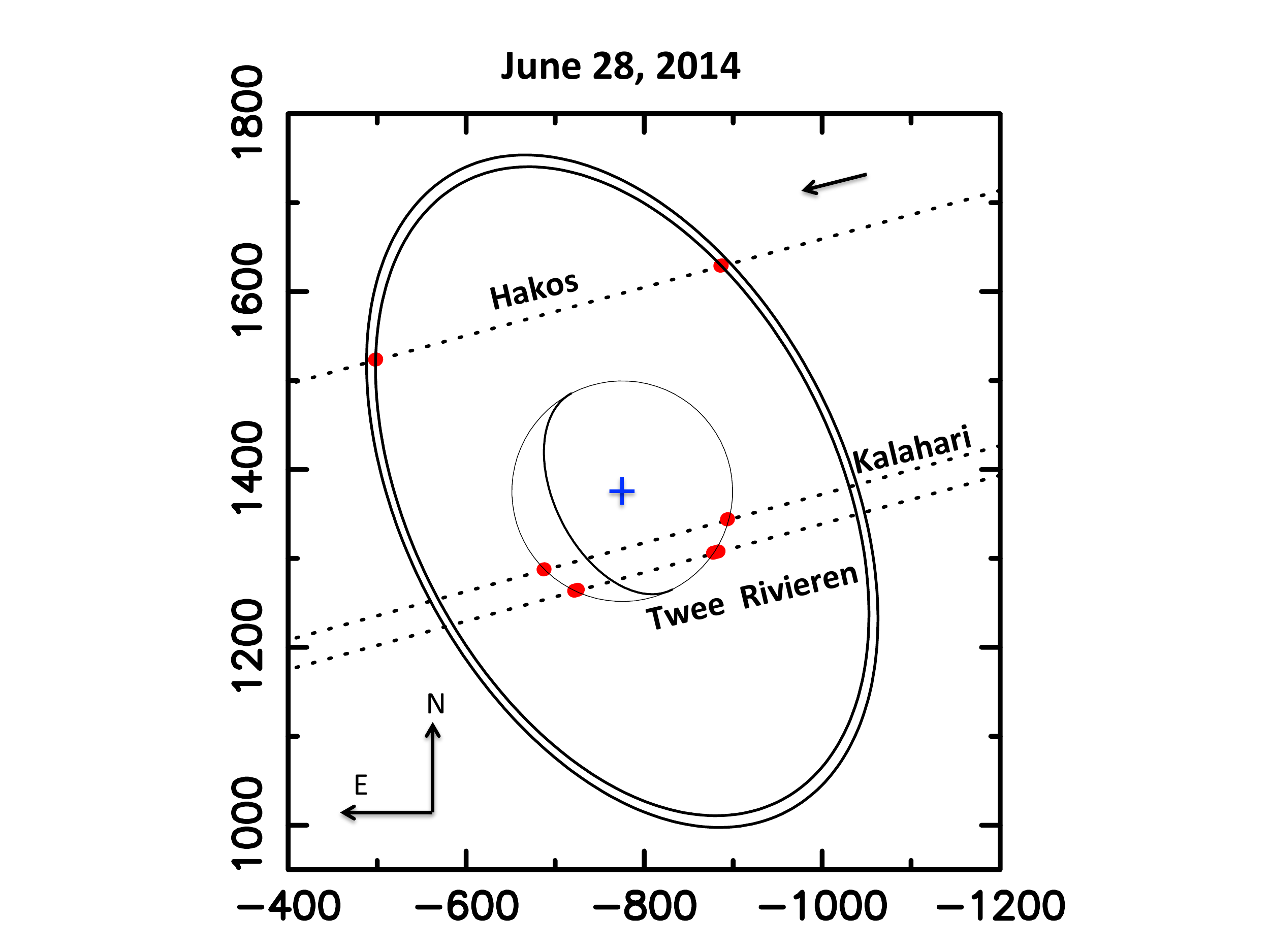} \\
\includegraphics[angle=0,scale=0.3,trim=10cm 0 2cm 0]{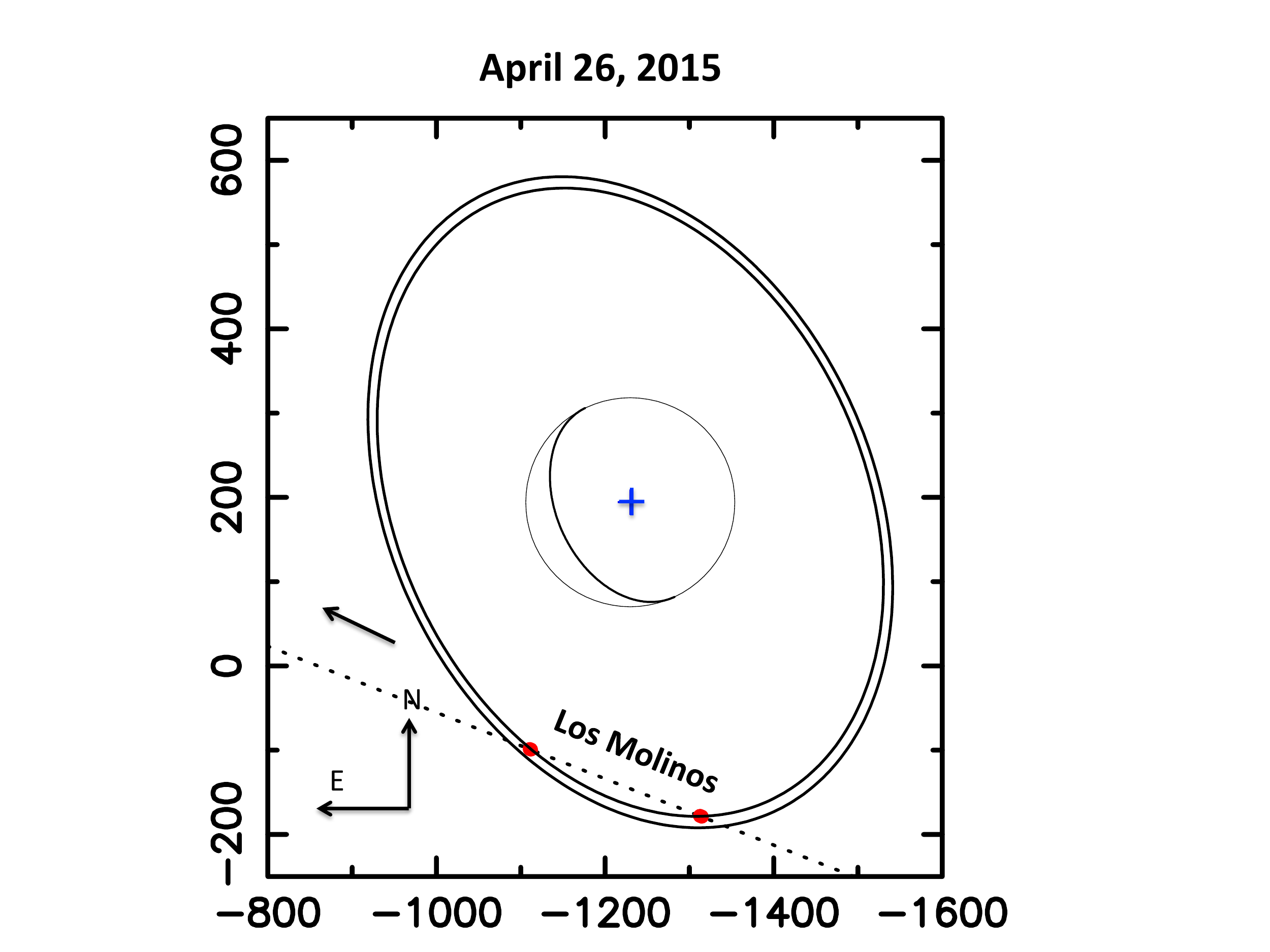}
\includegraphics[angle=0,scale=0.3,trim=4cm 0 2cm 0]{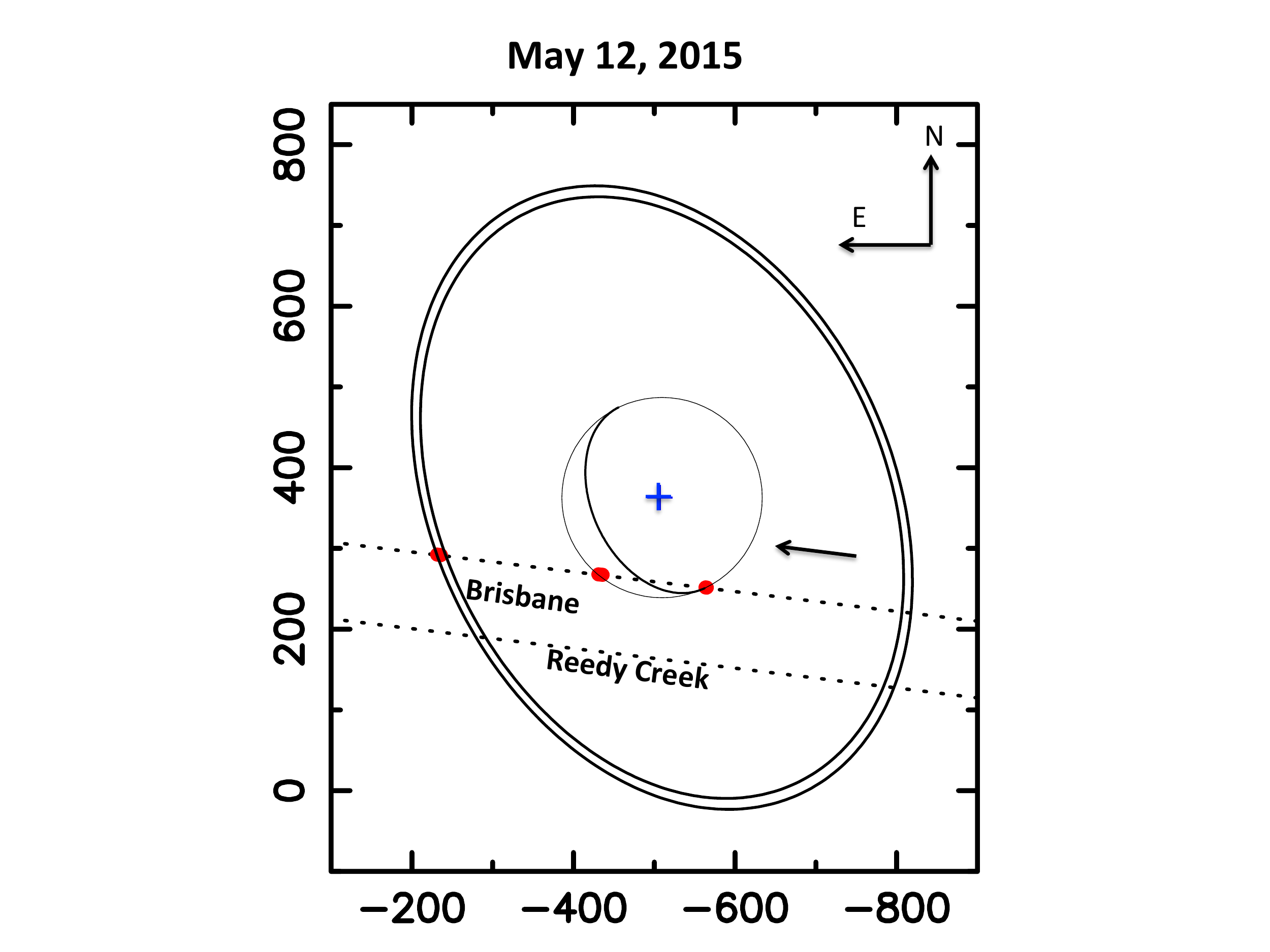}
\includegraphics[angle=0,scale=0.3,trim=4cm 0 10cm 0]{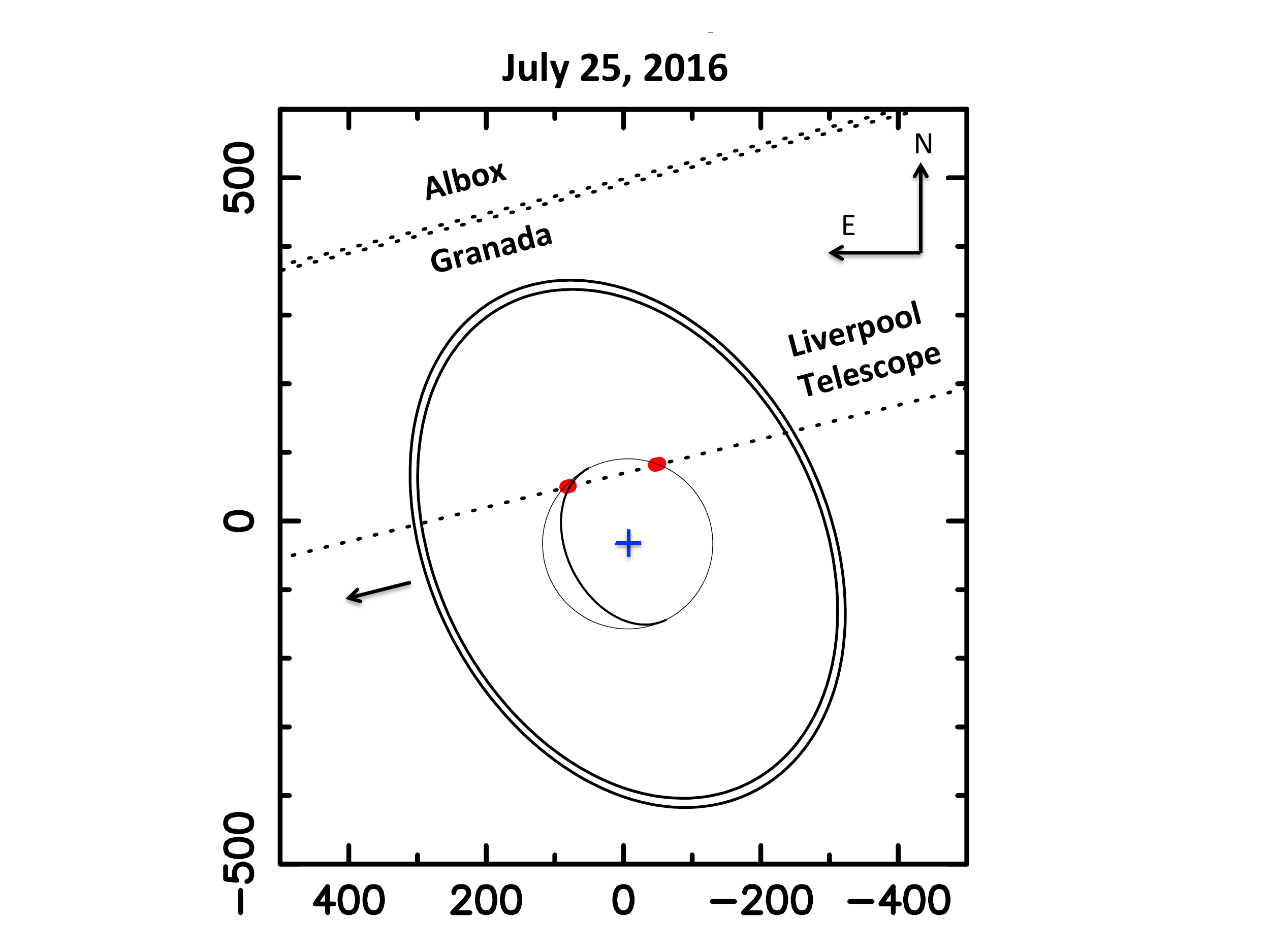} 
\\
\caption{
\footnotesize
Reconstructed geometries of the occultations. The dotted lines are the trajectories of the occulted star relative to Chariklo in the plane of sky as seen from each station (the arrow indicates the direction of the apparent movement of the star). The red segments are the $1\sigma$ level error bars on each chord extremity, derived from the corresponding error bars on timings (see Table~\ref{tab_param_rings},~\ref{tab_param_rings_unresolved} and~\ref{tab_param_body}). For those plots, we use the pole position and the radii from \cite{bra14}: $r_{C1R}=390.6$~km, $r_{C2R}=404.8$~km and $r_{Ck}=124$~km. The center of ring system (blue cross) in each panel represents the offset in right ascension and declination between the predicted and observed positions of the Chariklo relative to the occulted star, as given in Table~\ref{tab_circumstances_positives}. This offset was used to improve Chariklo's ephemeris. 
}
\label{fig_geometry}
\end{figure}

\begin{figure}[!htb]
\centering
\includegraphics[angle=0,scale=0.3,trim=10cm 0 1cm 0]{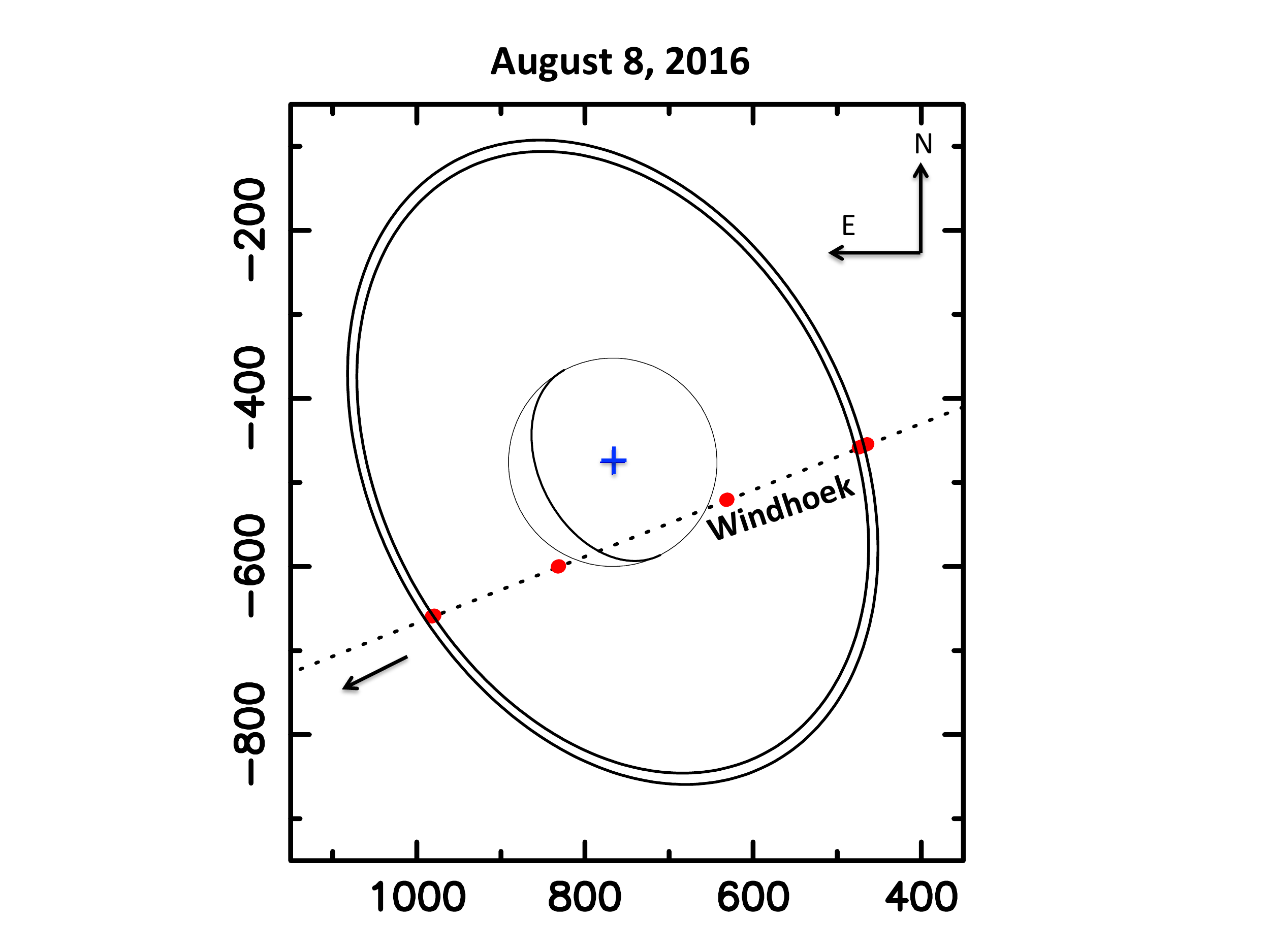}
\includegraphics[angle=0,scale=0.3,trim=5cm 0 1cm 0]{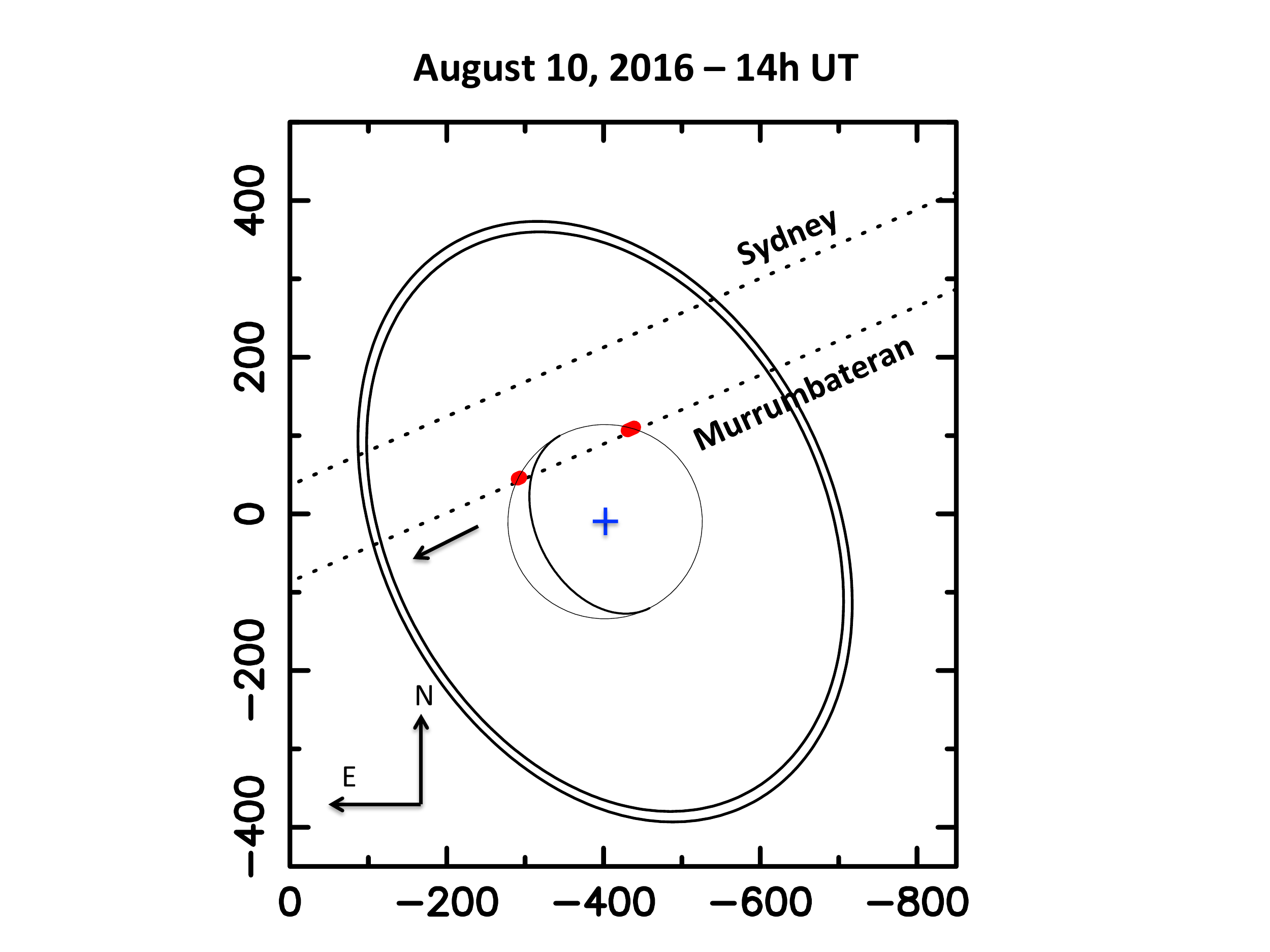}
\includegraphics[angle=0,scale=0.3,trim=5cm 0 10cm 0]{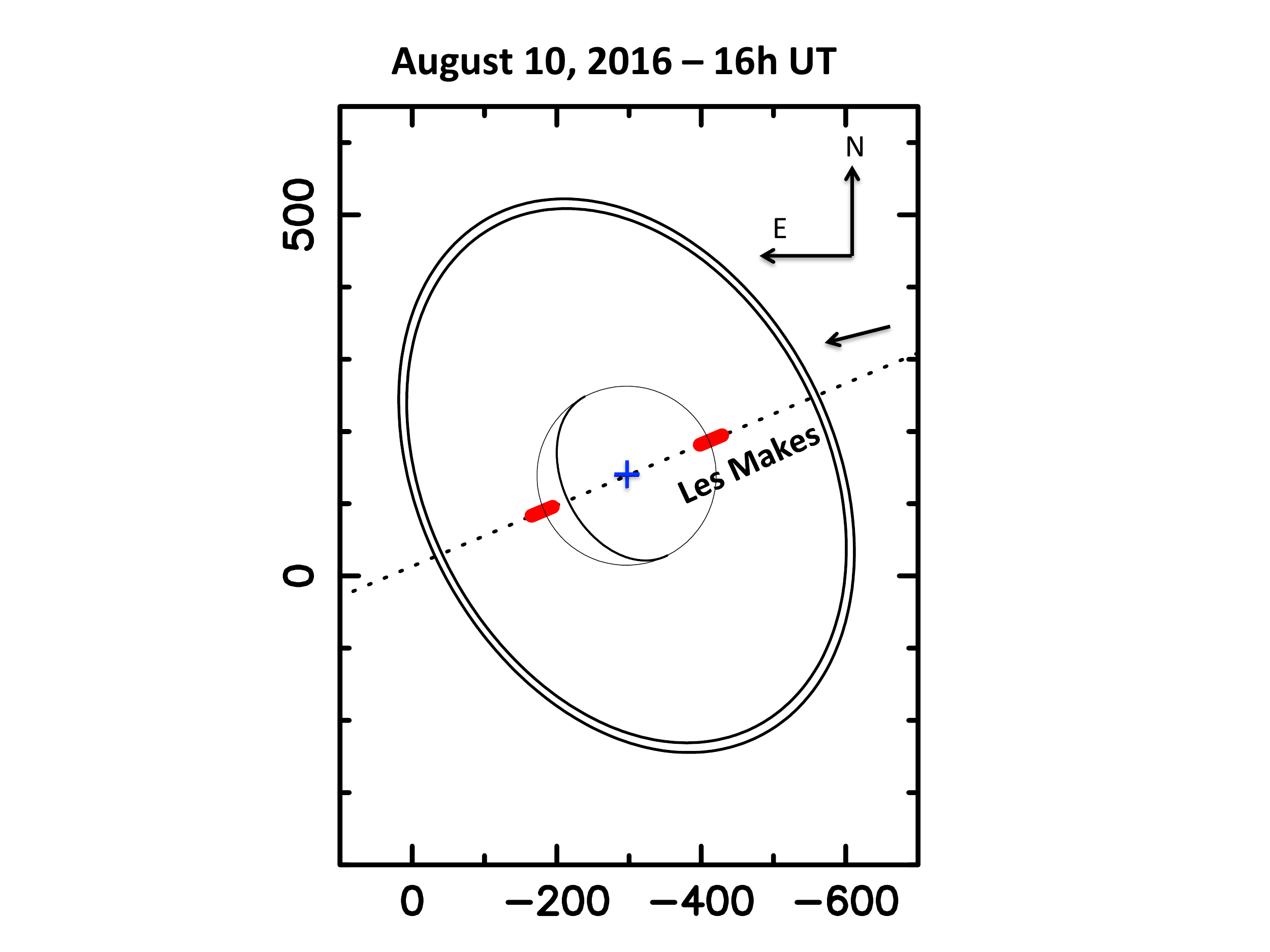}\\
\includegraphics[angle=0,scale=0.3,trim=10cm 0 1cm 0]{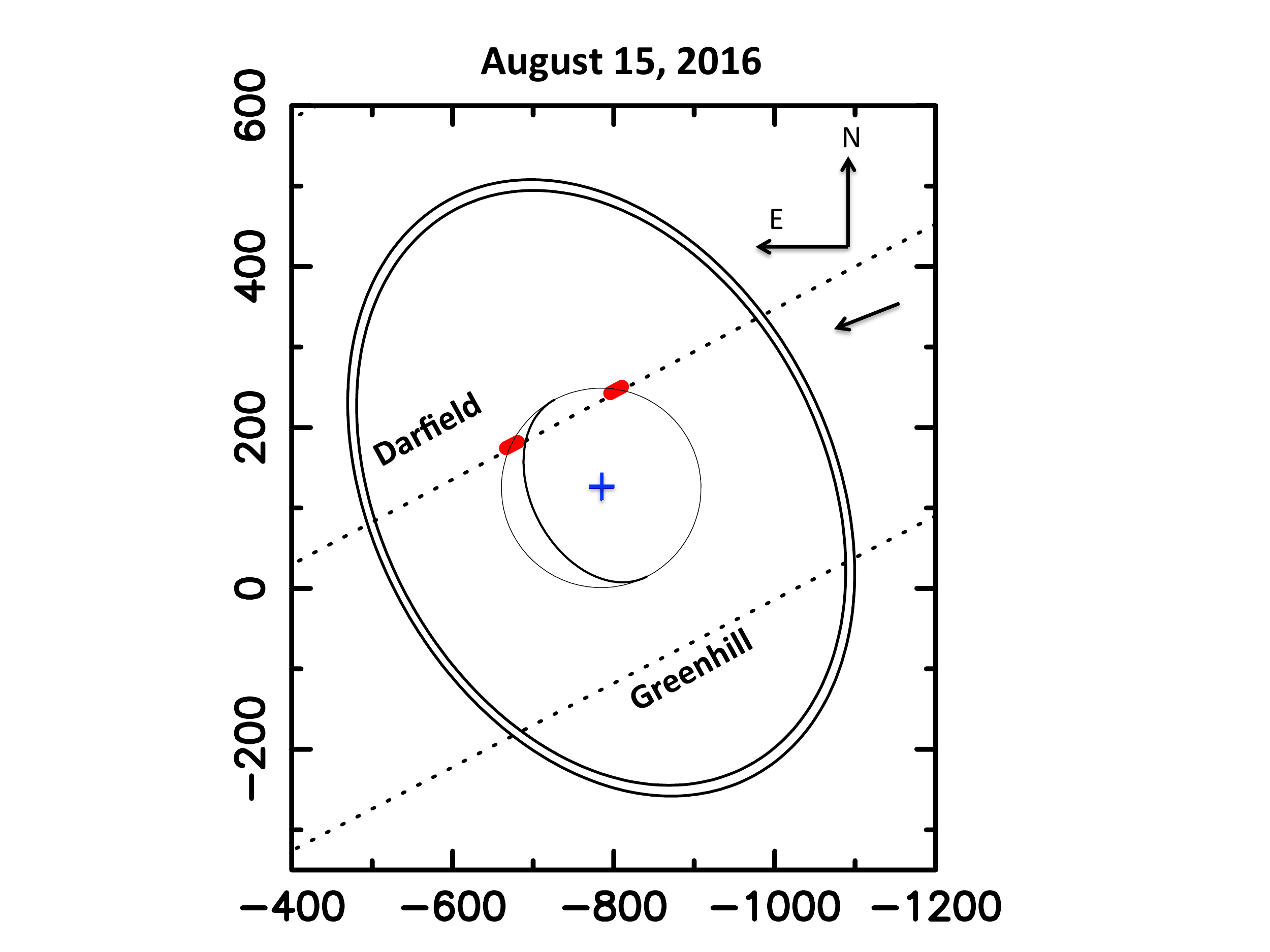}
\includegraphics[angle=0,scale=0.3,trim=5cm 0 10cm 0]{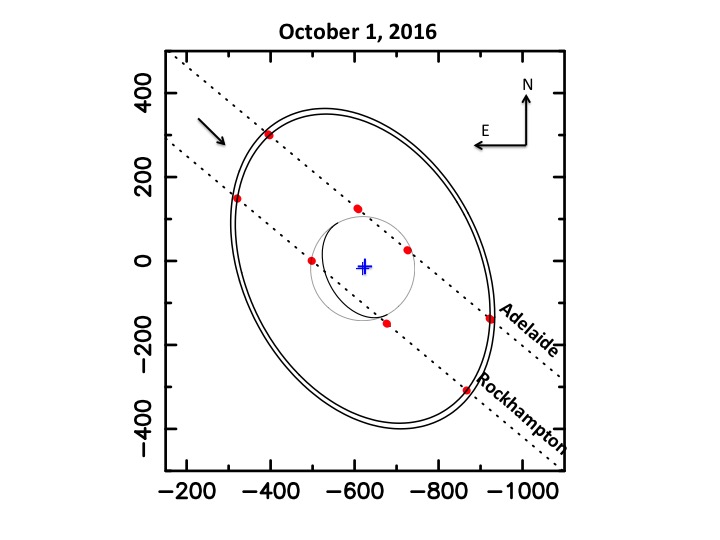}
\\
\caption{
Following Figure~\ref{fig_geometry}
}
\label{fig_geometry2}
\end{figure}


\begin{figure}[!htb]
\centering
\begin{tabular}{ccc}
\includegraphics[angle=0,scale=0.25,trim=6cm 2cm 1cm 0]{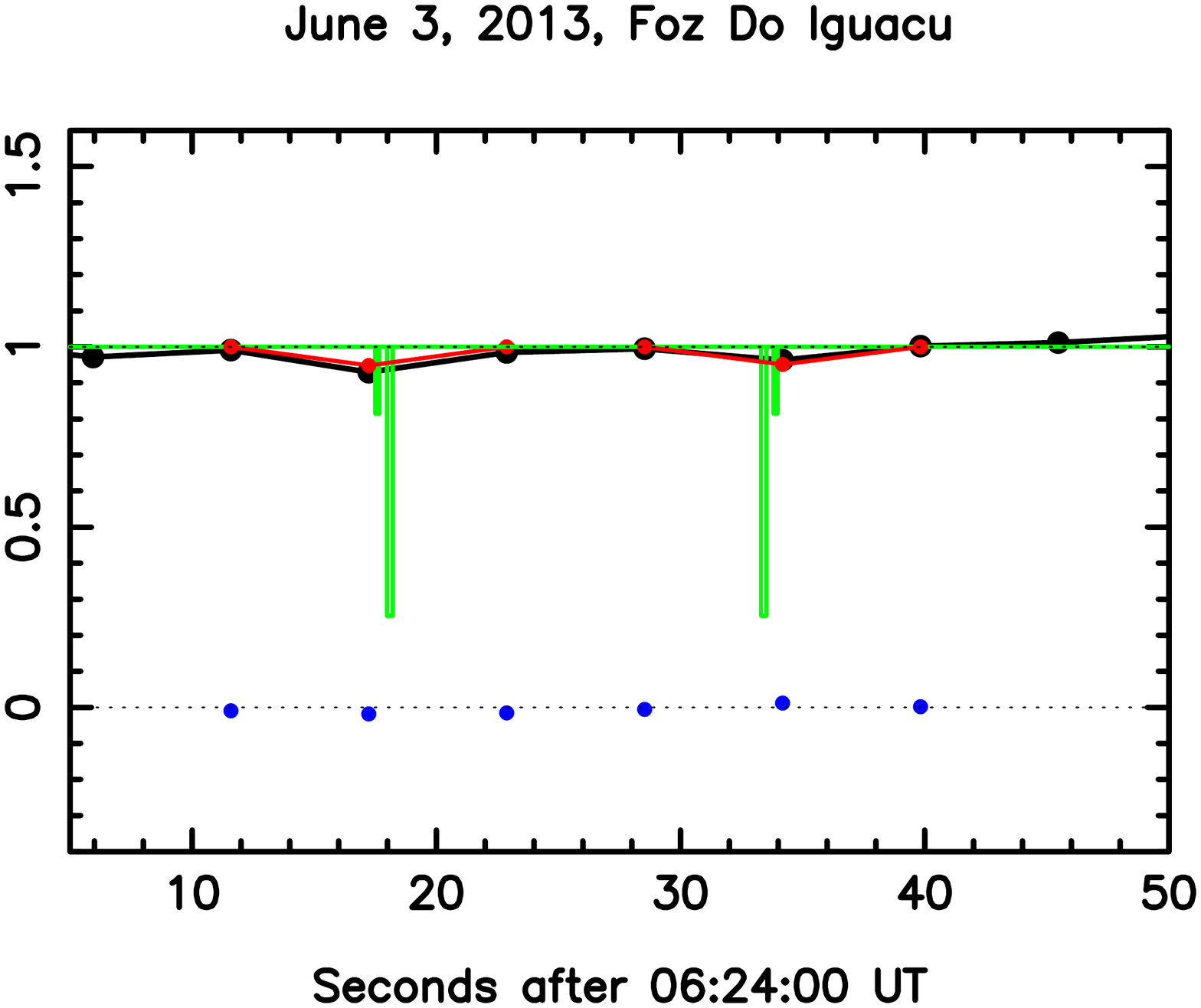} &
 \includegraphics[angle=0,scale=0.25,trim=6cm 2cm 1cm 0]{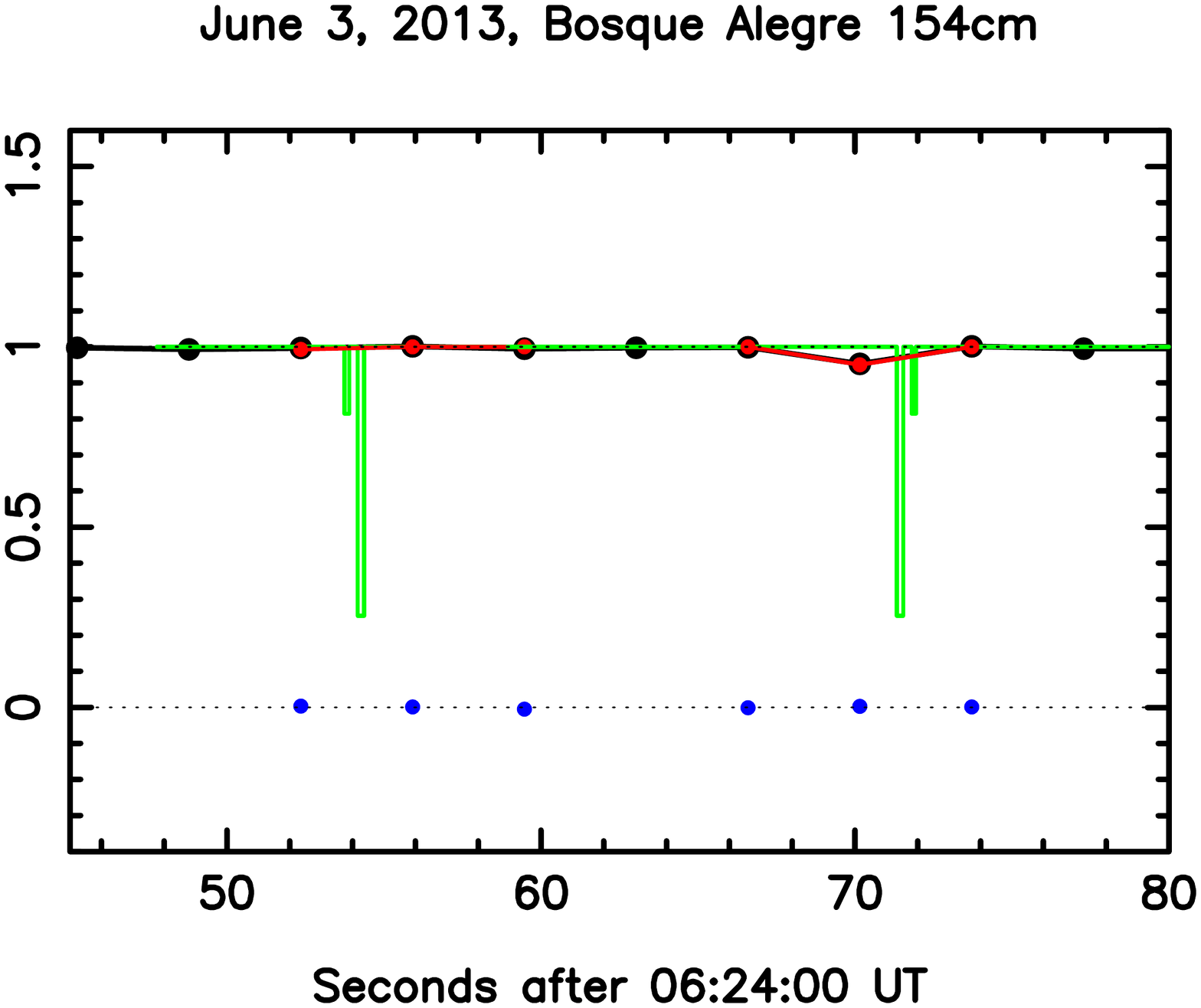} & 
 \includegraphics[angle=0,scale=0.25,trim=6cm 2cm 1cm 0]{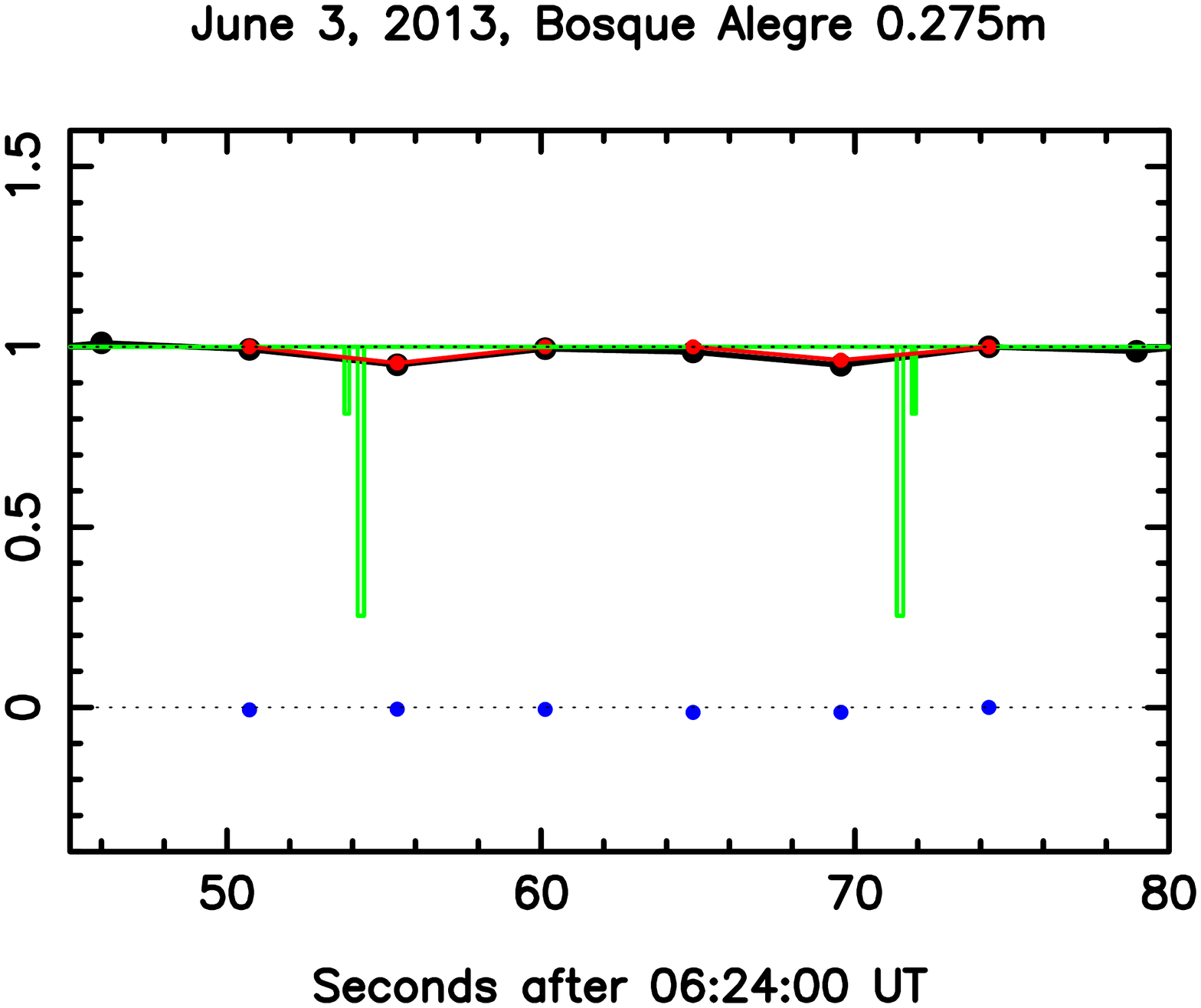}\\ 

\includegraphics[angle=0,scale=0.25,trim=6cm 2cm 1cm 0]{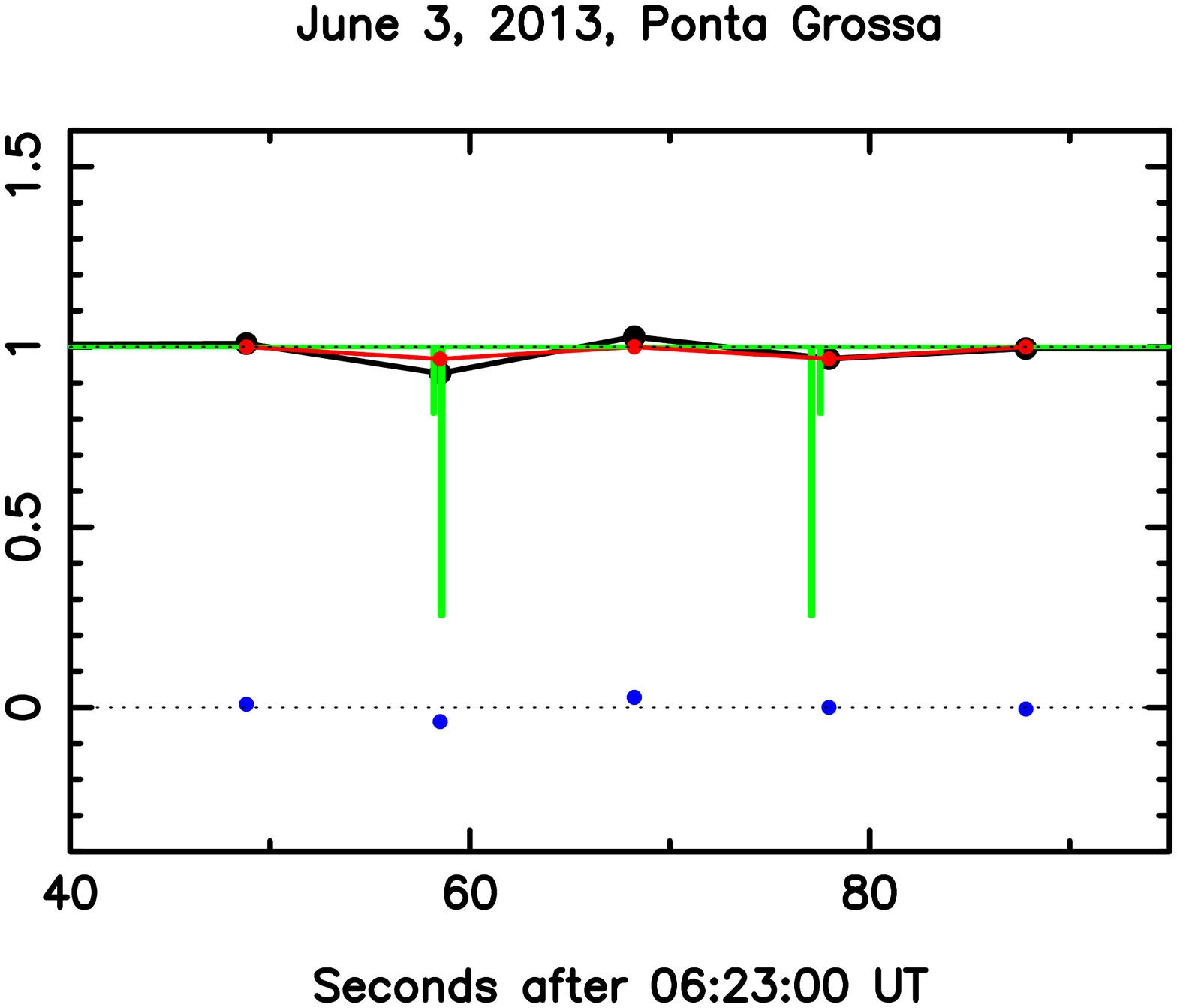}  &
\includegraphics[angle=0,scale=0.25,trim=6cm 2cm 1cm 0]{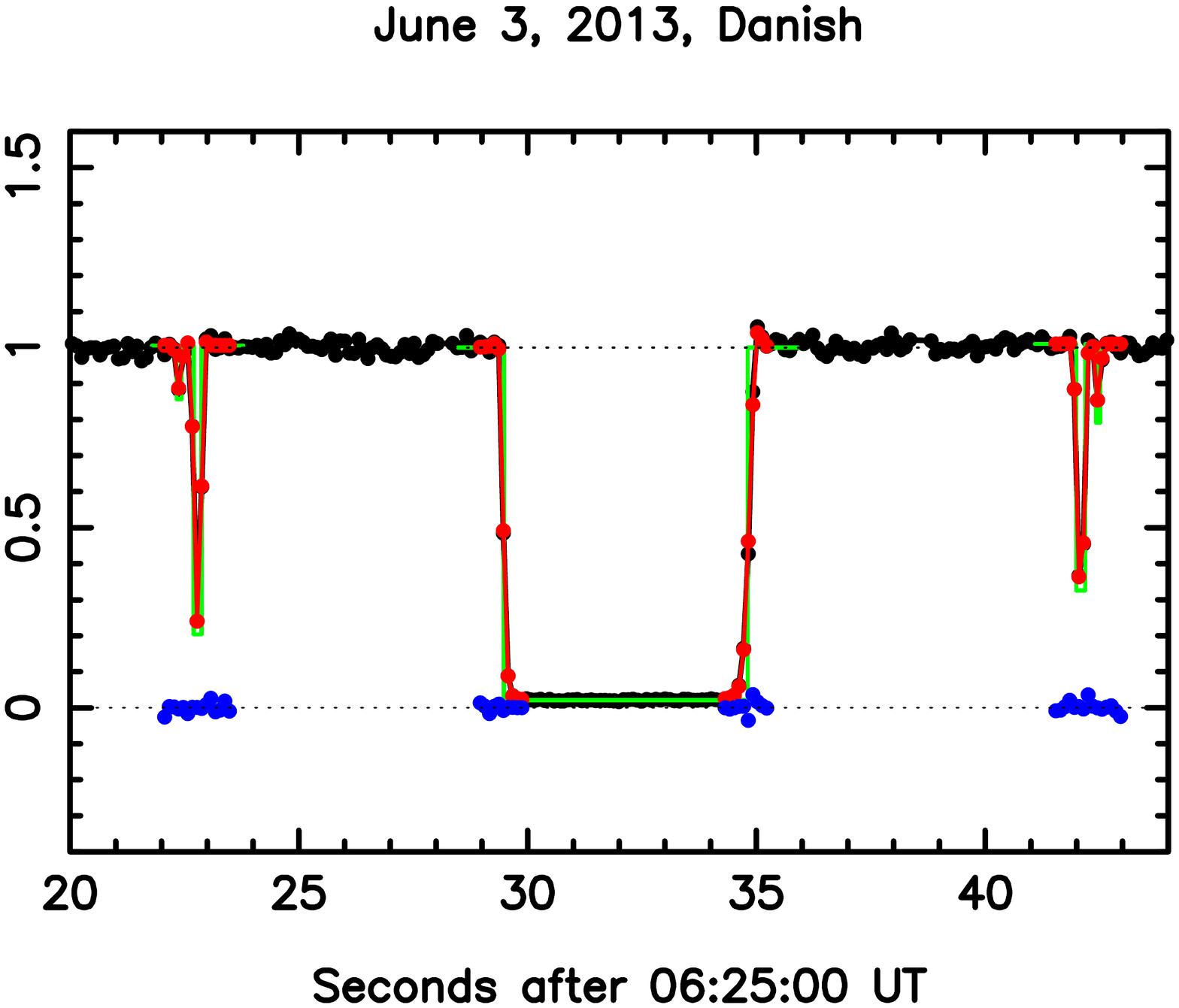} & 
\includegraphics[angle=0,scale=0.25,trim=6cm 2cm 1cm 0]{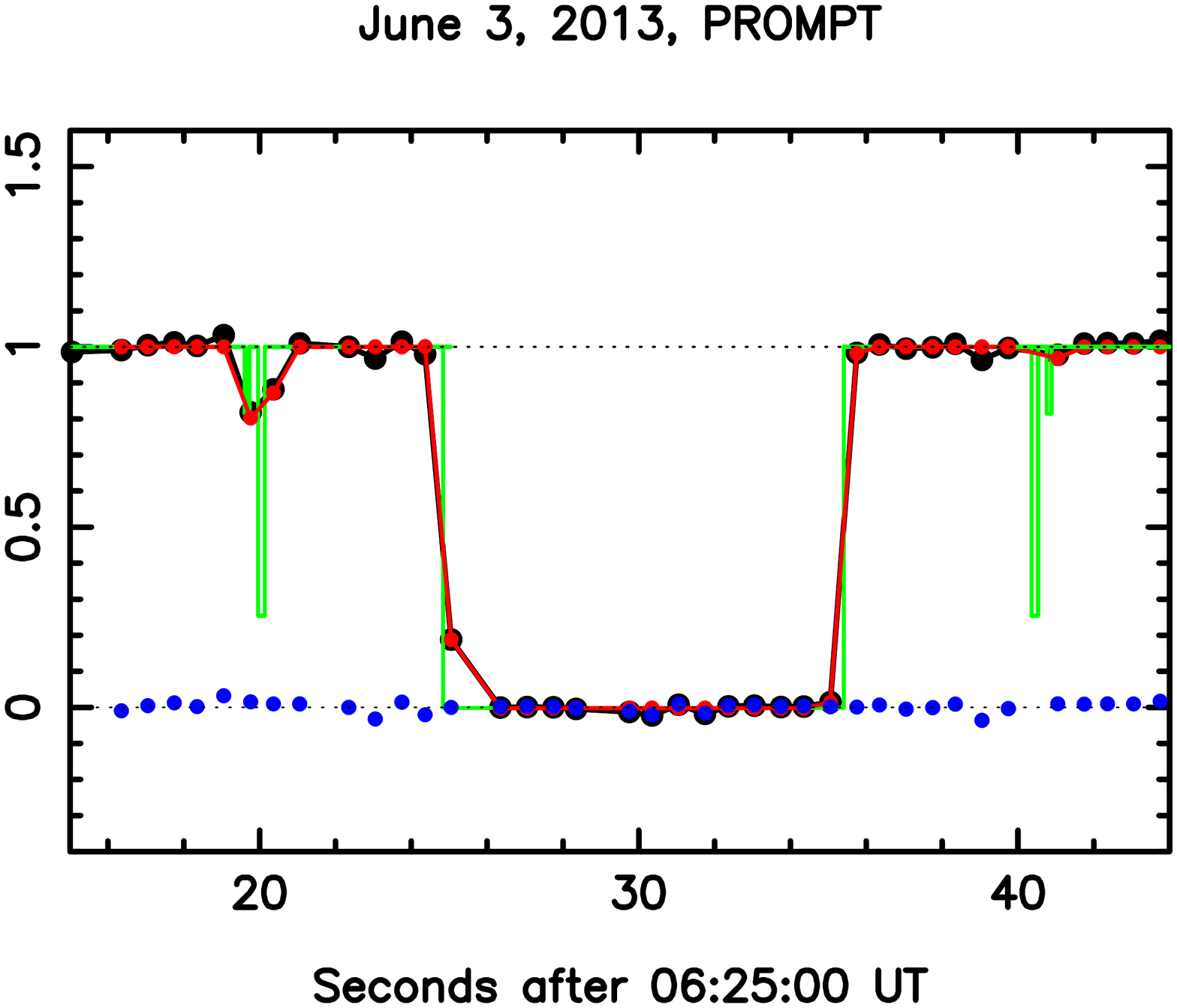} \\

\includegraphics[angle=0,scale=0.25,trim=6cm 2cm 1cm 0]{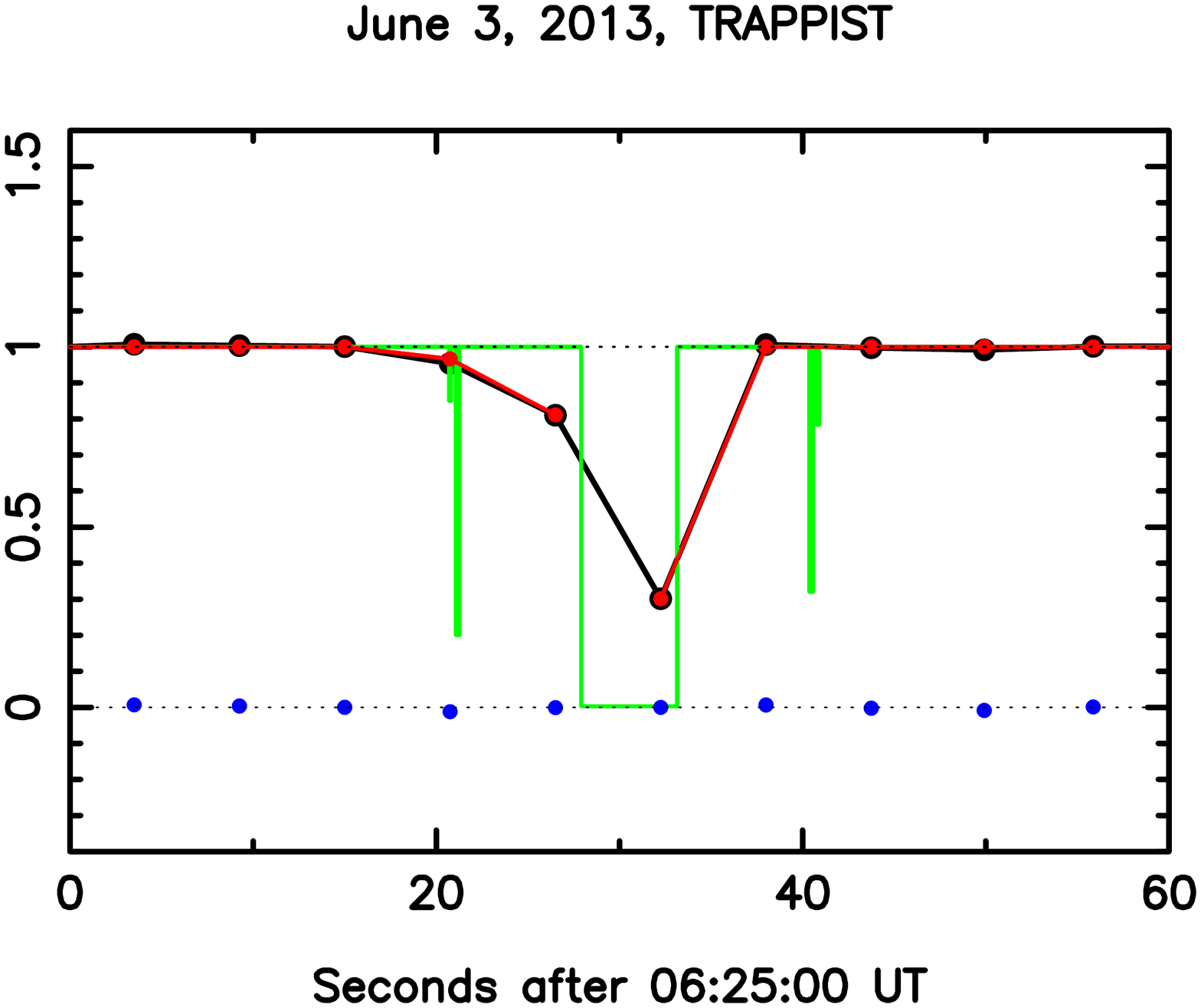} & 
\includegraphics[angle=0,scale=0.25,trim=6cm 2cm 1cm 0 ]{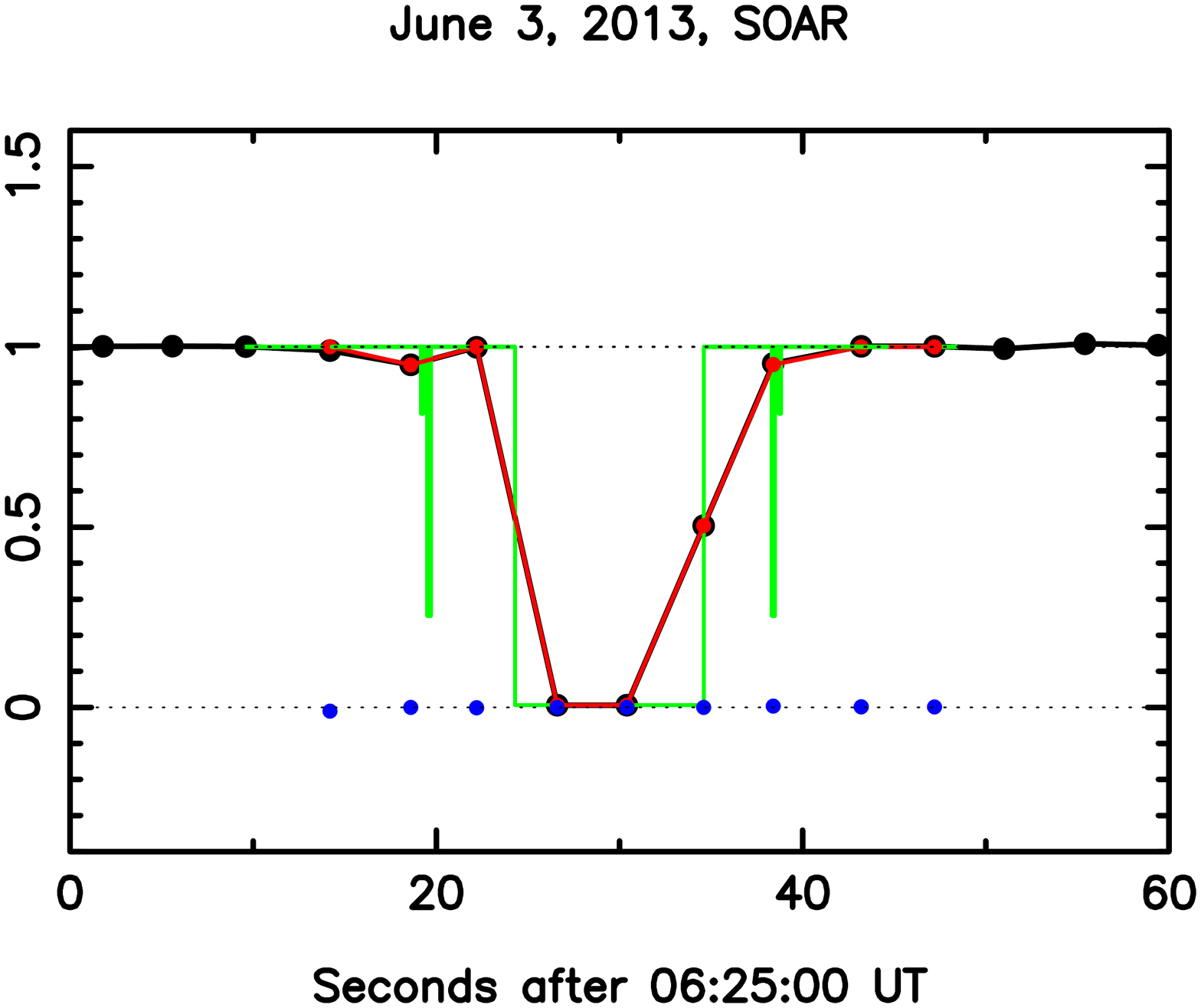}  &
\includegraphics[angle=0,scale=0.25,trim=6cm 2cm 1cm 0]{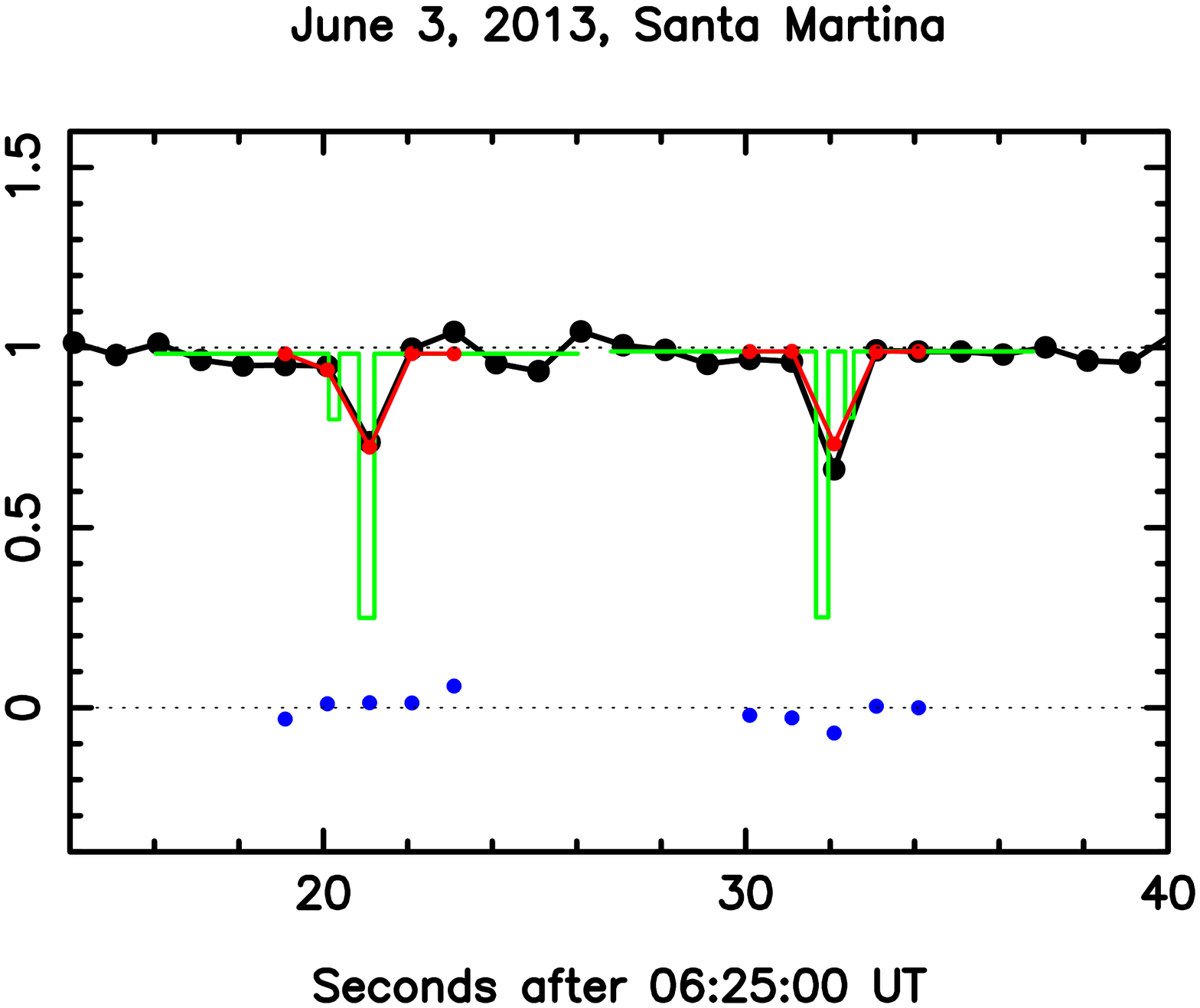} \\

\includegraphics[angle=0,scale=0.25,trim=6cm 2cm 1cm 0]{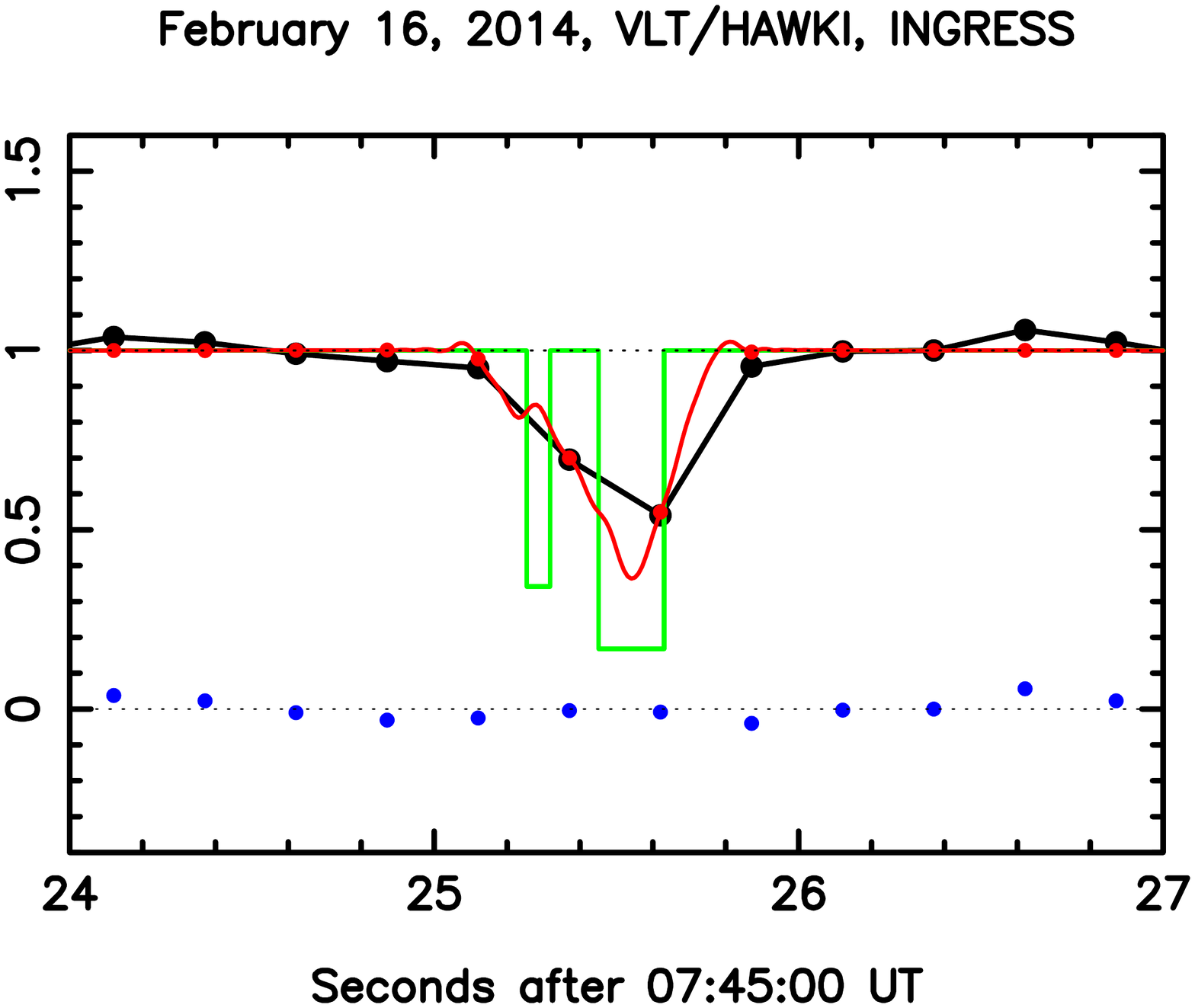} & 
\includegraphics[angle=0,scale=0.25,trim=6cm 2cm 1cm 0]{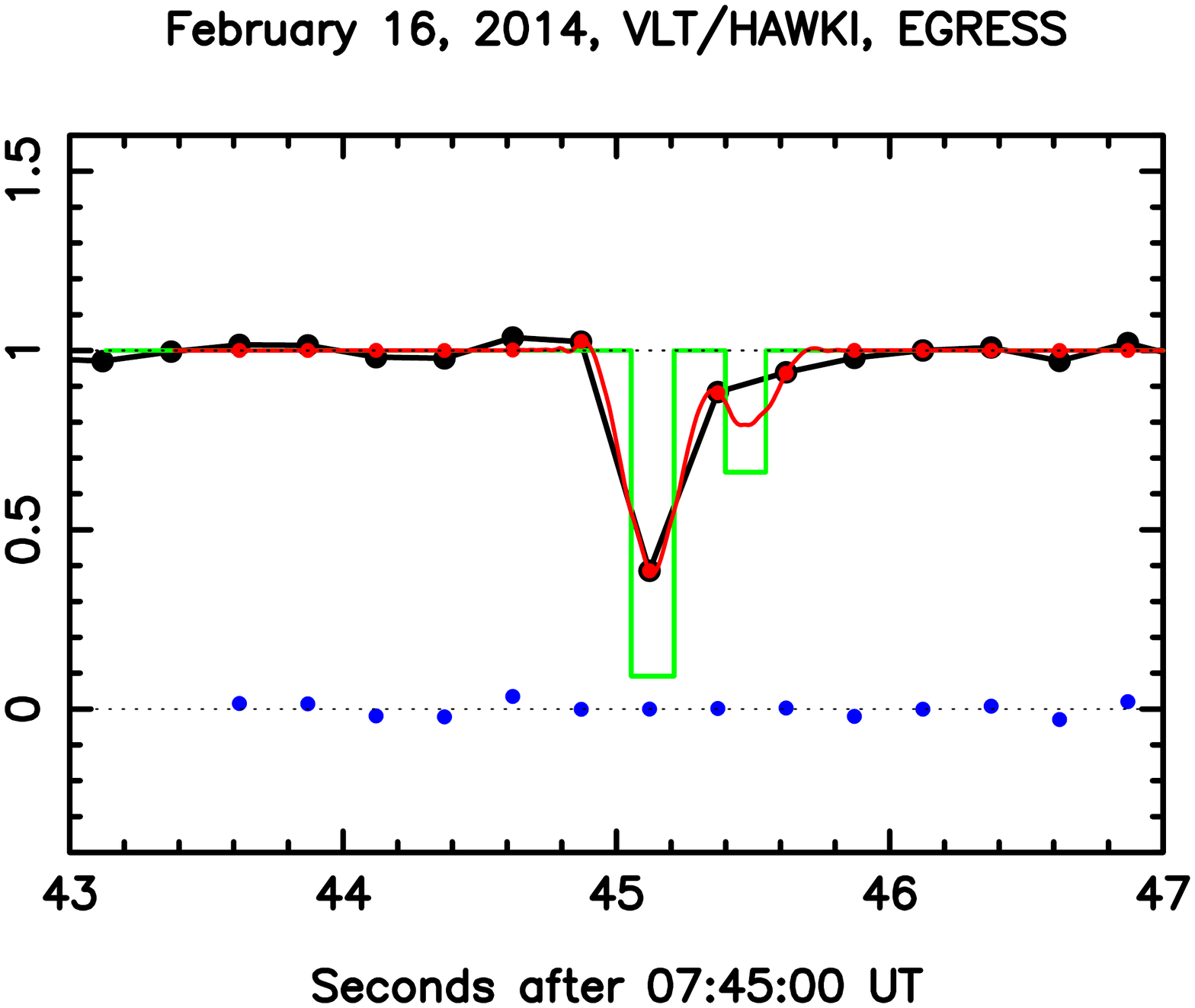} &
\includegraphics[angle=0,scale=0.25,trim=6cm 2cm 1cm 0]{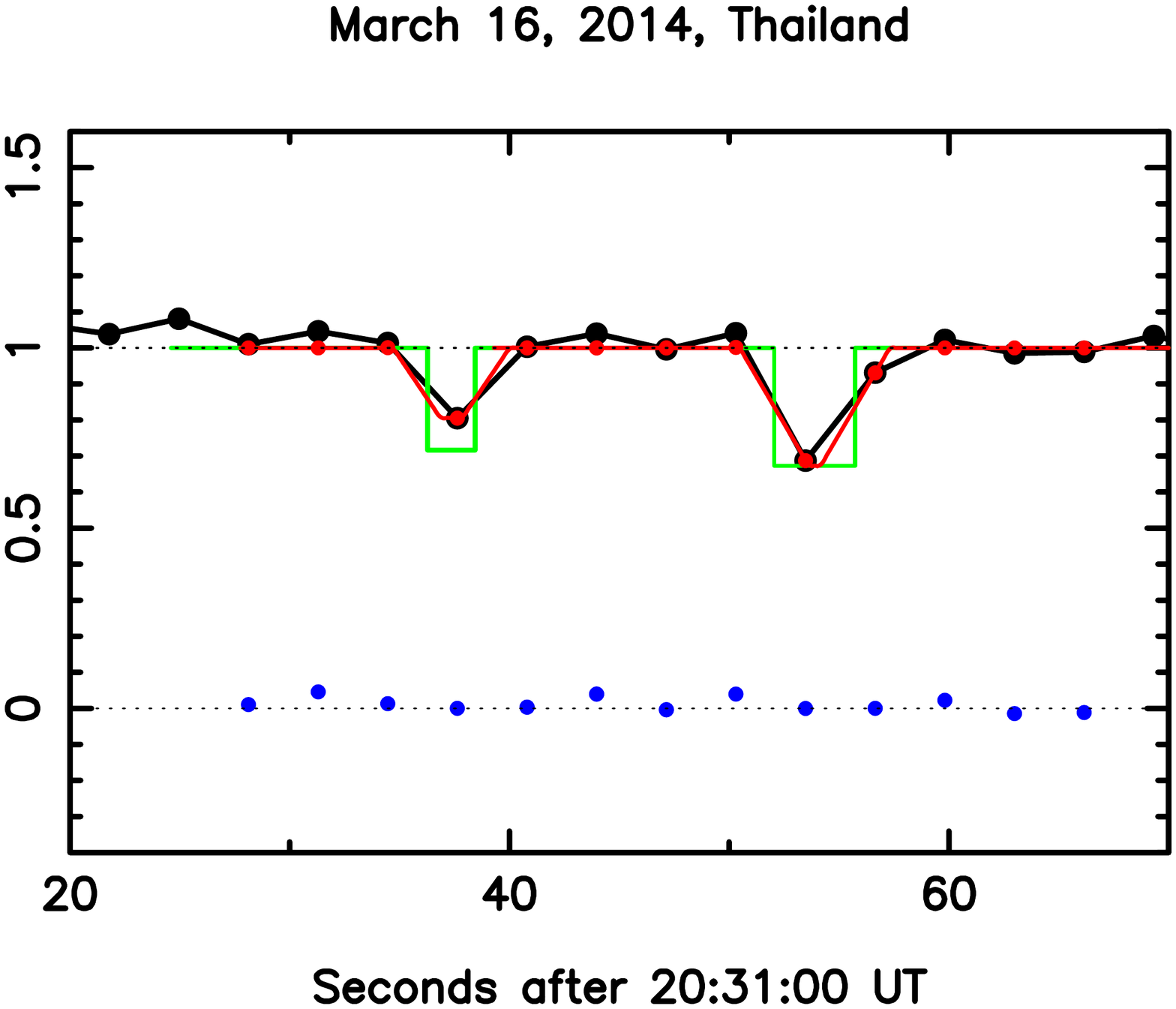} \\

\end{tabular}
\caption{ 
\footnotesize
Best fits to ring and main body occultations.
The black dots are the data points of the light curves (vertical axis represents the normalized flux). 
They are normalized between zero and unity.
The latter corresponding to the full flux from Chariklo and the star occulted.
The dotted lines correspond to the zero-level of the occulted star. 
The green curves are the best fitting square-well models used to generate the synthetic profiles, plotted in red. 
The physical characteristics of the rings extracted from these plots are listed 
in Tables~\ref{tab_param_rings} and \ref{tab_param_rings_unresolved}. 
The blue dots are the residual between the synthetic light curves and the data at each data point.
Figure~\ref{fig_fits_rings_2} shows the fits of remaining rings occultations.
}%
\label{fig_fits_rings_1}
\end{figure}

\begin{figure}[!htb]
\centering
\begin{tabular}{ccc}
\includegraphics[angle=0,scale=0.25,trim=6cm 2cm 1cm 0]{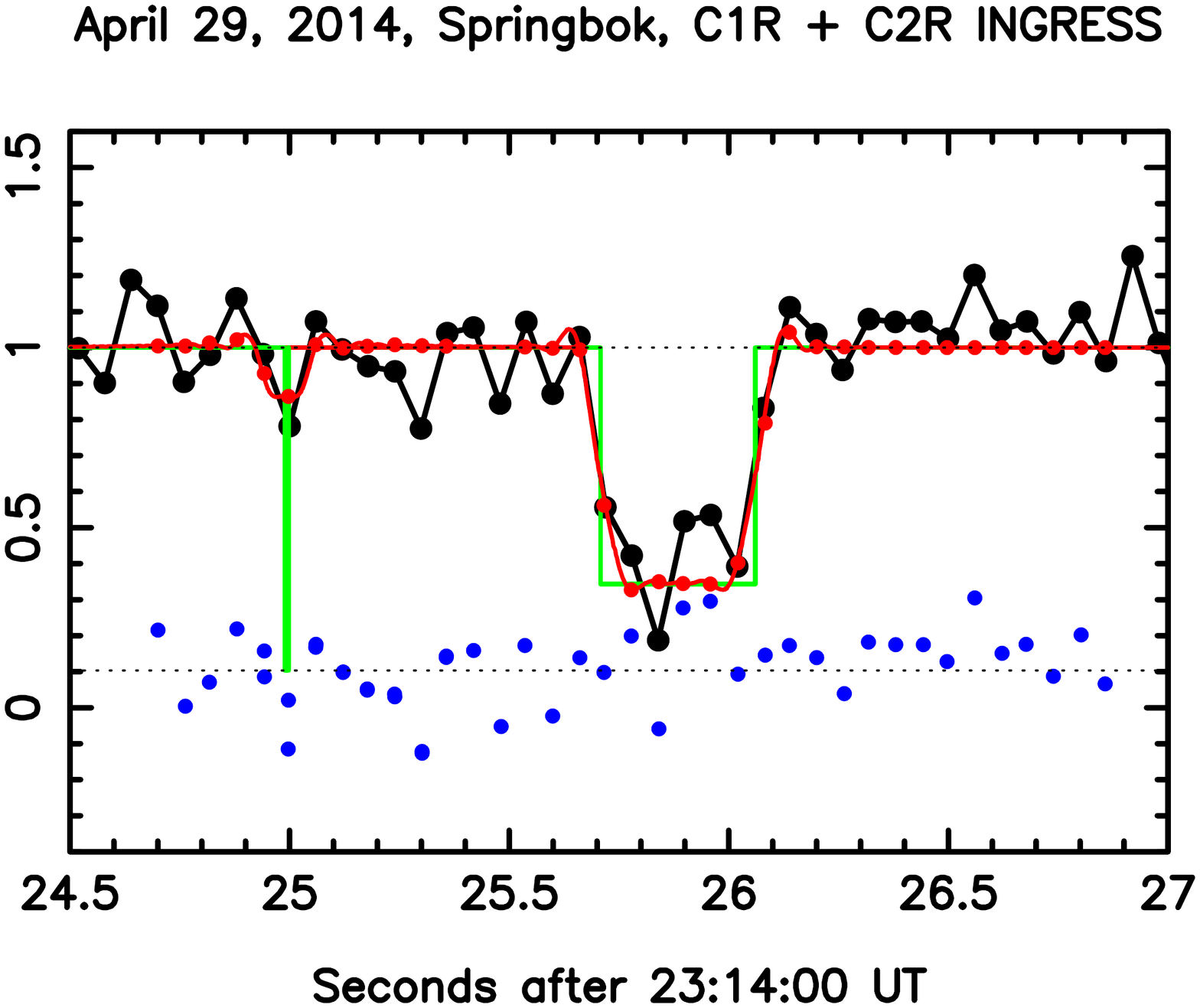} & 
\includegraphics[angle=0,scale=0.25,trim=6cm 2cm 1cm 0]{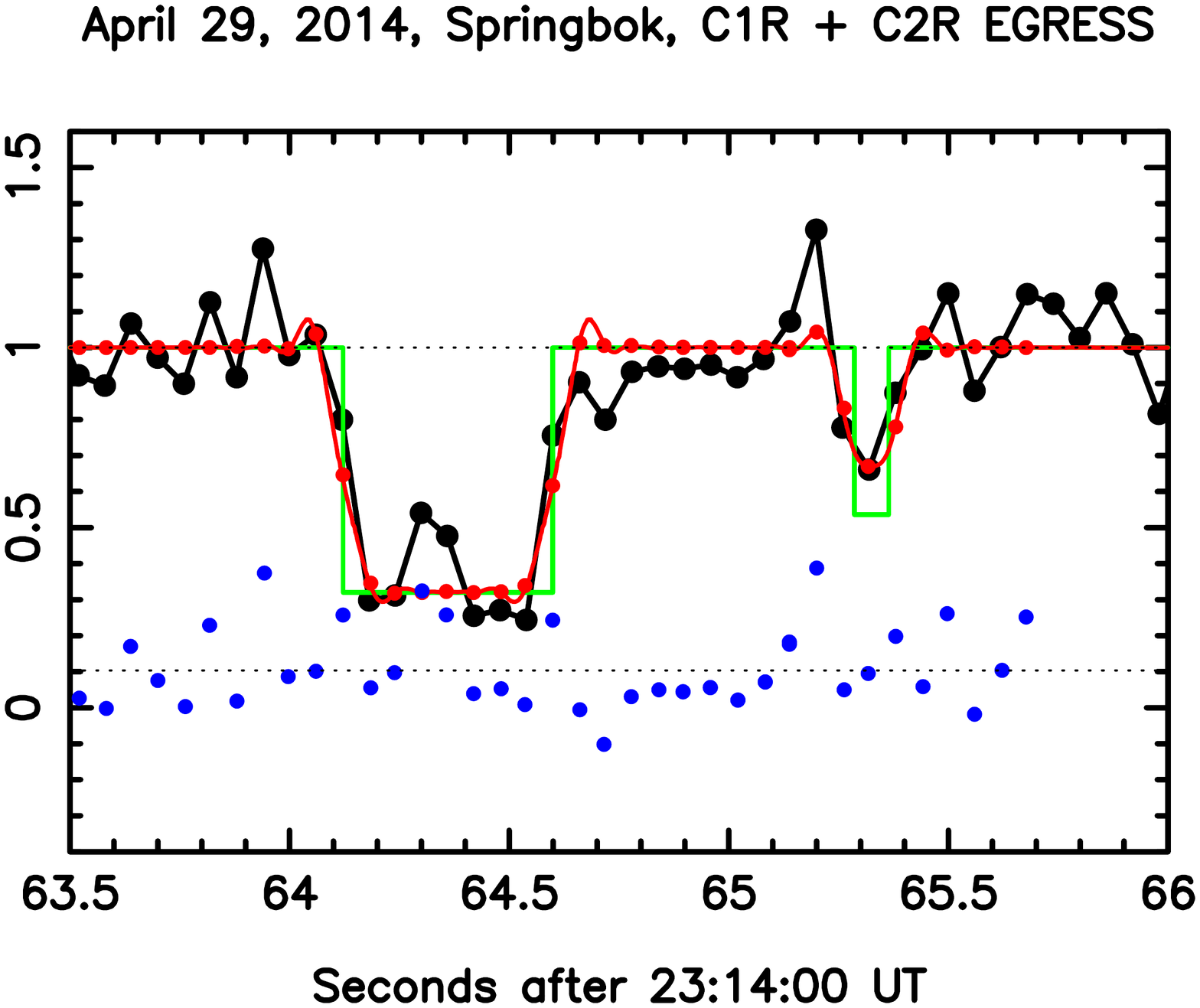} &
 \includegraphics[angle=0,scale=0.25,trim=6cm 2cm 1cm 0]{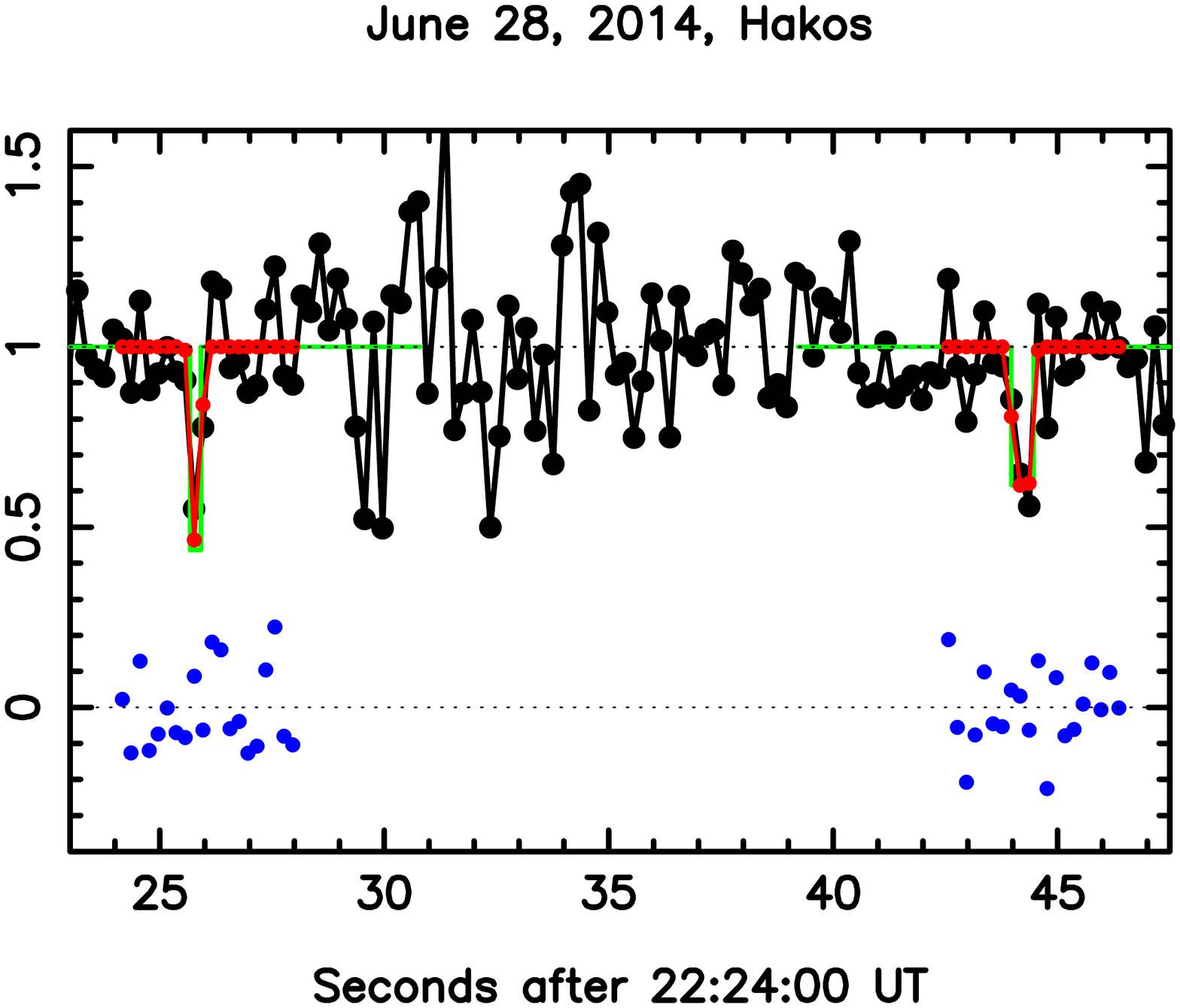} \\
 
\includegraphics[angle=0,scale=0.25,trim=6cm 2cm 1cm 0]{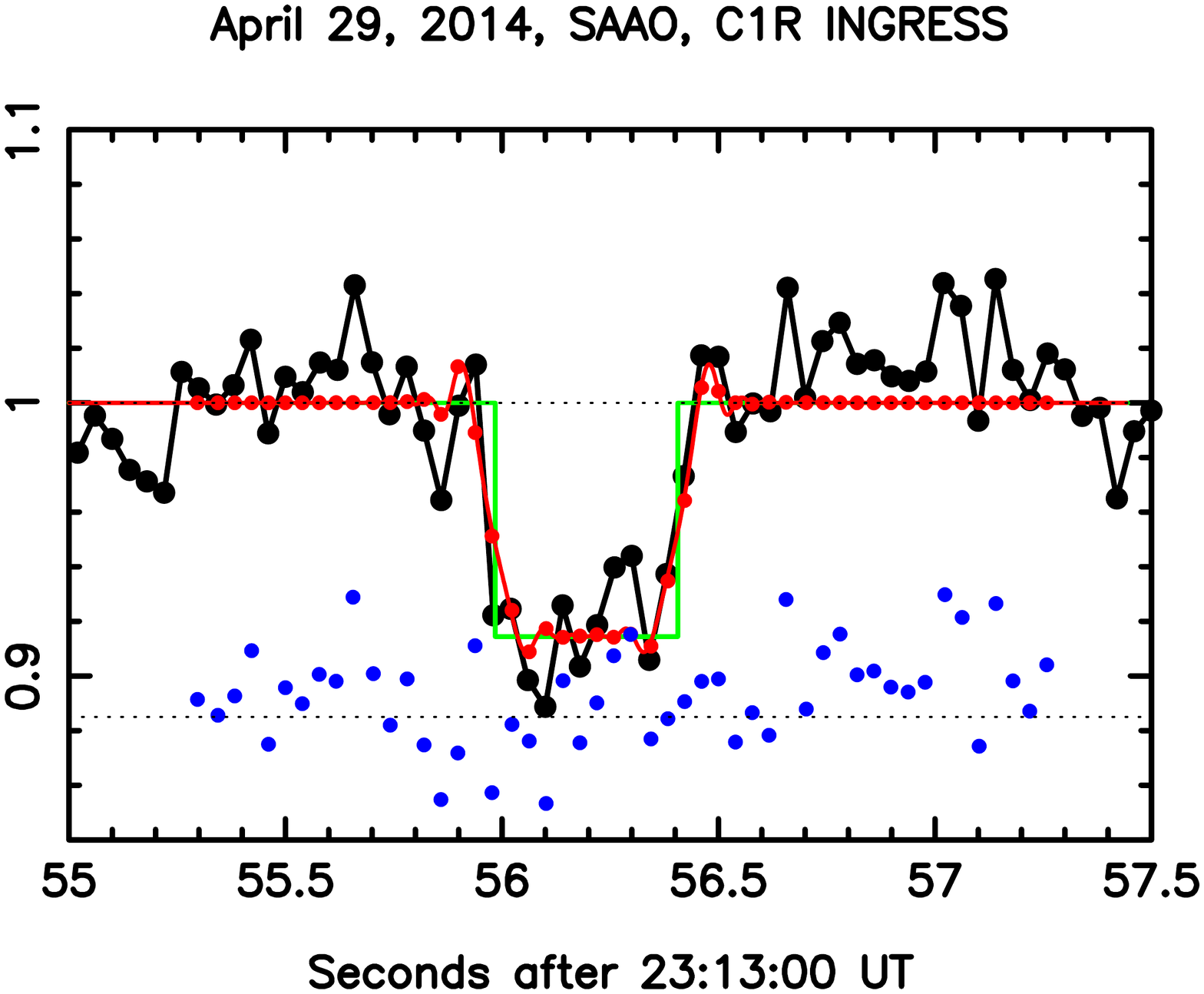}&
\includegraphics[angle=0,scale=0.25,trim=6cm 2cm 1cm 0]{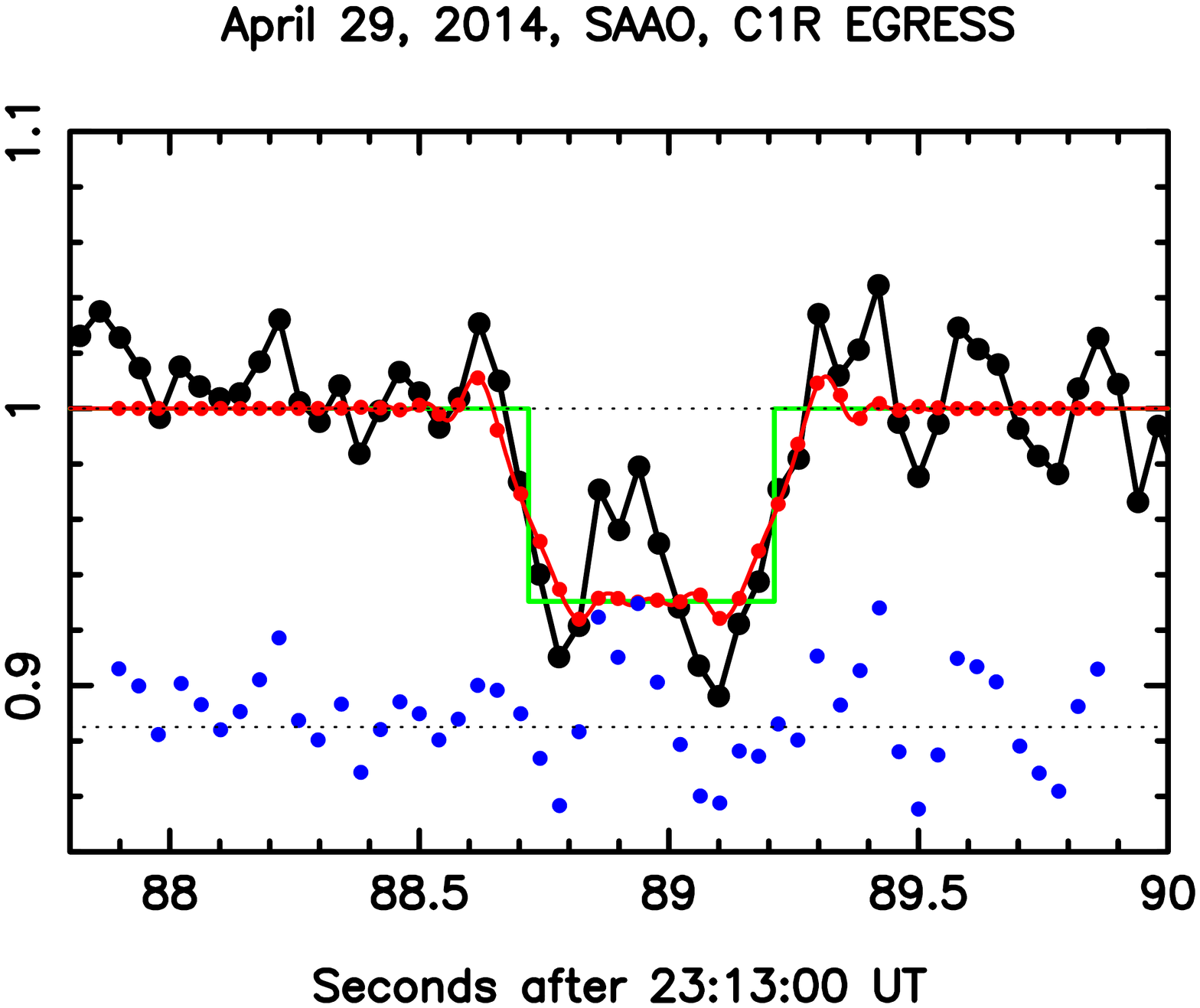}&
\includegraphics[angle=0,scale=0.25,trim=6cm 2cm 1cm 0]{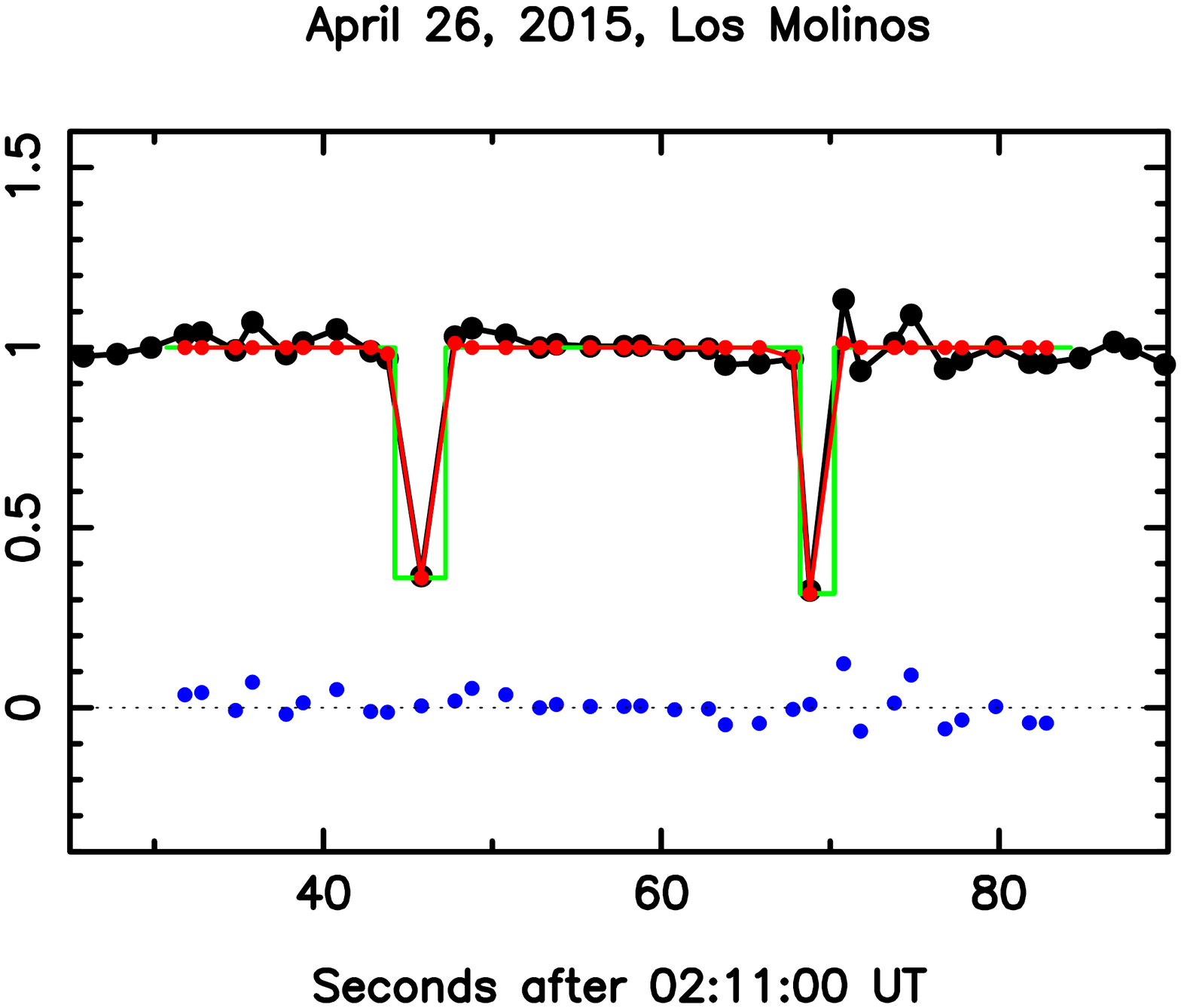}\\

  \includegraphics[angle=0,scale=0.25,trim=6cm 2cm 1cm 0]{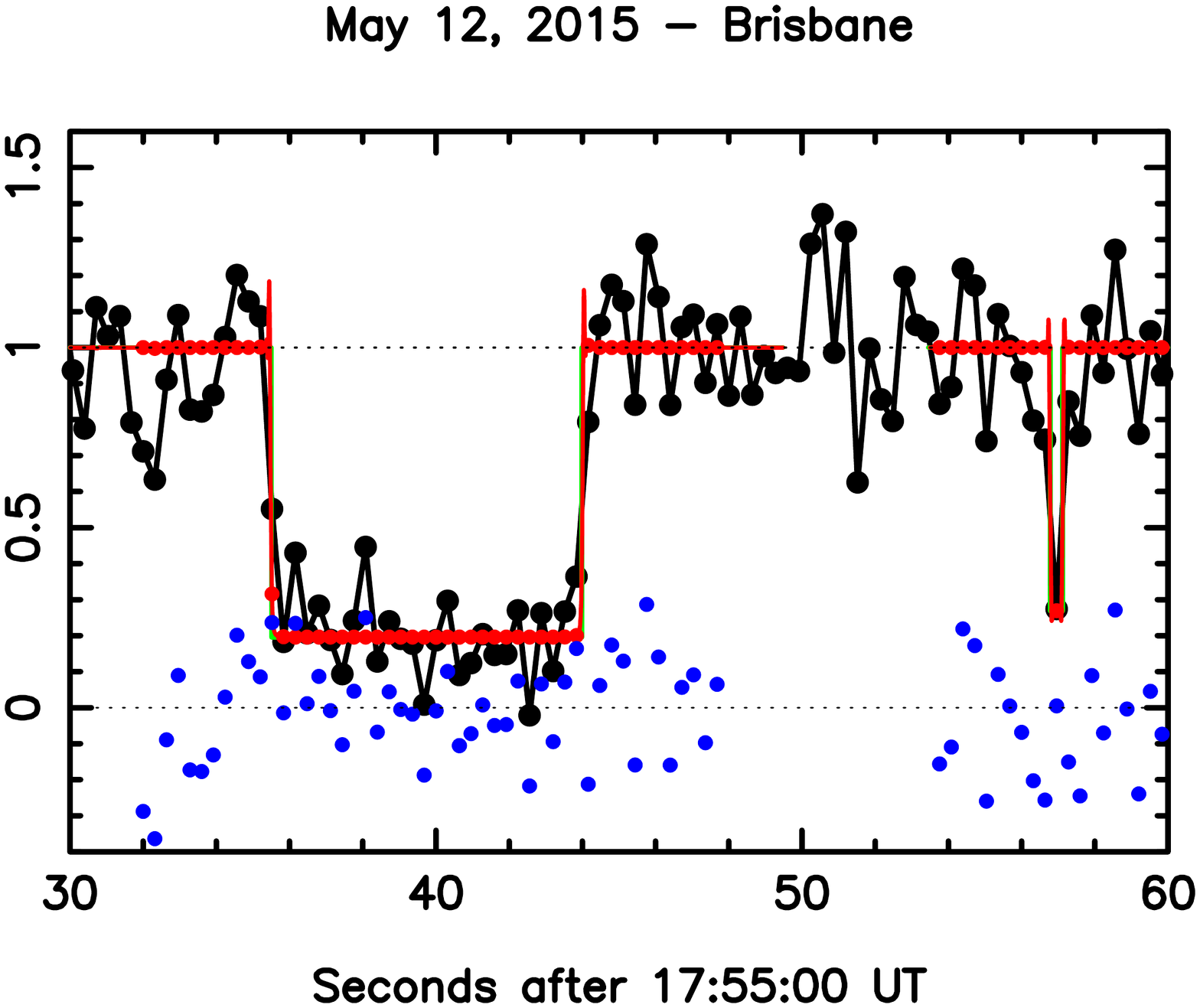} & 
  \includegraphics[angle=0,scale=0.25,trim=6cm 2cm 1cm 0]{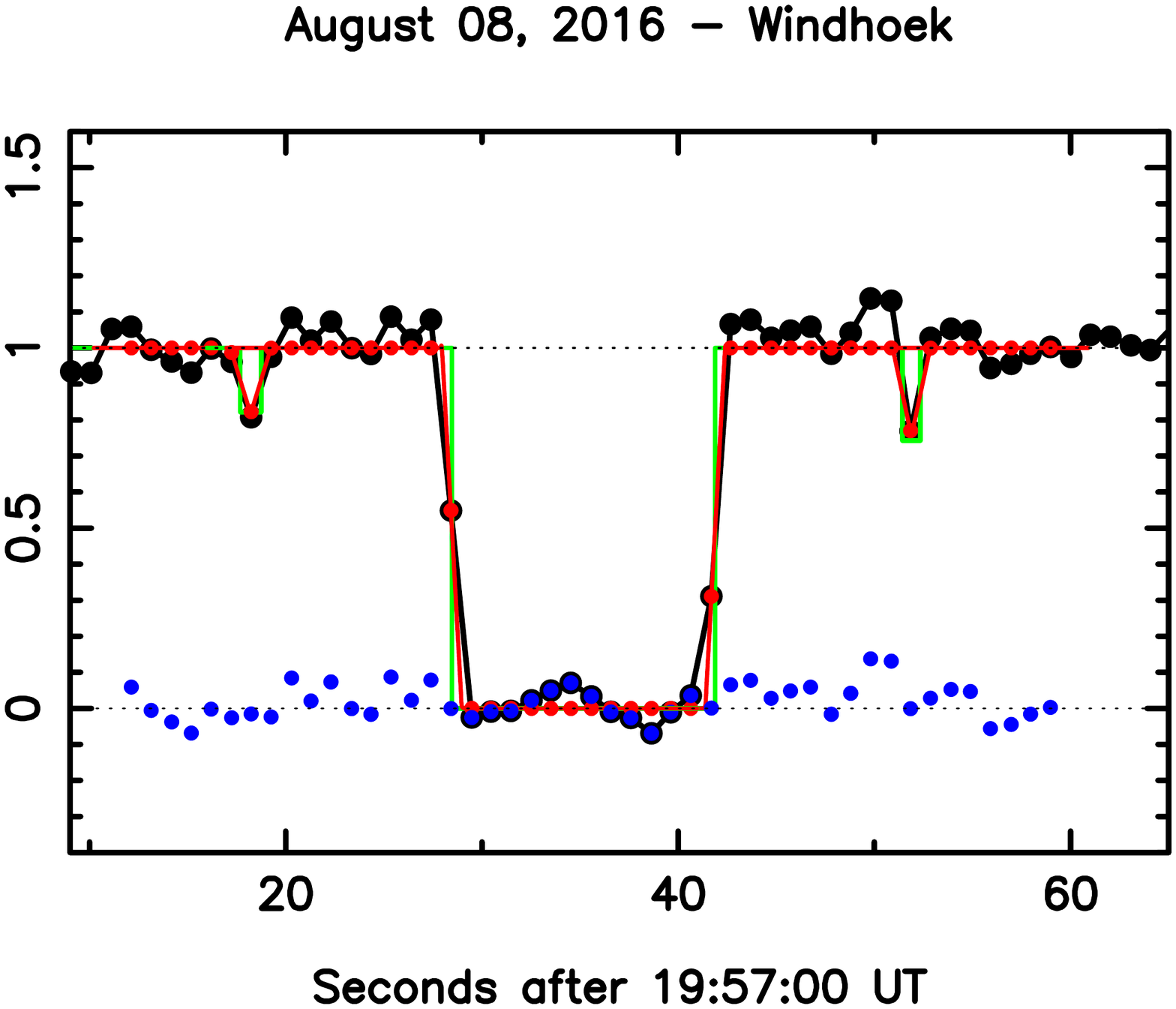} & \\
  
   \includegraphics[angle=0,scale=0.25,trim=6cm 2cm 1cm 0]{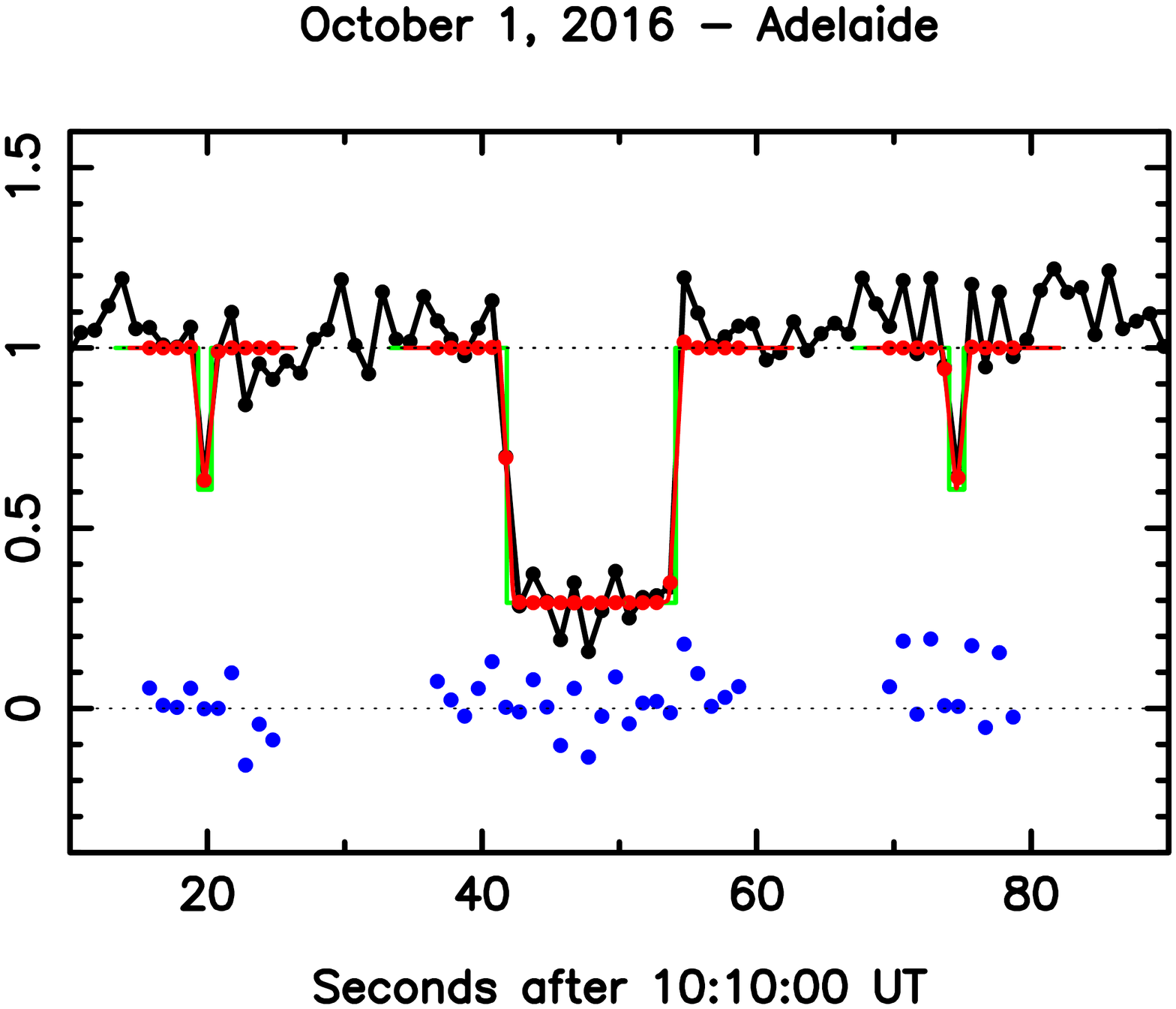} &
   \includegraphics[angle=0,scale=0.25,trim=6cm 2cm 1cm 0]{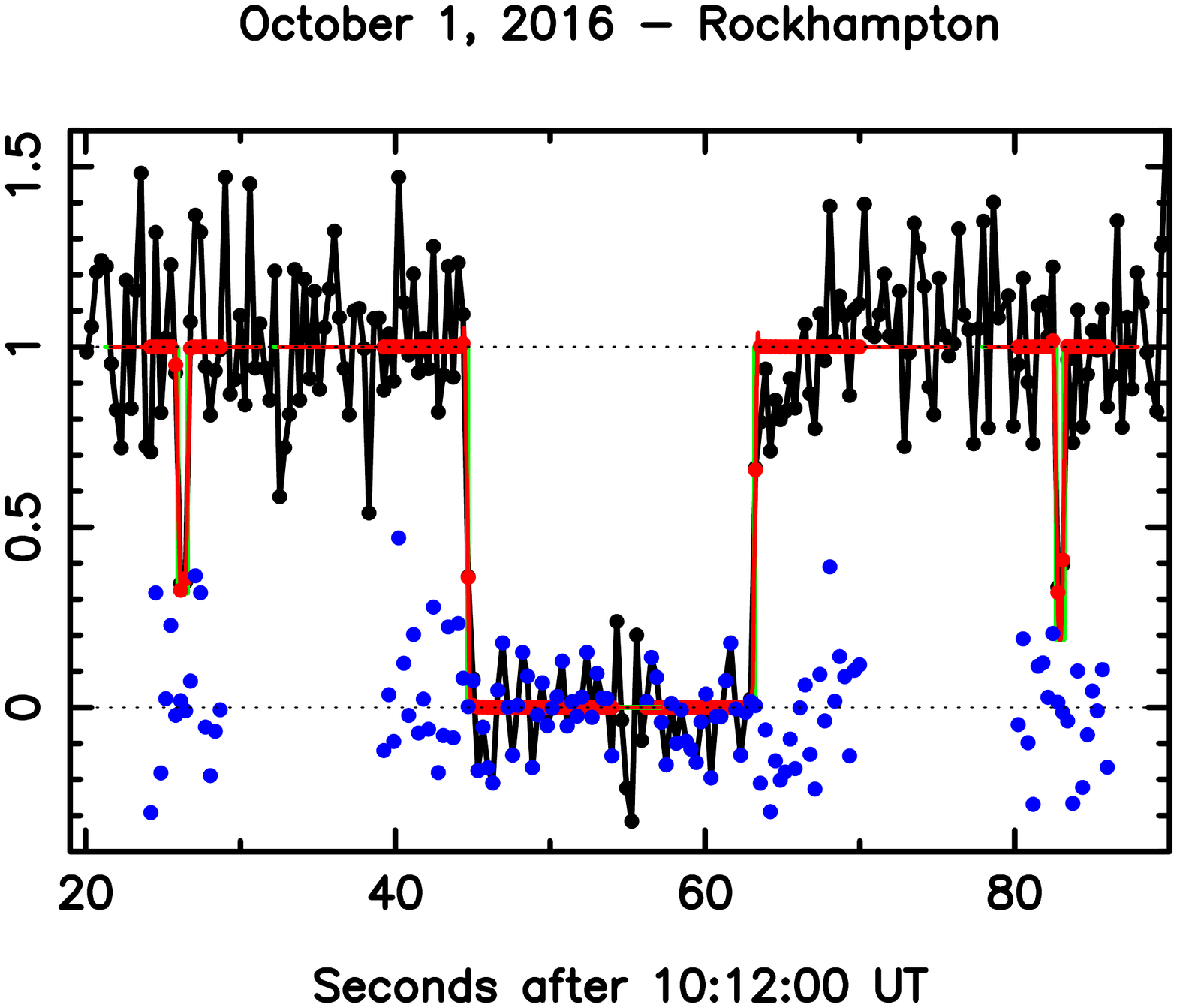} &  \\

\end{tabular}
\caption{ 
\footnotesize
Best fits to ring and main body occultations (following and completing Fig.~\ref{fig_fits_rings_1}).
Same legend as in Figure~\ref{fig_fits_rings_1},
except in the case of the occultation on April 29, 2014 where two stars were occulted. 
In this case unity corresponds to the flux of the two stars and Chariklo.
As SAAO observed an occultation of a secondary star (see Section~\ref{section_etoile}),
its vertical scale is different from other light curves, for better viewing.
}%
\label{fig_fits_rings_2}
\end{figure}


\begin{figure}[!htb]
\centering
\begin{tabular}{ccc}
 \includegraphics[angle=0,scale=0.27,trim=6cm 2cm 1cm 0]{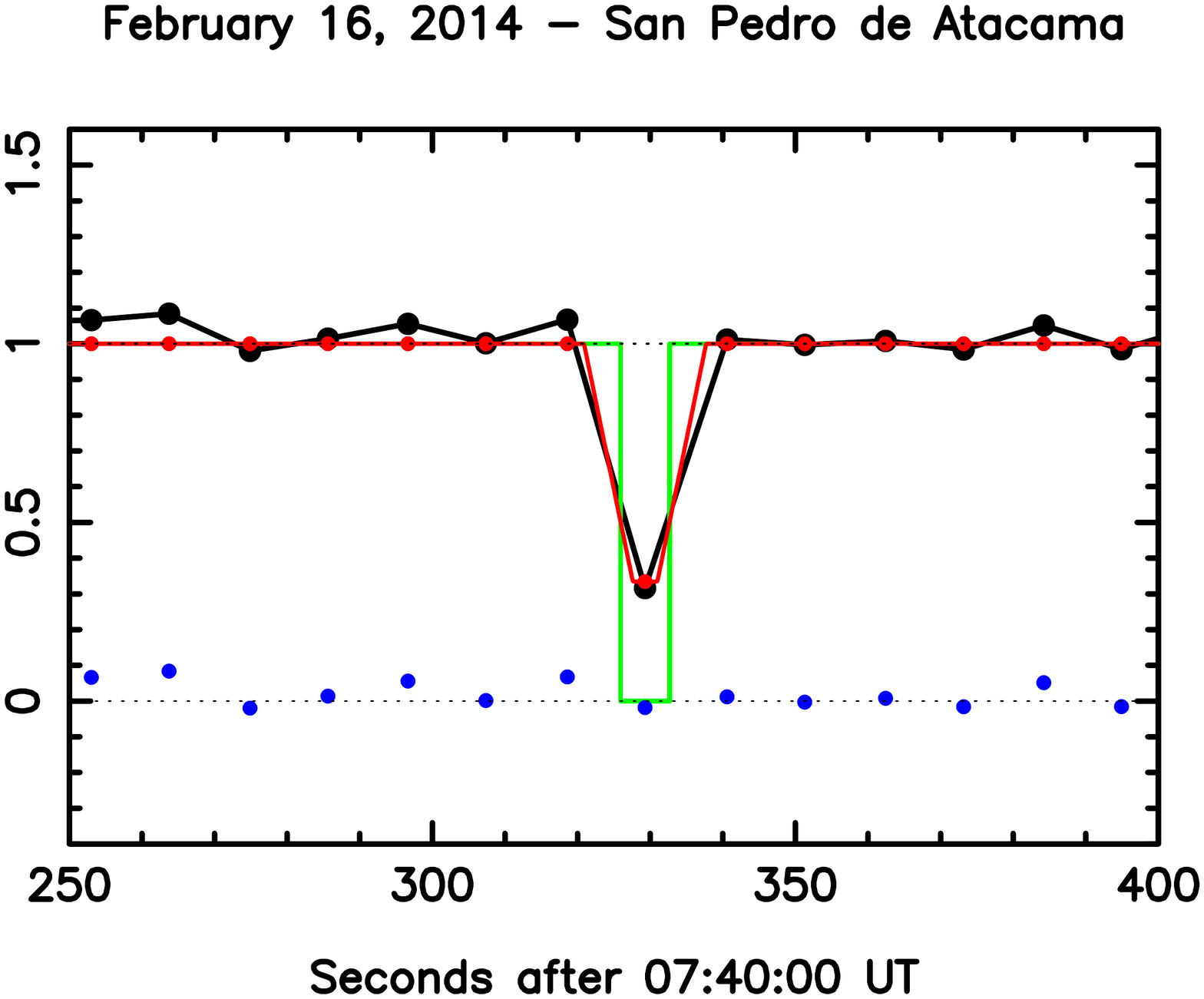} & 
  \includegraphics[angle=0,scale=0.27,trim=6cm 2cm 1cm 0]{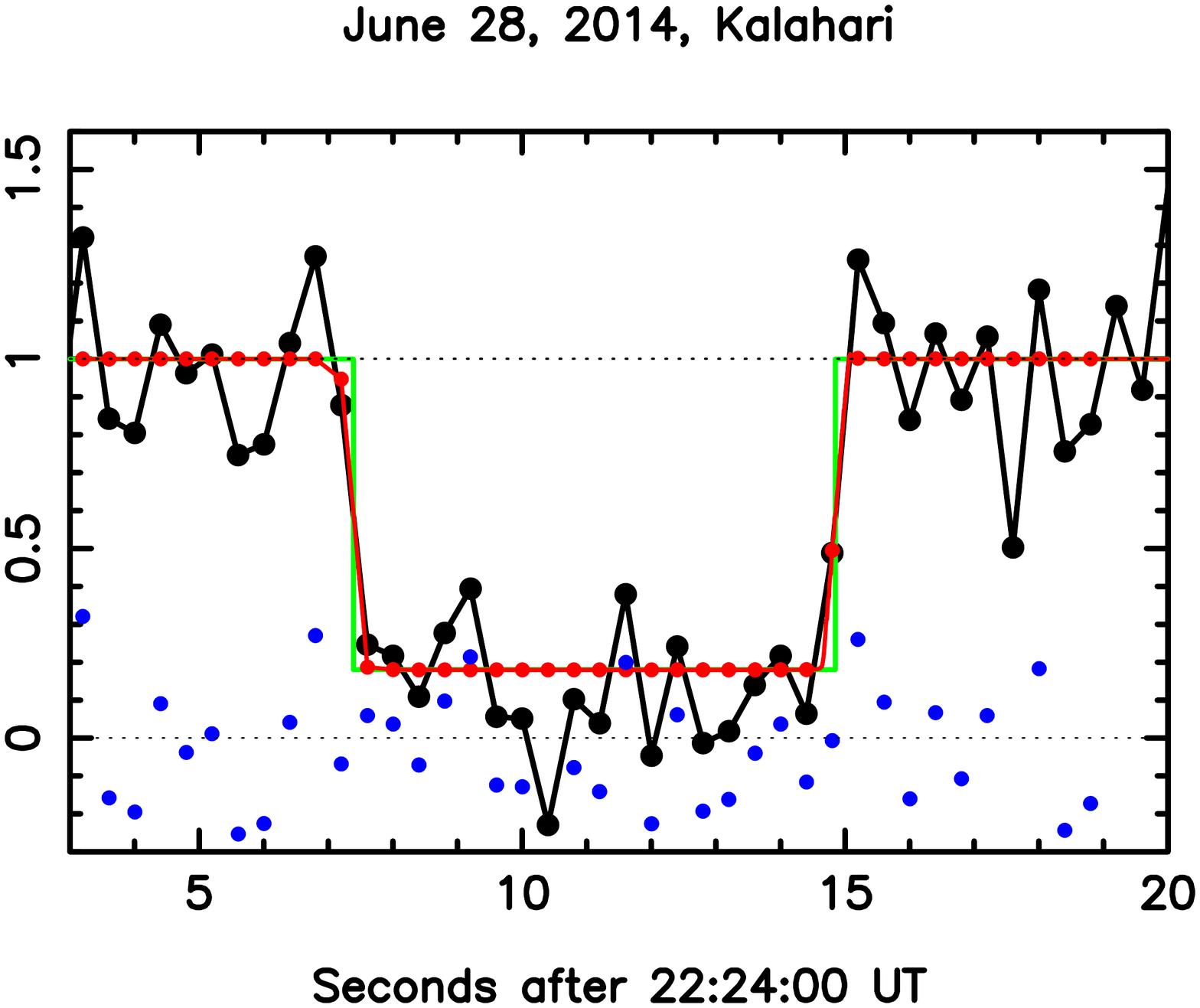} & 
  \includegraphics[angle=0,scale=0.27,trim=6cm 2cm 1cm 0]{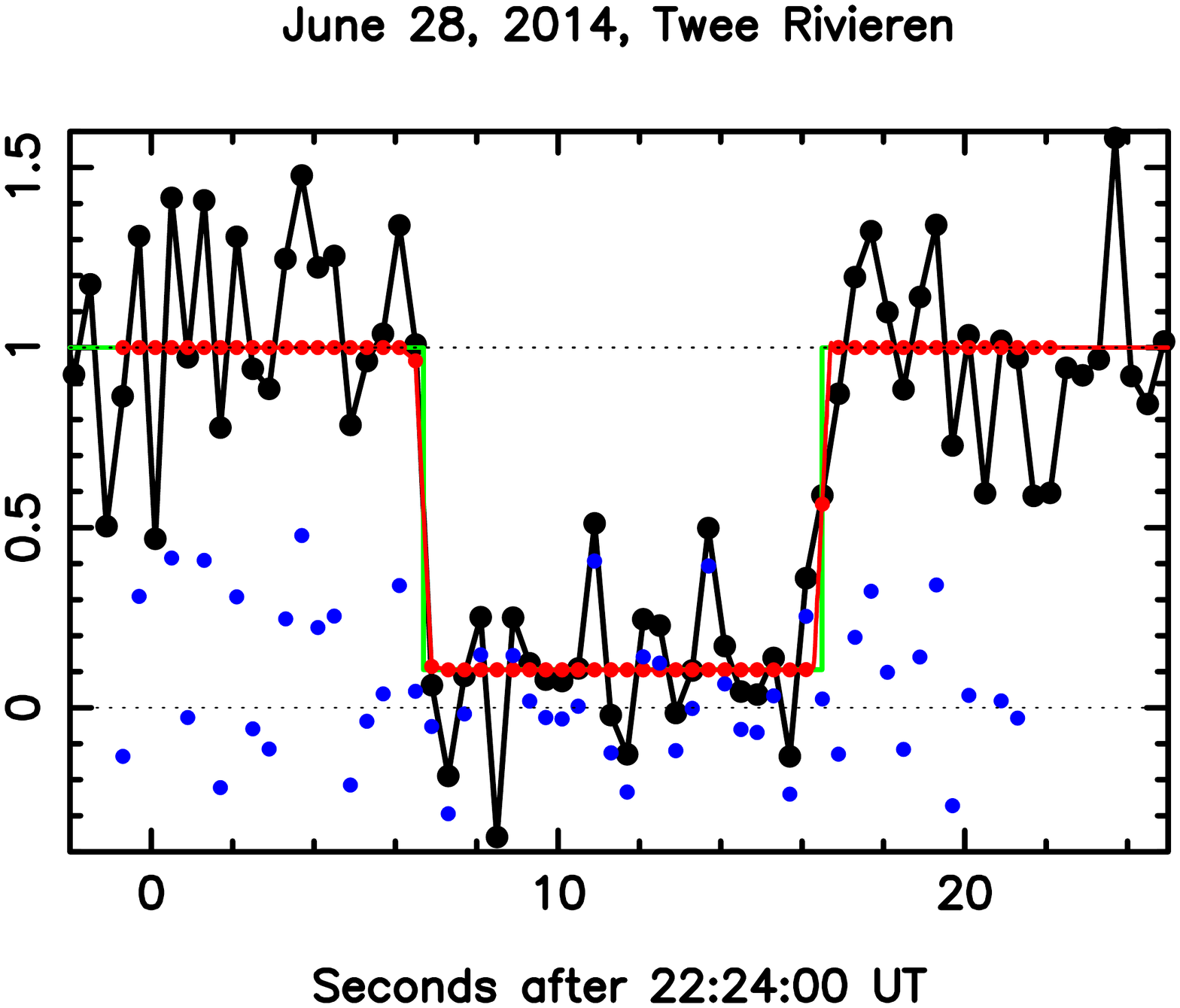}\\
 \includegraphics[angle=0,scale=0.27,trim=6cm 2cm 1cm 0]{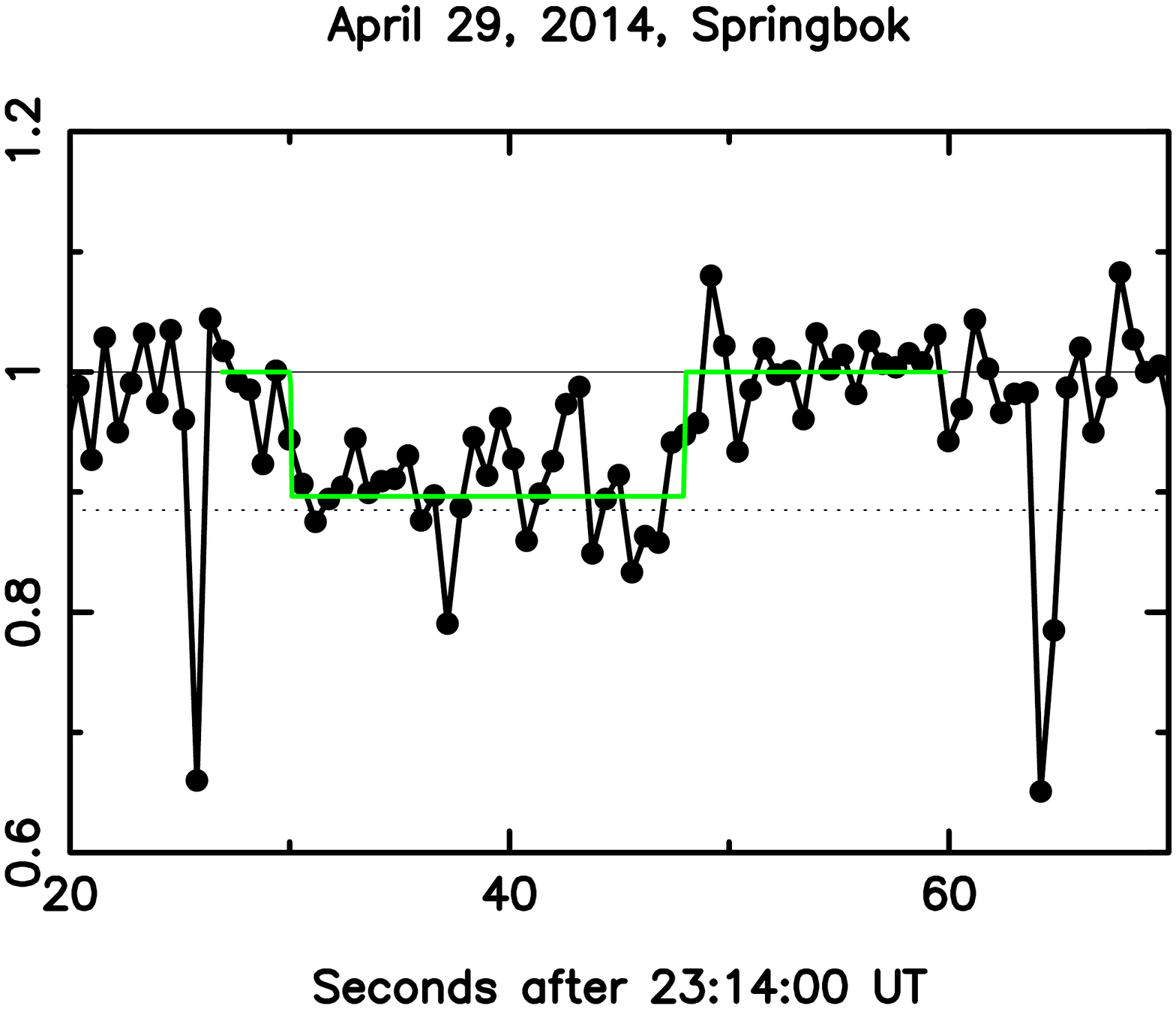} &
  \includegraphics[angle=0,scale=0.27,trim=6cm 2cm 1cm 0]{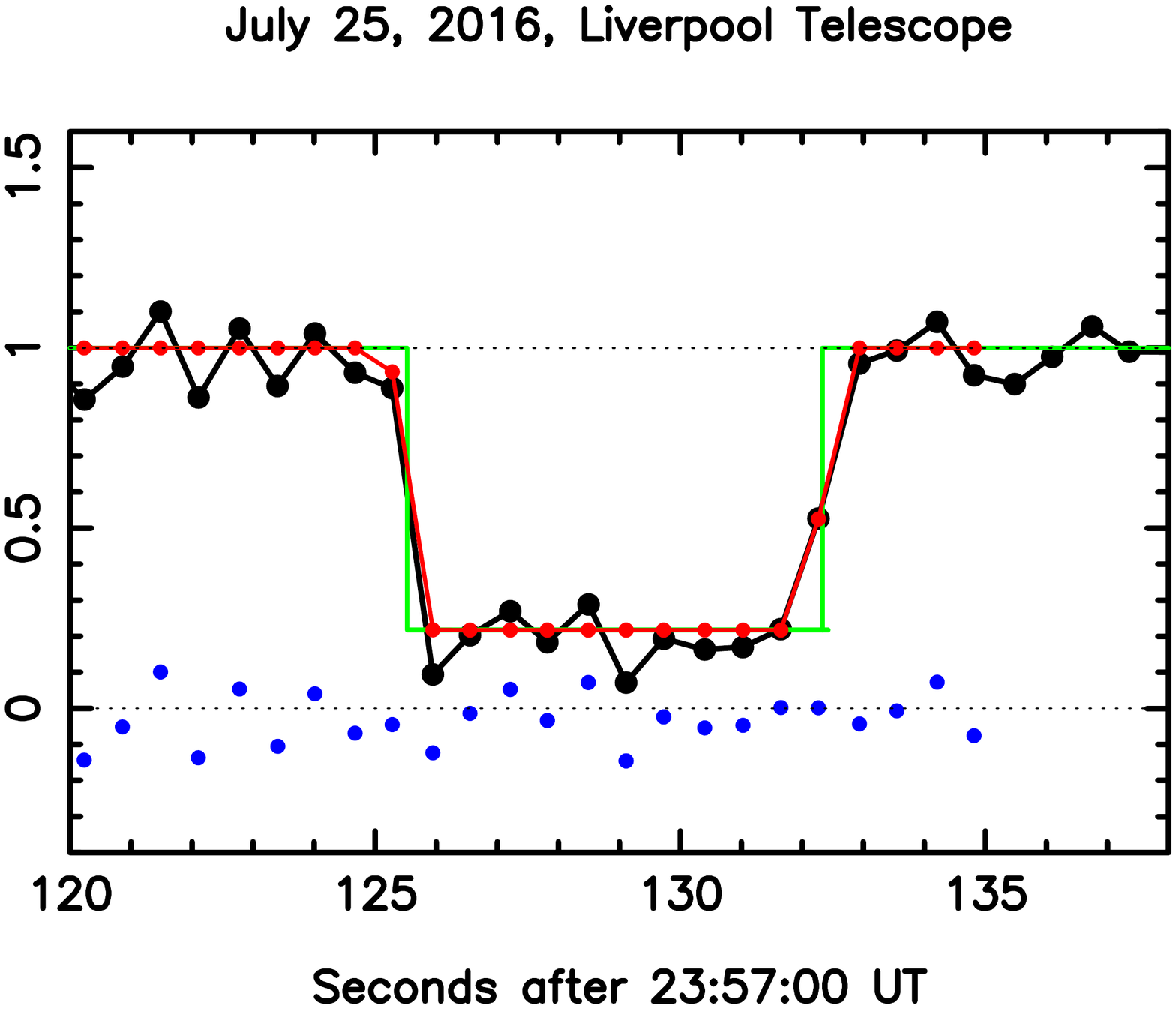} &
   \includegraphics[angle=0,scale=0.27,trim=6cm 2cm 1cm 0]{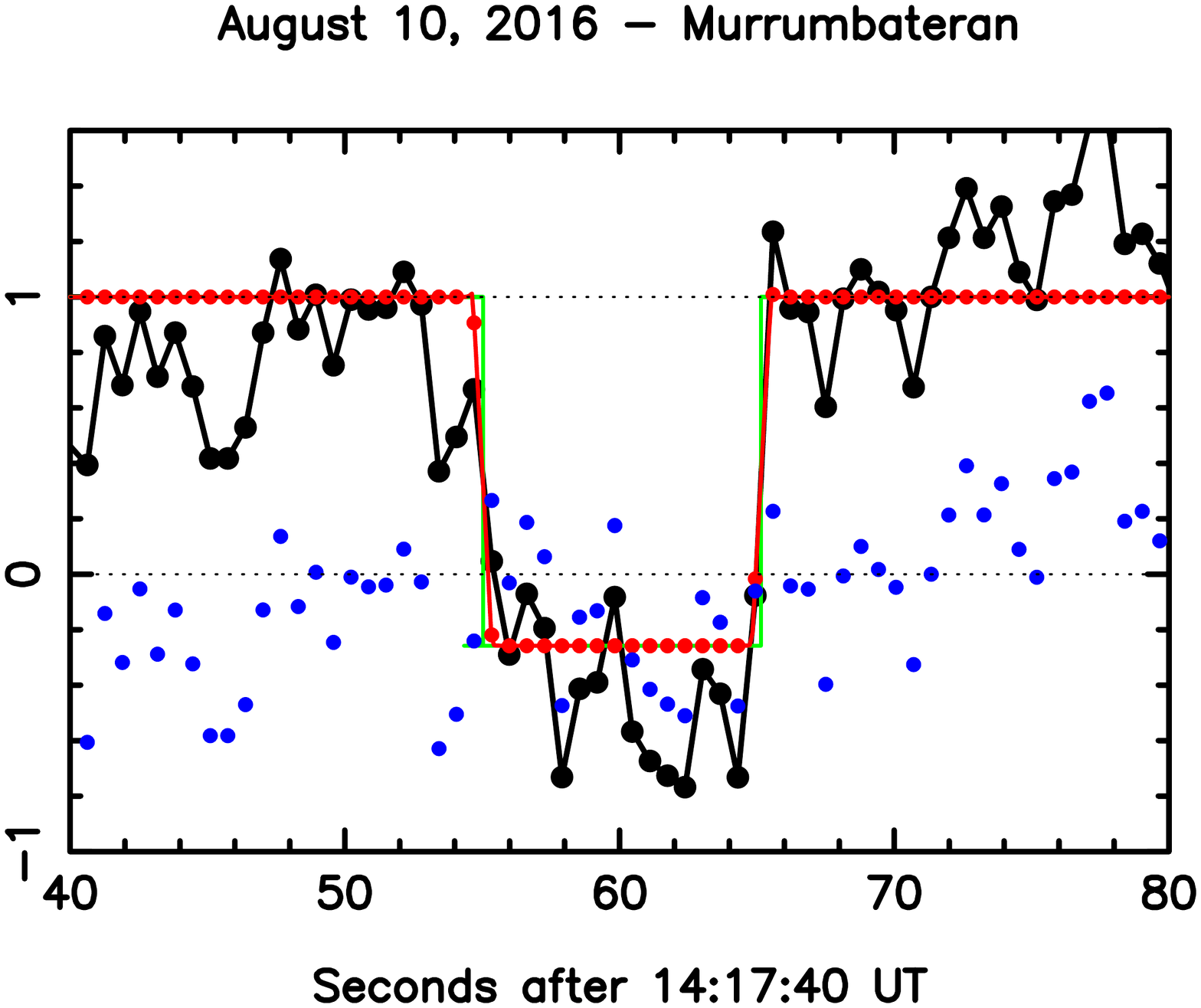} \\  
   \includegraphics[angle=0,scale=0.27,trim=6cm 2cm 1cm 0]{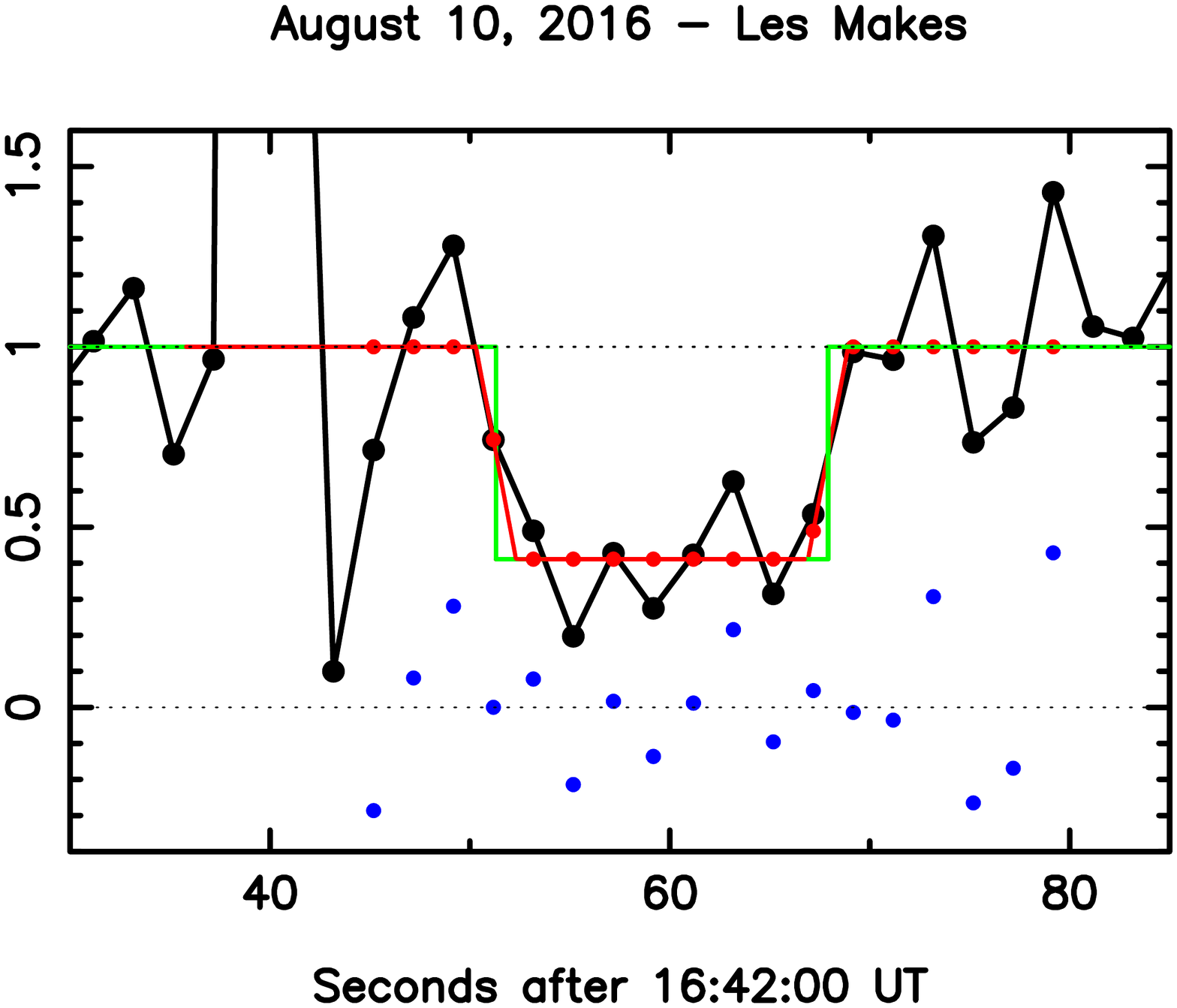}&
   \includegraphics[angle=0,scale=0.27,trim=6cm 2cm 1cm 0]{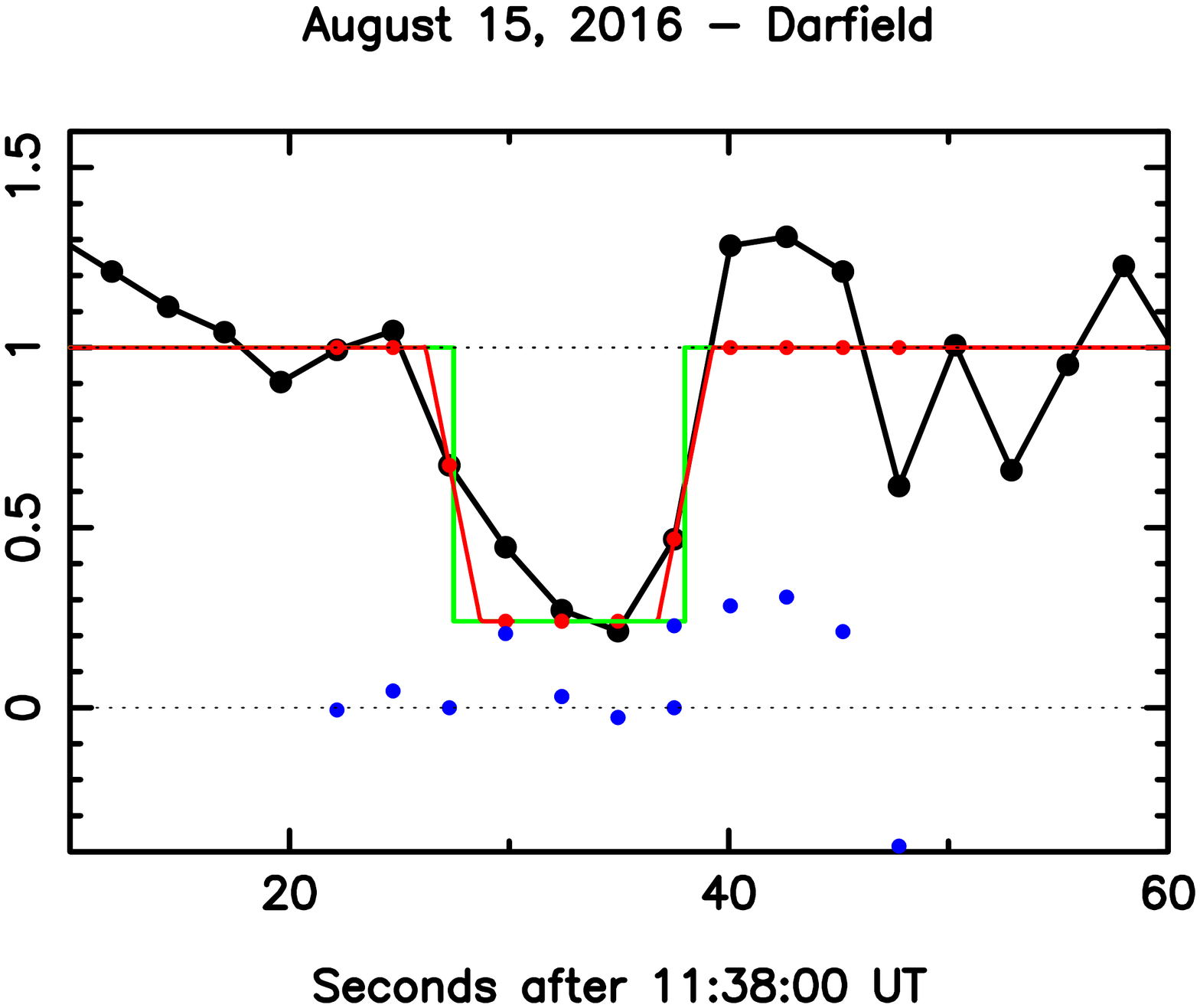}&
   \\
\end{tabular}
\caption{ 
\footnotesize
Same as Figure~\ref{fig_fits_rings_1}, but for events with only main body detections.} 
\label{fig_fits_body}
\end{figure}


\begin{figure}[!htb]
\centering
\includegraphics[angle=0,scale=0.5,trim=0 0 0 0]{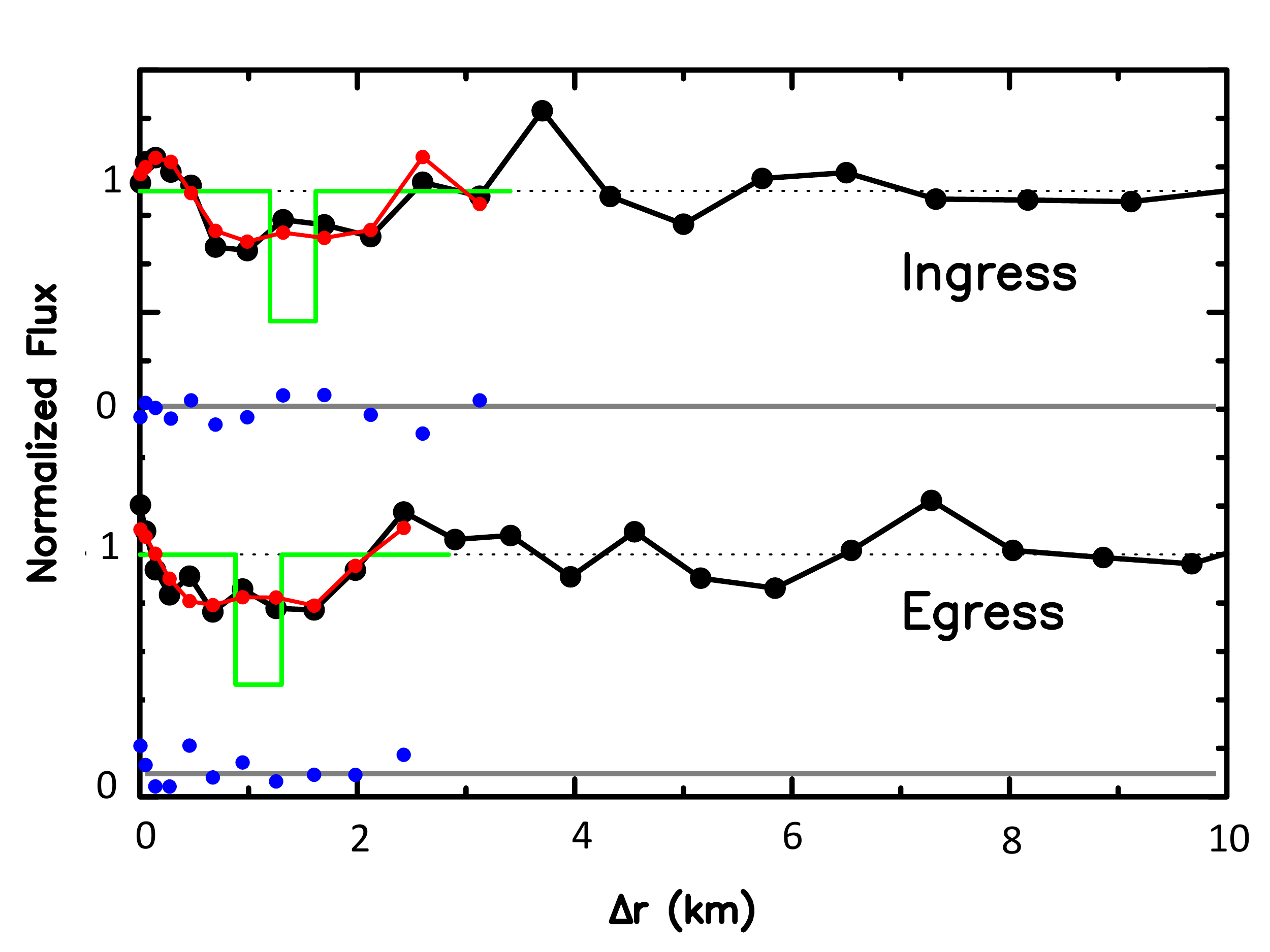} 
\caption{%
\footnotesize
Fits to grazing event in Gifberg (April 29, 2014) using a common width ($W_{\perp}~=0.422$~km and $p'=0.4$)
 for both rings into $1\sigma$ level (see Table~\ref{tab_param_rings}).
The star motion relative to C2R was grazing, so that its velocity 
perpendicular to the ring changed significantly during the occultation.
In this case, it is therefore necessary to express the flux against the distance 
to the point of closest approach to Chariklo's center in km in the sky plane, $\Delta r$.
Other than that, color conventions and vertical axis are the same as in Fig.~\ref{fig_fits_rings_1}.
}%
\label{fig_fits_gifberg}
\end{figure}


\begin{figure}[!htb]
\centering
\includegraphics[angle=0,scale=0.5,trim=0 0 0 0]{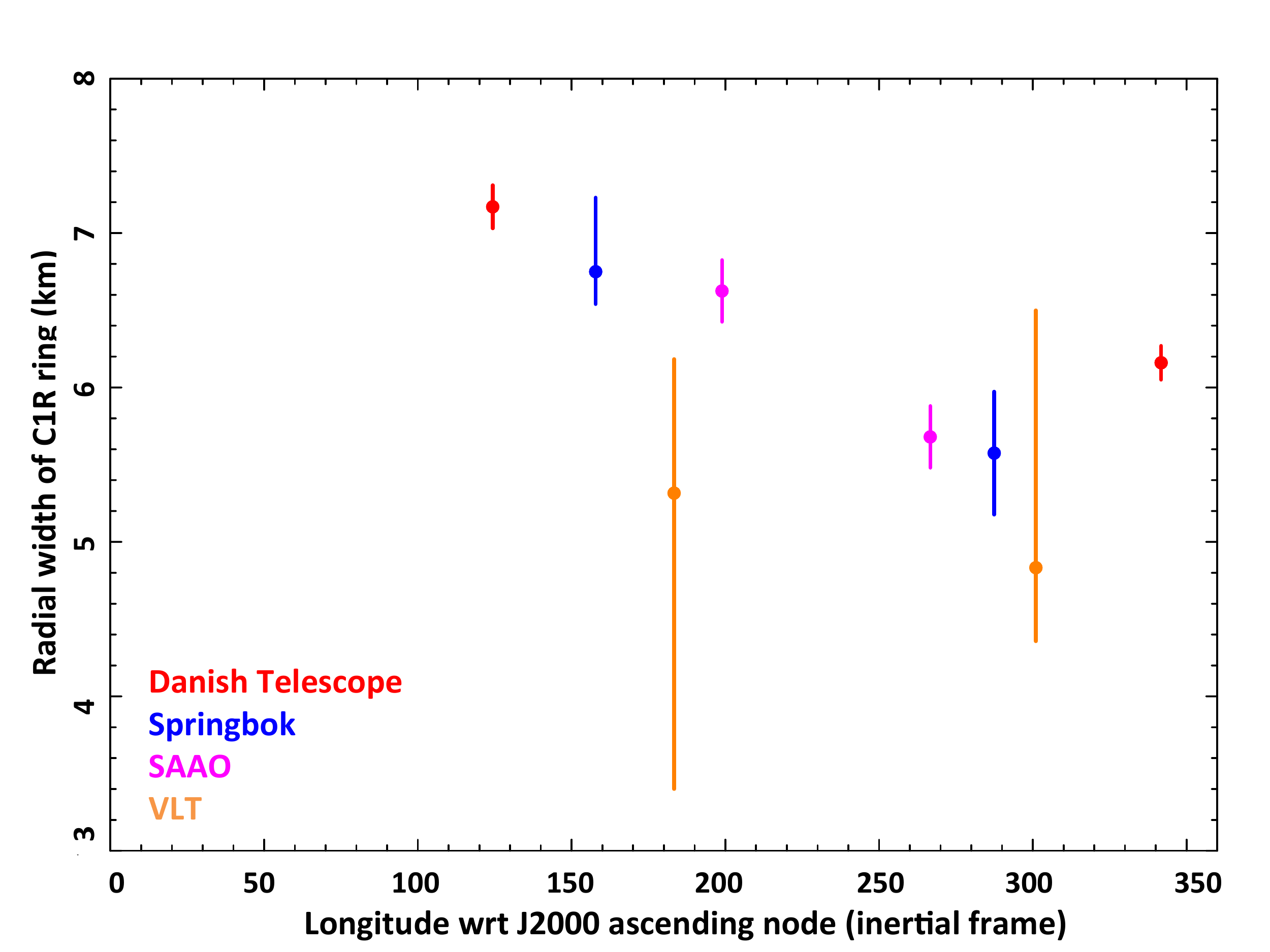} 
\caption{%
\footnotesize
Variation of C1R radial width (and $1\sigma$ error bars) with true longitude $L$ counted from the J2000 ring plane ascending node for resolved events.
}%
\label{fig_W_vs_L}
\end{figure}


\begin{figure}[!htb]
\centering
\includegraphics[angle=0,scale=0.7,trim=0 0 0 0]{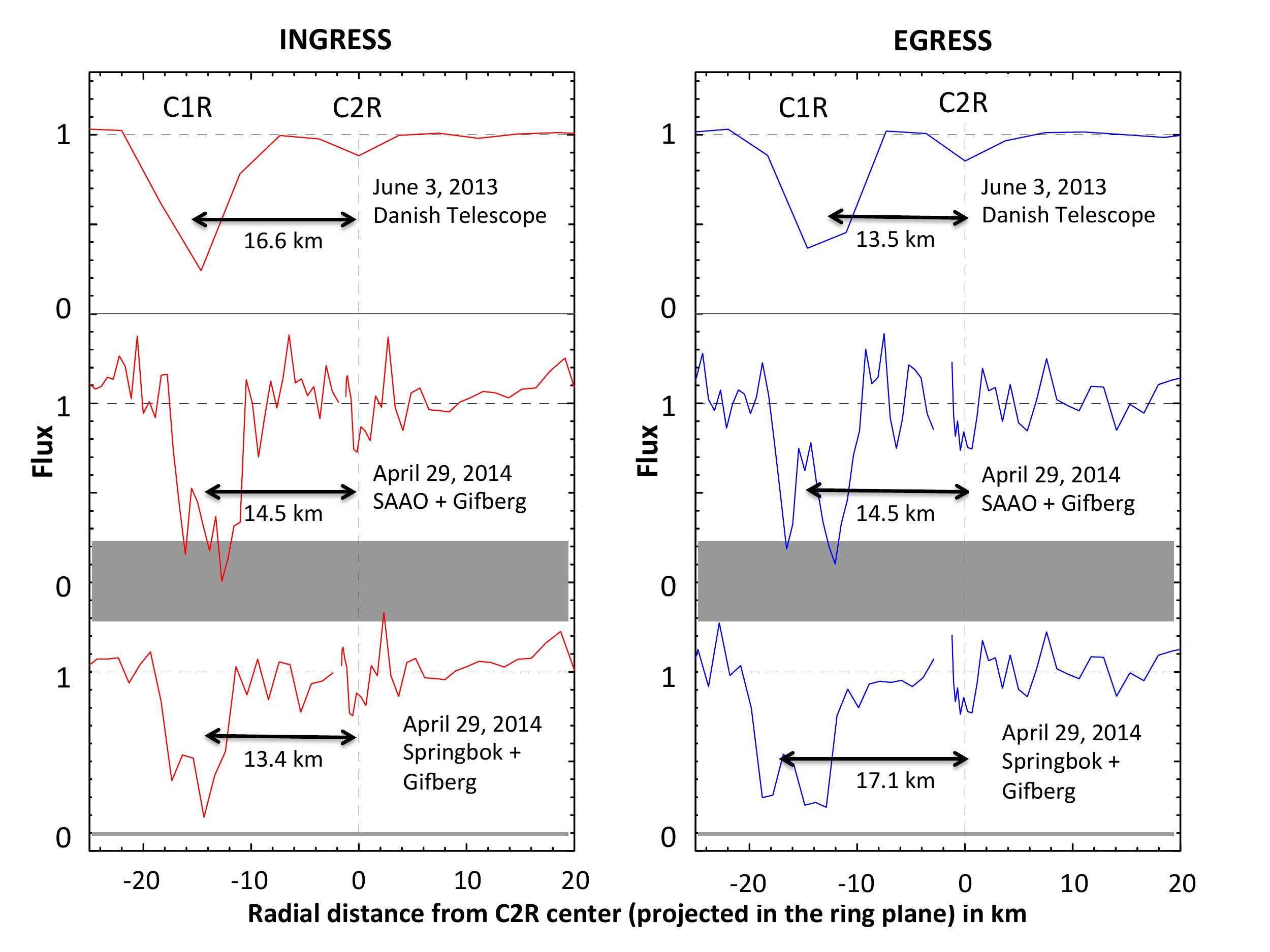}
\caption{%
\footnotesize
Best radial profiles of rings C1R and C2R.
The profiles have been plotted arbitrarily against the radial distance (in the ring plane) 
to the center of the C2R profile, using the pole position of \cite{bra14}.
This choice enhances possible changes in the relative distances of the two rings, 
due for instance to eccentricities of C1R and/or C2R.
The horizontal dashed lines correspond to the unocculted star + Chariklo flux.
The horizontal gray boxes correspond to the respective zero stellar fluxes.
The thickness of a gray box indicates the uncertainty of the photometric calibrations,
see text for details (Section~\ref{section_etoile}).
The left (resp. right) panel corresponds to ingress (resp. egress).
Top panels: 
the June 3, 2013 profiles from the Danish Telescope.
Middle and bottom panels: 
montages constructed from the April 29, 2014 event.
The Gifberg profiles showing C2R have been combined with 
SAAO light curve (middle panel) and 
Springbok light curve (bottom panel). 
%
}%
\label{fig_profile_C2R_centre}
\end{figure}


\begin{figure}[!htb]
\centering
\begin{tabular}{c}
 \includegraphics[angle=0,scale=0.5,trim=0 0 0 0]{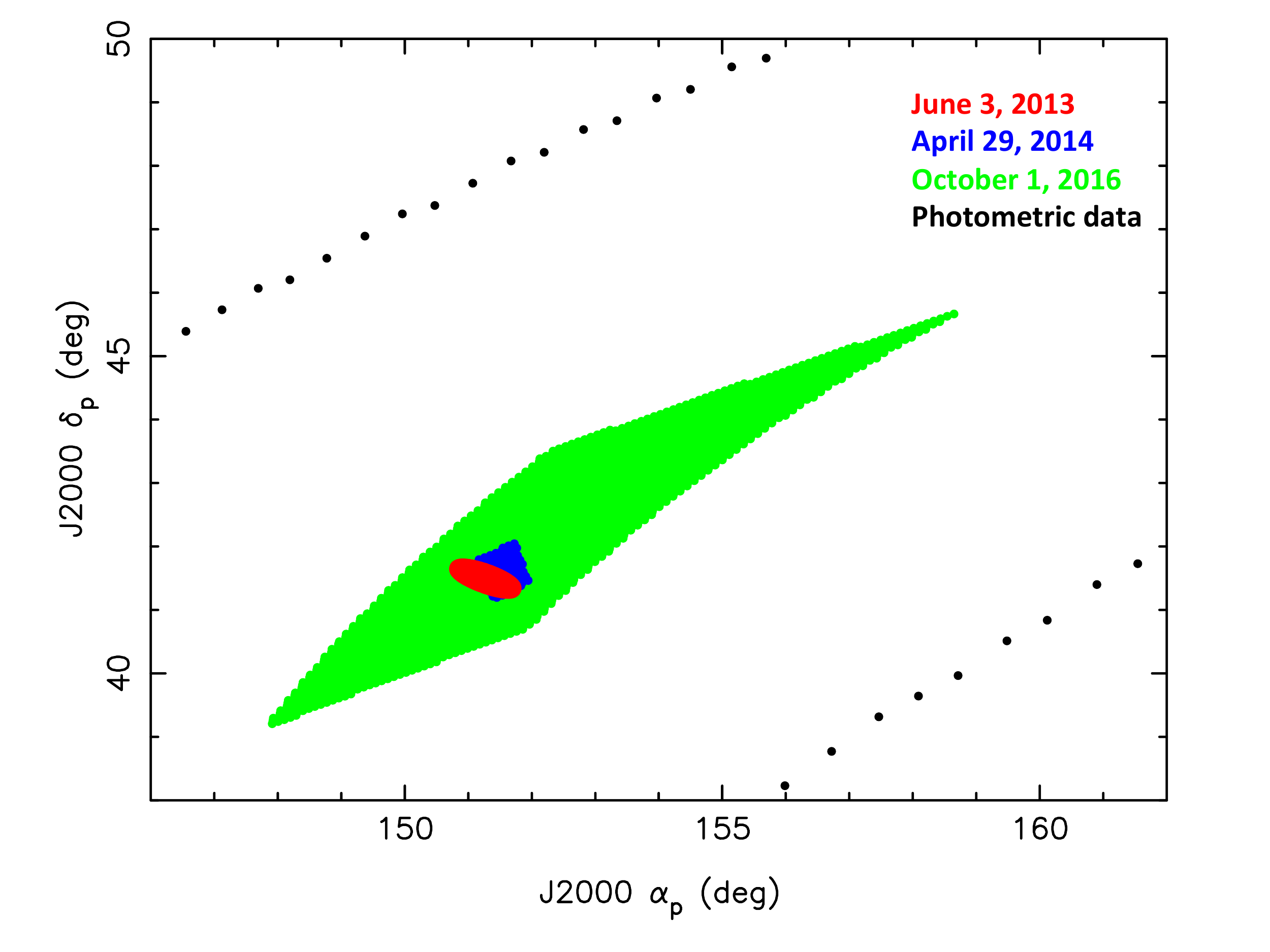} 
\end{tabular}
\caption{ 
\footnotesize
Constraints on ring pole. 
The uncertainty domains (1$\sigma$ level) on the pole position ($\alpha_p,\delta_p$) for the event on 
June 3, 2013, April 29, 2014 and October 1, 2016 are plotted in red, blue and green, respectively. 
The black dots outline the uncertainty domain derived from the long term variations in Chariklo's photometry \citep{duf14}.
}
\label{fig_pole_circular}
\end{figure}



\begin{figure}[!htb]
\centering
\begin{tabular}{c}
 \includegraphics[angle=0,scale=0.5,trim=0 0 0 0]{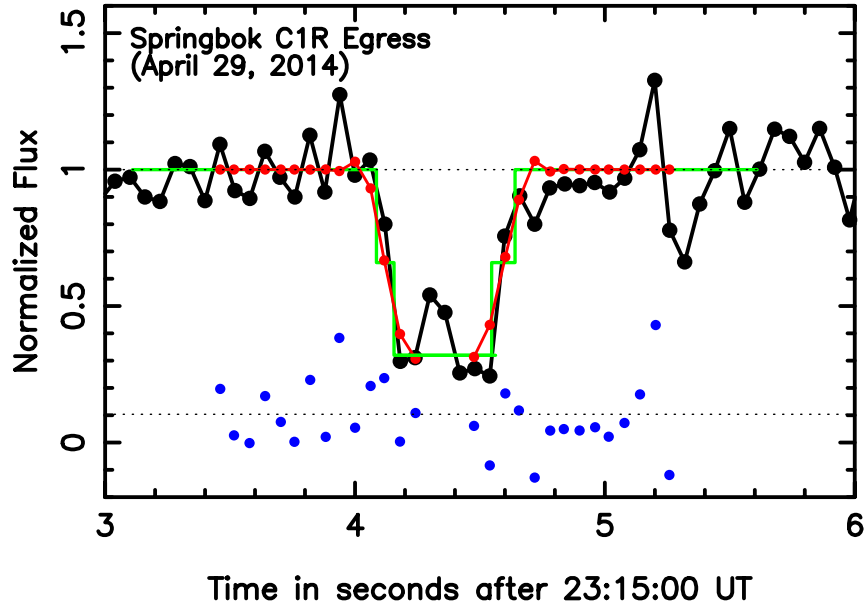} 
\end{tabular}
\caption{%
\footnotesize
Measurement of C1R's edges sharpnesses
with an example taken from the Springbok egress profile (April 29, 2014).
The green line is the step-wise model of width $\Delta w_r$ described in Section~\ref{section_sharpness}.
The red dots are the resulting synthetic points (the blue dots showing the residuals).
The sharpness parameters $ \Delta w_r$ shown here are the maximum values that are
compatible with the data at the 1$\sigma$ level, 
with values
$\Delta w_r=1.2$ km for the left (inner) edge and
$ \Delta w_r=1.5$ km for the right (outer) edge.
Table~\ref{tab_sharpness} lists the values of  $ \Delta w_r$ obtained with the 
other resolved C1R profiles.
}%
\label{fig_sharpness}
\end{figure}

\begin{figure}[!t]
\centering
\begin{tabular}{cc}
\includegraphics[angle=0,scale=0.3,trim=0 0 0 0]{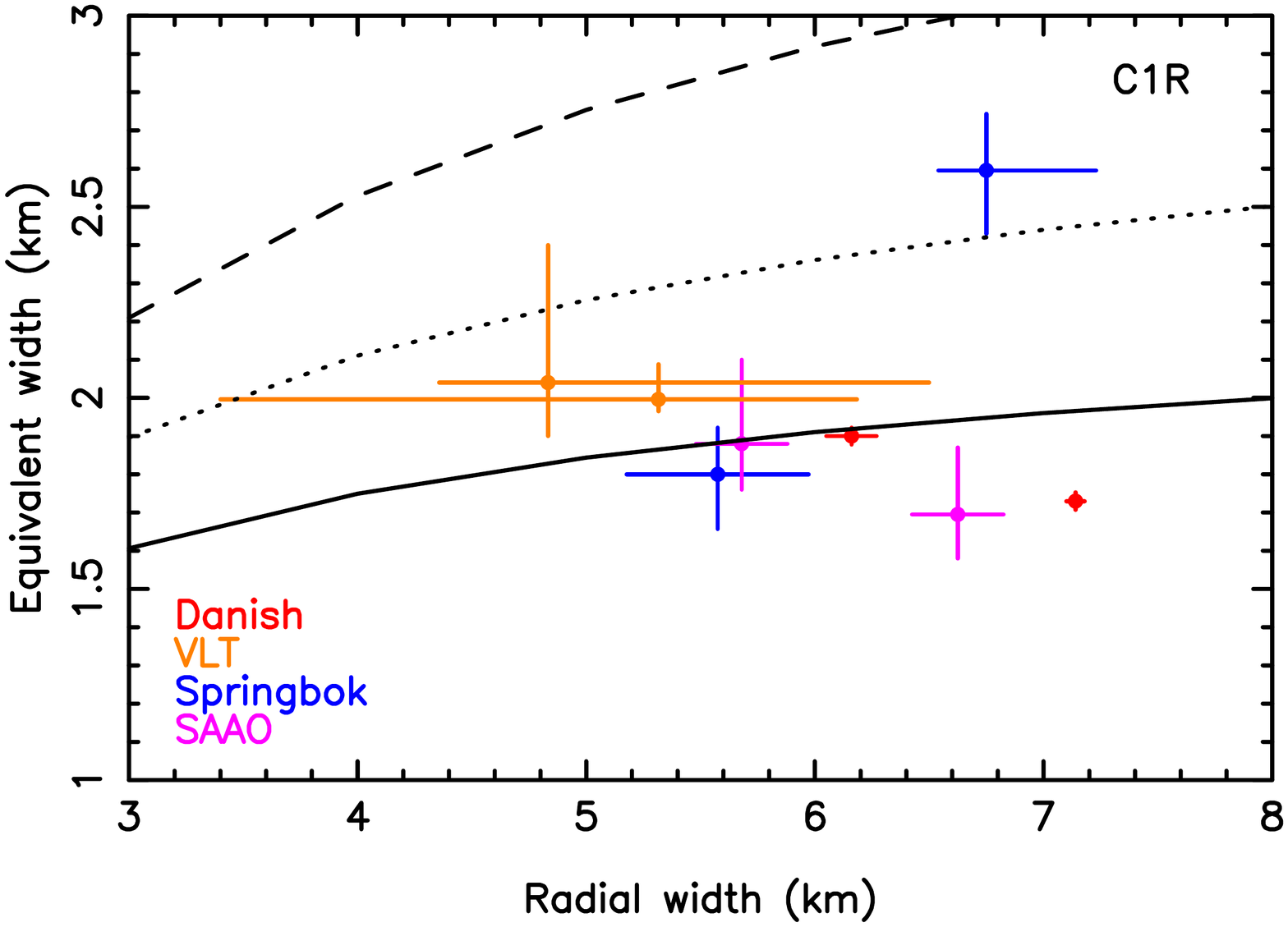} & \includegraphics[angle=0,scale=0.3,trim=0 0 0 0]{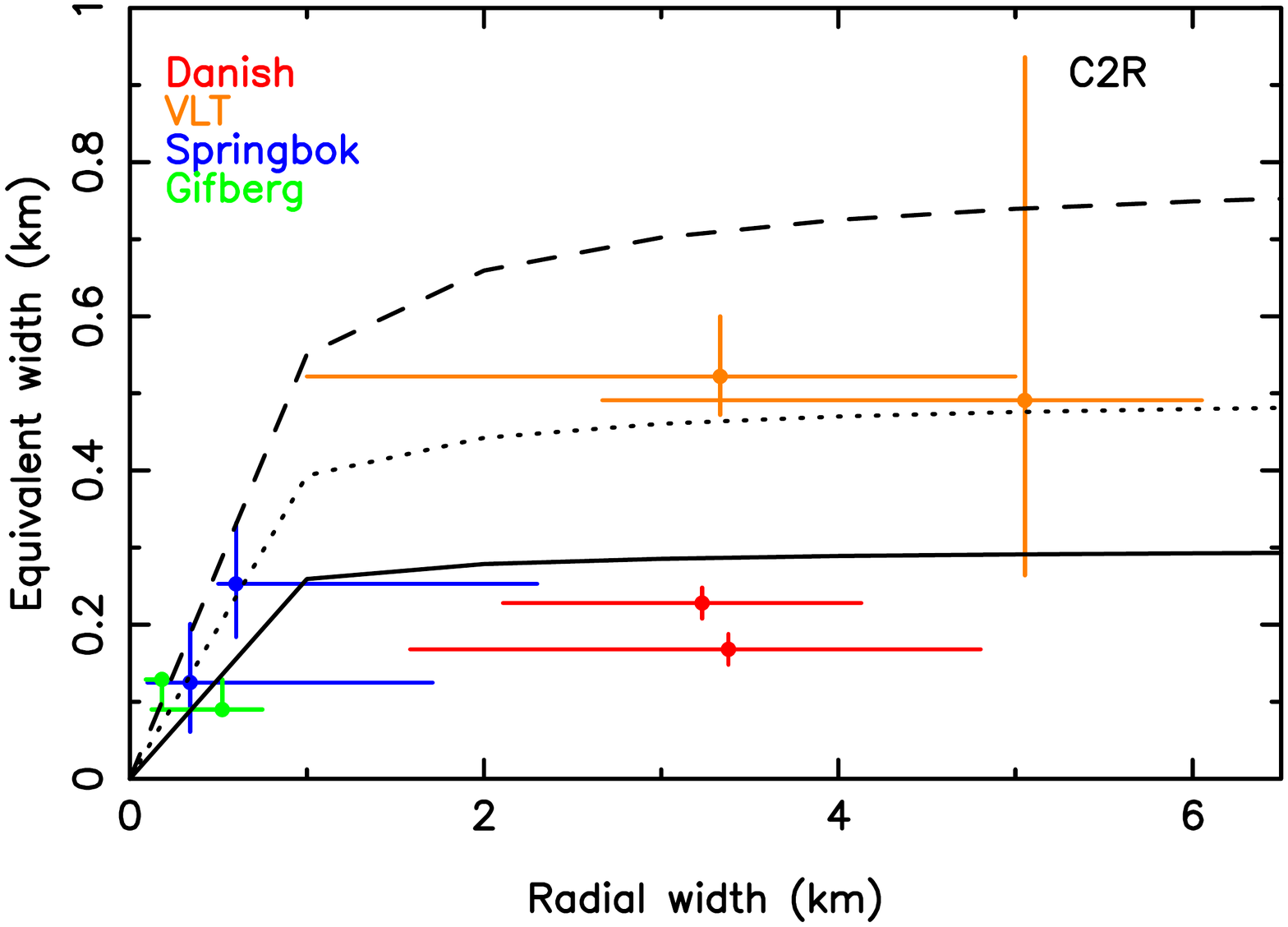}\\
\multicolumn{2}{c}{%
\includegraphics[angle=0,scale=0.3,trim=0 0 0 0]
{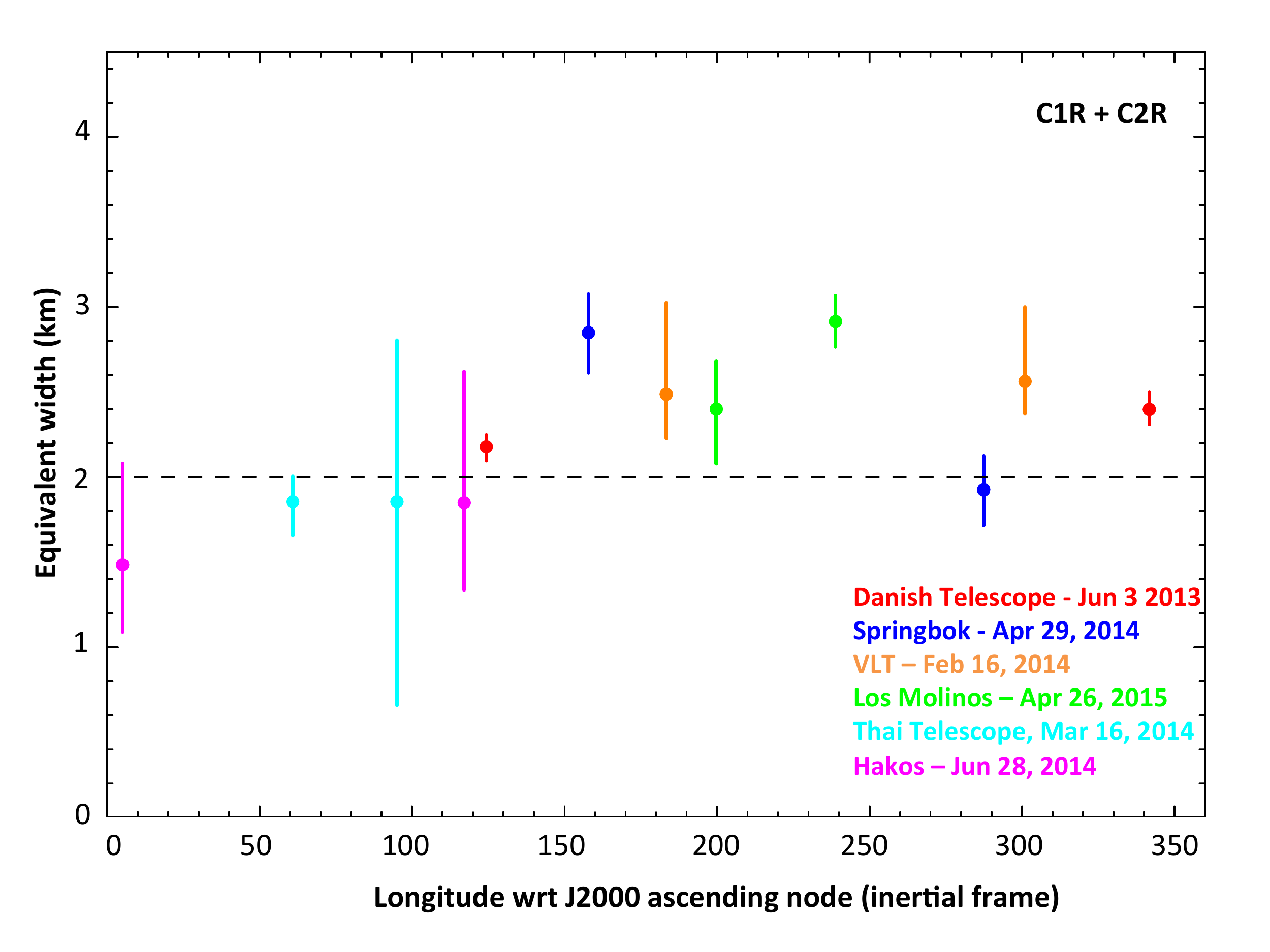}}\\

\end{tabular}
\caption{ 
\footnotesize
\textit{Top left:} 
Equivalent width $E_{p}$ (using Eq.~\ref{eq_E_square}) of C1R versus the radial width for resolved events.
For information, theoretical lines $E_p$ vs $W_r$ expected from a polylayer ring,  see Eq~\ref{eq:Ep_Wr}, have been plotted in black:
with $\bar A_{\tau}=1.15$~km (solid line), $\bar A_{\tau}=1.5$~km (dotted line), $\bar A_{\tau}=2.$~km (dashed line).
\textit{Top right}: 
Same for C2R.
The black lines are now with $\bar A_{\tau}=0.15$~km (solid line), $\bar A_{\tau}=0.25$~km (dotted line), $\bar A_{\tau}=0.40$~km (dashed line).
\textit{Bottom}: 
the integrated equivalent width $E_p(1+2)$ of C1R and C2R
versus the true longitude $L$, counted from the J2000 ring plane ascending node, 
from our best events (resolved or not).
As SAAO detected only a C1R occultation, and Gifberg only a C2R event, they have been removed from the plot.
The dashed line indicates the mean value of the data points.
}
\label{fig_E_A_C1R}
\end{figure}


\begin{figure}[!t]
\centering
\includegraphics[angle=0,scale=0.3,trim=0 0 0 0]{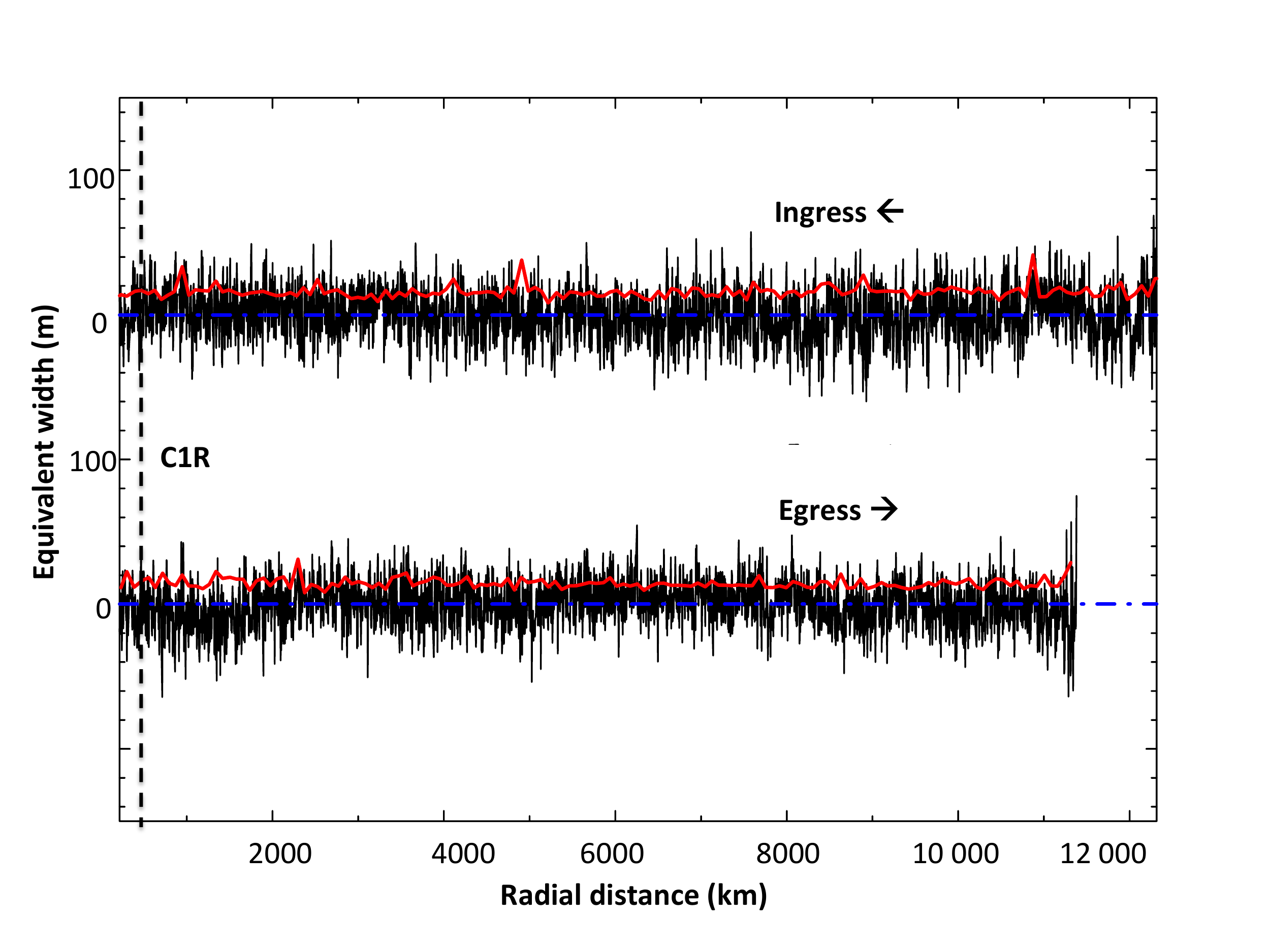} \\
\caption{%
\footnotesize
Search for faint ring material using the Danish light curve (June 3, 2013 event).
Black solid lines: 
the equivalent width $E_p$ of possible ring material (Eq.~\ref{eq_Er}) vs. 
the radial distance (in the ring plane) to Chariklo's center.
The data points corresponding to the detections of 
the main body and C1R and C2R have been removed for clarity
(for comparison $E_{p,C1R}\sim 2$~km and $E_{p,C2R}\sim 500$~m - 
see Table~\ref{tab_param_rings}).
The black vertical dotted line indicates the location of C1R, and
the horizontal dash-dotted blue lines mark the zero level for $E_p$.
Red solid lines: 
standard deviation ($1\sigma$ level) of $E_p(i)$ estimated in bins
of width 60 km, see text for details.
%
}%
\label{fig_r_Er}
\end{figure}

\end{document}